\documentclass[useAMS,usenatbib]{mn2e}

\usepackage{amssymb}
\usepackage{amsmath}
\usepackage{graphicx}
\usepackage{epstopdf}
\usepackage{verbatim}
\usepackage{latexsym}
\usepackage{theorem}
\usepackage{fixltx2e}

\numberwithin{equation}{section}

\begin{document}
\title[]{The effect of $^{12}$C + $^{12}$C rate uncertainties on the evolution and nucleosynthesis of massive stars}

\author[M. E. Bennett et al.]{M. E. Bennett$^{1,10}$\thanks{Email: meb@astro.keele.ac.uk}, R. Hirschi$^{1,2,10}$, M. Pignatari$^{3,10}$, S. Diehl$^{4,10}$, C. Fryer$^{5,10}$,
\newauthor 
F. Herwig$^{6,10}$, A. Hungerford$^{5,10}$, K. Nomoto$^7$, G. Rockefeller$^{5,10}$, F. X. Timmes$^{8,9,10}$
\newauthor 
 and M. Wiescher$^8$. \\
$^1$ Astrophysics Group, Keele University, ST5 5BG, UK \\
$^2$ IPMU, University of Tokyo, Kashiwa, Chiba 277-8582, Japan \\
$^3$ Department of Physics, Basel University, Klingelbergstrasse 82, 4056, Basel, Switzerland \\
$^4$ Theoretical Astrophysics (T-6), LANL, Los Alamos, NM, 87545, USA \\
$^5$ Computational Physics and Methods (CCS-2), LANL, Los Alamos, NM, 87545, USA \\
$^6$ Dept. of Physics \& Astronomy, Victoria, BC, V8W 3P6, Canada \\
$^7$ Institute for Physics and Mathematics of the Universe, University of Tokyo, Kashiwa, Chiba 277-8583, Japan \\
$^8$ Joint Institute for Nuclear Astrophysics, University of Notre Dame, IN, 46556, USA \\
$^9$ School of Earth and Space Exploration, University of Arizona, Tempe, AZ 85287, USA \\
$^{10}$ NuGrid collaboration}

\date{Accepted 2011 November 13.  Received 2011 November 11; in original form 2011 August 19}


\maketitle

\begin{abstract}
Over the last forty years, the $^{12}$C + $^{12}$C fusion reaction has been the subject of considerable experimental efforts to constrain uncertainties at temperatures relevant for stellar nucleosynthesis.  Recent studies have indicated that the reaction rate may be higher than that currently used in stellar models.  In order to investigate the effect of an enhanced carbon burning rate on massive star structure and nucleosynthesis, new stellar evolution models and their yields are presented exploring the impact of three different $^{12}$C + $^{12}$C reaction rates.  Non-rotating stellar models considering five different initial masses, 15, 20, 25, 32 and 60$M_{\sun}$, at solar metallicity, were generated using the Geneva Stellar Evolution Code (GENEC) and were later post-processed with the NuGrid Multi-zone Post-Processing Network tool (MPPNP).  A dynamic nuclear reaction network of $\sim 1100$ isotopes was used to track the s-process nucleosynthesis.  An enhanced $^{12}$C + $^{12}$C reaction rate causes core carbon burning to be ignited more promptly and at lower temperature.  This reduces the neutrino losses, which increases the core carbon burning lifetime.  An increased carbon burning rate also increases the upper initial mass limit for which a star exhibits a convective carbon core (rather than a radiative one).  Carbon shell burning is also affected, with fewer convective-shell episodes and convection zones that tend to be larger in mass.  Consequently, the chance of an overlap between the ashes of carbon core burning and the following carbon shell convection zones is increased, which can cause a portion of the ashes of carbon core burning to be included in the carbon shell.  Therefore, during the supernova explosion, the ejecta will be enriched by s-process nuclides synthesized from the carbon core s process.  The yields were used to estimate the weak s-process component in order to compare with the solar system abundance distribution.  The enhanced rate models were found to produce a significant proportion of Kr, Sr, Y, Zr, Mo, Ru, Pd and Cd in the weak component, which is primarily the signature of the carbon-core s process.  Consequently, it is shown that the production of isotopes in the Kr-Sr region can be used to constrain the $^{12}$C + $^{12}$C rate using the current branching ratio for $\alpha$- and p-exit channels.
\end{abstract}



\section{Introduction}\label{sec:Intro}

Despite the limitations of 1D stellar models, their capability to reproduce several observables makes them a fundamental tool to understand stellar nucleosynthesis sites in the galaxy.  Calculated stellar abundances can be compared with observed abundances from meteoritic data or stellar spectra.  In massive stars ($M > 8 M_{\sun}$) the presence of advanced burning stages during their evolution and their final fate as a supernova explosion provides a useful test-bed for many sensitivity studies, which are important to constrain uncertainties in input physics.  In particular, nuclear reaction rates are often found to be sources of uncertainty as the task of experimentally determining precise cross sections at astrophysically relevant energies is often difficult.  The $^{12}$C + $^{12}$C reaction is a good example where, despite over four decades of research, the reaction rate still carries substantial uncertainties because of the nuclear structure and reaction dynamics governing the low energy cross section of the fusion process \citep{2010JPhCS.202a2025S}. The extrapolation of the laboratory data into the stellar energy range - Gamow peak energies ($E_0 \simeq 1.5$ MeV, or $T \simeq 0.5$ GK) - depends critically on a reliable theoretical treatment of the reaction mechanism. Present model extrapolations differ by orders of magnitude; this affects directly the reaction rate with significant impact on a number of stellar burning scenarios \citep{2007PhRvC..76c5802G}.

The $^{12}$C + $^{12}$C reaction cross section is characterized complex resonance structure associated either with scattering states in the nucleon-nucleon potential or with quasimolecular states of the compound nucleus $^{24}$Mg \citep{1968PL..27B..267Im}, which at low energies can be described by a resonant-part superimposed on a non-resonant part, where the latter is also rather uncertain \citep{2010PhRvC..82d4609Y}.  A theory that predicts the location and strength of the resonant-part has not yet been proposed \citep{2008JPhG...35a4009S}, but resonance characteristics can be determined either by coupled-channel calculations or optical model potentials based on, for example, $\alpha$-particle condensates or cluster structures \citep[][and references therein]{2010PhRvC..81e4319X,1997RPPh...60..819B}.  Resonances have consequently been predicted by both approaches at energies $\sim 2$ MeV \citep{1972PhRvC...5..350M, 2006PAN....69.1372P} and it was shown that the experimentally observed data could be reasonably well reproduced in the framework of these models \citep{1978PTP...59..465}. Yet, none of these models provides the quantitative accuracy in resonance parameter predictions, required for a reliable extrapolation of the data into the stellar energy range. Complementary to the classical potential model approach, dynamic reaction theories are being developed.  They have been tested successfully for fusion of spherical nuclei like $^{16}$O + $^{16}$O \citep{2007PhLB..652..255D}, but the theoretical treatment of fusion reactions of two deformed $^{12}$C nuclei requires a non-axial symmetric formalism for a fully reliable treatment \citep{2008PhRvL.101l2501D}.

Taking a phenomenological approach a resonance with strength $(\omega \gamma) \simeq 3.4 \times 10^{-7}$ eV has been invoked to correct the ignition depth of neutron star superbursts \citep{2009ApJ...702..660C}, which are believed to be caused by ignition of carbon-burning reactions, triggering a thermonuclear runaway in the crust of a neutron star.  Type Ia supernovae should also exhibit changes to the ignition characteristics, but these conditions (other than central density) are less sensitive to an enhancement in the carbon burning rate \citep{2009ApJ...702..660C,2010A&A...512A..27I}. The possible existence of such a resonance, associated with a pronounced $^{12}$C + $^{12}$C cluster structure of the compound nucleus $^{24}$Mg, represents a source of uncertainty.

Alternatively, the reaction rate may not be dominated by resonances at lower energies because of predictions that the cross section drops much steeper than usually anticipated due to a fusion hindrance reported in heavy-ion reactions \citep[see for example, ][]{2004PhRvC..69a4604J, 2007PhRvC..75a5803J}.  The consequences of the hindrance phenomenon for the $^{12}$C+ $^{12}$C reaction in astrophysical scenarios was examined by \citet{2007PhRvC..76c5802G}, where it was demonstrated that hindrance is much more significant in the pycnonuclear regime than the thermonuclear regime, but does exhibit a noticeable effect on the yields of massive stars.  The reduced rate, by approximately a factor of 10-100 at carbon burning temperatures (see their Fig. 1), increases the temperature with which carbon burning occurs and therefore affects the nucleosynthesis.  Changes in the yields were generally rather small, but some specific isotopes, such as $^{26}$Al, $^{40}$Ca, $^{46}$Ca, $^{46}$Ti, $^{50}$Cr, $^{60}$Fe, $^{74}$Se, $^{78}$Kr and $^{84}$Sr, exhibited larger changes most likely due to the increased neutron density exhibited by the burning of neutron sources at higher temperatures.

The wide range of presently discussed model predictions requires new experimental effort to reduce the uncertainty range. However, the measurements towards low energies are extremely difficult, because the low cross section ($\sigma \ll 1$ nbarn) limits the experimental yield to an event rate below the natural and beam induced background events in the detectors.  Particle measurements are difficult because of the limited energy resolution of the particle detectors which makes a separation of the particle groups extremely difficult at the low count rate conditions. Beam induced background from reactions on target impurities is therefore difficult to distinguish from the actual reaction products \citep{2010nuco.confE..19Z}. The measurement of secondary gamma radiation associated with the particle decay is also handicapped by natural and cosmic ray induced background radiation \citep{2010JPhCS.202a2025S}.  While recent experiments suggest an increase in the low energy S-factor indicating the possibility of narrow resonances at lower energies \citep{2006NuPhA.779..318B,2006PhRvC..73f4601A,2007PhRvL..98l2501S}, the confirmation of the results and the experimental pursuit towards lower energies is stalled due to the present inability to differentiate the reaction data from the different background components \citep{2010nuco.confE..19Z}. Improved experimental conditions requires the preparation of ultra-pure target materials for experiments in an cosmic ray shielded underground environment \citep{2010JPhCS.202a2025S}.

The three dominant carbon burning reactions, with $Q$-values, are
\begin{align}
^{12}\textrm{C}(^{12}\textrm{C}&, \alpha)^{20}\textrm{Ne}, && Q = +4.617 \\
^{12}\textrm{C}(^{12}\textrm{C}&, \textrm{p})^{23}\textrm{Na}, && Q = +2.240 \\
^{12}\textrm{C}(^{12}\textrm{C}&, \textrm{n})^{23}\textrm{Mg}, && Q = -2.599.
\end{align}

During carbon-burning, the $\alpha$- and p-channels dominate with the n-channel making up less than 1 per cent of all $^{12}$C + $^{12}$C reactions \citep{1977NuPhA.279...70D}.  At this stage, the composition of the star is largely $^{12}$C and $^{16}$O, with the initial ratio of $^{12}$C to $^{16}$O at this stage largely governed by the $^{12}$C($\alpha, \gamma$)$^{16}$O reactions occurring during helium-core burning.  Carbon-core burning occurs at a central temperature $\sim 0.7$ GK and produces mainly $^{20}$Ne and $^{24}$Mg, since $\sim 99$ per cent of $^{23}$Na synthesised through the p-channel is destroyed via efficient $^{23}$Na(p, $\alpha$)$^{20}$Ne and $^{23}$Na(p, $\gamma$)$^{24}$Mg reactions \citep{1985ApJ...295..589A}.  Carbon-core burning, which is convective for stars with initial mass $M \lesssim 20 M_{\sun}$ and radiative for $M \gtrsim 20 M_{\sun}$ \citep[see for example][]{2005A&A...433.1013H}, is followed by convective carbon-shell burning episodes at temperatures $\sim 0.8 - 1.4$ GK.  The number of episodes and the spatial extent of each shell differs between massive stars of different initial mass as the development of the carbon shells is sensitive to the spatial $^{12}$C profile at the end of helium-core burning; the formation of a convective carbon-shell often lies at the same spatial coordinate as the top of the previous convective shell \citep{1972ApJ...176..699A, 2004ApJ...611..452E}.  The presence of a convective carbon core depends on the CO core mass as both the neutrino losses and energy generation rate depend on the density, which decreases with increasing CO core mass \citep{1972ApJ...176..699A,1986ARA&A..24..205W,2000ApJS..129..625L}.  Consequently, mechanisms that affect the CO core mass or the carbon burning energy budget, such as rotation \citep{2004A&A...425..649H} and the $^{12}$C abundance following helium burning \citep{2001ApJ...558..903I,2009SSRv..147....1E}, will affect the limiting mass for the presence of a convective core.

Massive stars are a site for the s process, which starts during helium-core burning and also occurs during the following carbon burning stages.  S-process nucleosynthesis also occurs in the helium-shell via the $^{22}$Ne neutron source, but this process is marginal compared to the s process operating in the helium-core or the carbon shells \citep[see for example][]{2007ApJ...655.1058T}.  Beyond carbon burning, the temperature becomes high enough in the interior ($\sim 2$ GK) for photodisintegration reactions to destroy heavy nuclides.  Because the s process can probably occur during both central and shell carbon-burning, one can expect that changes in the $^{12}$C + $^{12}$C rate affect the stellar structure and nucleosynthesis and therefore also the s process.

The $^{22}$Ne neutron source, which is formed during helium burning via the $^{14}$N($\alpha, \gamma$)$^{18}$F($\beta^+$)$^{18}$O($\alpha, \gamma$)$^{22}$Ne reaction chain is the main neutron source \citep{1968ApJ...154..225P,1974ApJ...190...95C,1977ApJ...217..213L}.  As the temperature approaches $0.25$ GK near the end of helium-burning, $^{22}$Ne($\alpha$, n)$^{25}$Mg reactions become efficient \citep{1985A&A...151..205B,1991ApJ...367..228R}.  During this phase a $25 M_{\sun}$ star, for example, has a neutron density $n_n \sim 10^6$ cm$^{-3}$ and a neutron exposure $\tau_n \sim 0.2$ mb$^{-1}$ \citep[see for instance][and references therein]{2010ApJ...710.1557P}.  The $^{22}$Ne source becomes efficient in a convective environment and heavy elements formed through neutron captures are mixed out from the centre of the star.  Some of these abundances will be modified by further explosive nucleosynthesis later in the evolution, but will otherwise survive long enough to be present in the supernova ejecta and contribute to the total yields of the star.  Consequently, $^{22}$Ne in massive stars is the dominant neutron source responsible for the classical weak-s-process component \citep{1977ApJ...216..797T, 1987ApJ...315..209P, 1989RPPh...52..945K, 1991ApJ...371..665R}.

Any remaining $^{22}$Ne present at the end of helium-core burning is later reignited during carbon-shell burning resulting in an s-process with a higher neutron density and a lower neutron exposure \citep[$n_n \sim 10^{11-12}$ cm$^{-3}$ and $\tau_n \sim 0.06$ mb$^{-1}$;][]{1991ApJ...371..665R}.  The increased neutron density is responsible for changing the branching ratios of unstable isotopes, which is particularly important for branching isotopes, such as $^{69}$Zn, $^{79}$Se and $^{85}$Kr, since they inhabit positions in the isotope chart of nuclides where different s-process paths across the valley of stability are available \citep{1989RPPh...52..945K}.  The increase in neutron density is responsible for opening the s-process path so that the carbon-shell burning contribution to specific isotopes, such as $^{70}$Zn, $^{86}$Kr and $^{80}$Se, may be relevant \citep[see for example][]{1991ApJ...371..665R,2007ApJ...655.1058T}.

Another potential neutron source is $^{13}$C, which is formed through the $^{12}$C(p,$\gamma$)$^{13}$N($\beta^+$)$^{13}$C reaction chain \citep{1969ApJ...157..339A}.  During carbon-core burning this neutron source, via the $^{13}$C($\alpha$,n)$^{16}$O reaction, becomes efficient which results in an s-process in the carbon-core with a typical neutron density $n_n = 10^7$ cm$^{-3}$ \citep{1985ApJ...295..589A, 1998ApJ...502..737C}.  The abundance of $^{13}$C is dependent on the $^{13}$N($\gamma$,p)$^{12}$C reaction, which dominates the depletion of $^{13}$N at temperatures above $0.8$ GK.  The $^{22}$Ne neutron source is the dominant neutron source when the temperature rises above such a temperature, although the $^{13}$C neutron source may also provide an important contribution to the total neutron exposure \citep{1968psen.book.....C, 1991A&A...249..134A}.  In any case, the carbon-core s process occurs primarily in radiative conditions with a relatively small neutron exposure and any heavy elements synthesised via the ensuing neutron-captures usually remain in the core (see however the discussion on overlapping convection zones in \S \ref{sec:Nucleosynthesis}); photodisintegration and the supernova explosion process will ensure that these elements are not present in the final ejecta and do not contribute to the final yields of the star \citep[see for example,][]{1998ApJ...502..737C}.

A preliminary study (\citealt{2010JPhCS.202a2023B}a) found that changes to the total $^{12}$C + $^{12}$C rates within a factor of 10 affect the convection zone structure and nucleosynthesis of a 25 $M_{\sun}$ star at solar metallicity.  The main conclusions were an increase in the carbon-burning shell contribution to the s-process abundances by two different scenarios.  The first, applicable to the case where the rate was increased by a factor of 10, was due to the presence of large carbon-burning shells that `overlapped'.  In this situation, the second carbon-burning shell was polluted with ashes from the first carbon-burning shell, modifying the overall composition.  The second scenario, applicable to the case where the rate was reduced by a factor of 10, was an increase in neutron density associated with the neutron source, $^{22}$Ne, burning at a higher temperature in the convective shell.  The overall increase in the abundances of most isotopes with $60 < A < 90$ was approximately $0.1$ to $0.4$ dex.  Strongly enhanced rates were also investigated (\citealt{2010arXiv1012.3258B}b), which show that the presence of a larger convective core has a significant impact on the total yields, since the convective core adds an additional neutron exposure towards the total contribution of s-process yields; abundances of many heavy nuclides increased by up to $\sim 2$ dex.  However, no comparison could be made with observations as a 25 $M_{\sun}$ stellar model (at solar metallicity) was the only one considered.

In this paper, a sensitivity study is made over a set of massive star models, at solar metallicity, to determine whether a comparison between the yields and the solar system abundances can constrain the $^{12}$C + $^{12}$C rate.  \S \ref{sec:Comp_approach} explains the models and the choice of input physics in the simulations.  In \S \ref{sec:Stellarstructure}, the changes in stellar structure are analysed.  \S \ref{sec:Nucleosynthesis} describes the nucleosynthesis, focusing on the s process during carbon-core and carbon-shell burning.  \S \ref{sec:Yields} presents the yields.  The discussion and conclusions can be found in \S \ref{sec:Discussion} and \S \ref{sec:Conclusions} respectively.


\section{Computational Approach}\label{sec:Comp_approach}

\subsection{The $^{12}$C + $^{12}$C reaction rates}\label{sec:c12rate}

\begin{figure}
\includegraphics[width=0.5\textwidth]{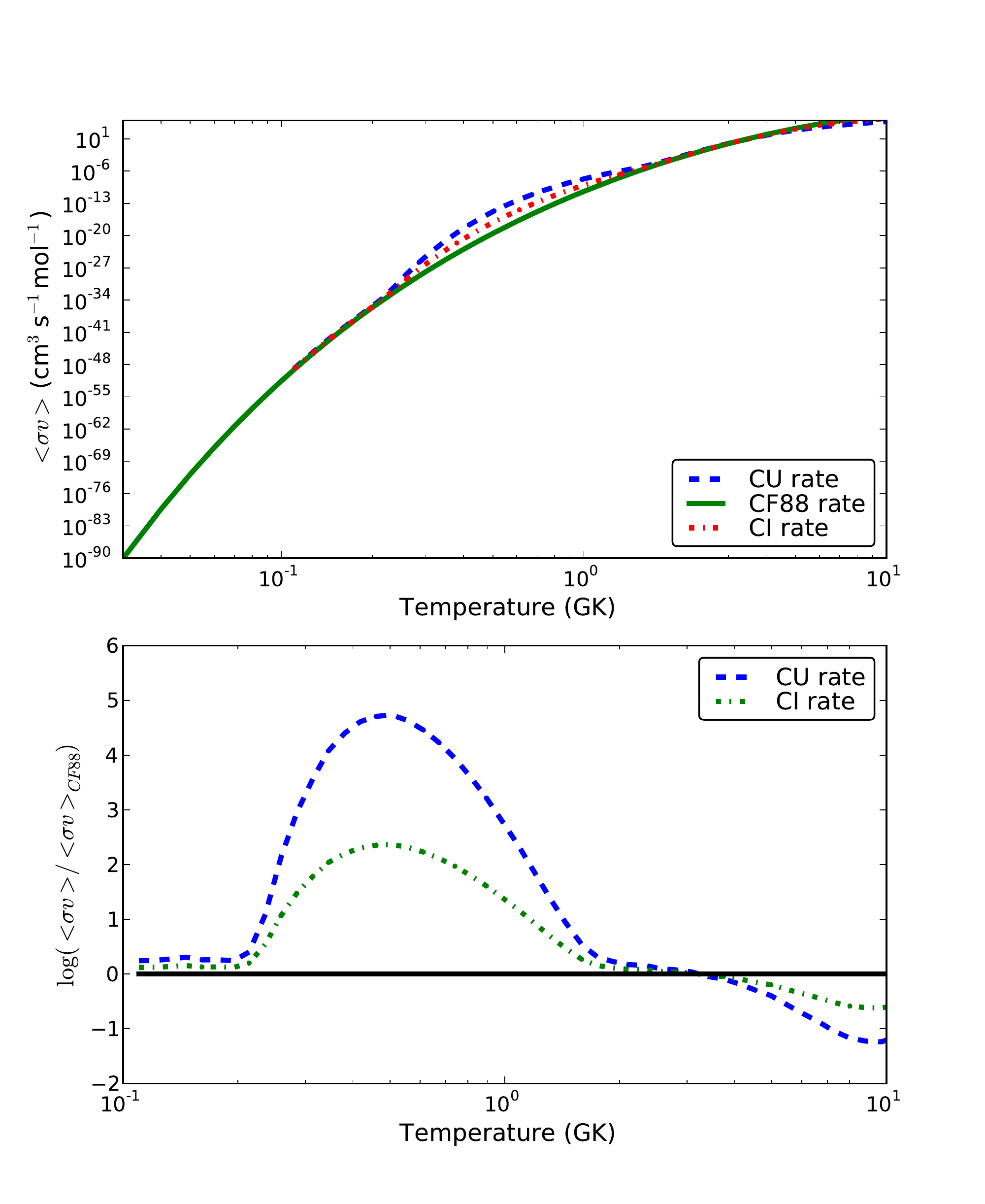}
\caption{\emph{Top panel:} Maxwellian-averaged cross-sections for $^{12}$C + $^{12}$C rates used in (\citealt{2010arXiv1012.3258B}b) and also in this study.  The three rates are the Caughlan \& Fowler (1988) `standard' rate (ST), an upper limit rate (CU) and an intermediate rate (CI).  The CI rate is a geometric mean of the ST and CU rates.  \emph{Bottom panel:} The Maxwellian-averaged cross-sections relative to the ST rate.}
\label{fig:C12rates}
\end{figure}

We build on the previous work (\citealt{2010arXiv1012.3258B}b) where three carbon burning rates in a $25 M_{\sun}$ star were considered.  These are the \citet{1988ADNDT..40..283C} `standard' rate (ST) and two enhanced rates: an `upper limit' rate (CU) and an intermediate rate (CI), the latter of which is a geometric mean of the ST and CU rates.  The CU rate is the ST rate including a reasonance of strength $(\omega \gamma) = 6.8 \times 10^{-5}$ eV at a centre-of-mass energy $E_{\rm com} = 1.5$ MeV.  This choice of resonance originates from a preliminary particle spectroscopy experiment on $^{12}$C + $^{12}$C obtained at the CIRCE radioactive beam facility in Caserta/Napoli, Italy \citep{2007NIMPB.259...14T}.  Although the CI rate was determined via a geometric mean, a resonance that would replicate the peak at $1.5$ MeV for this rate would have a magnitude of $(\omega \gamma) \simeq 3.4 \times 10^{-7}$ eV.  The top panel of Fig. \ref{fig:C12rates} shows the Maxwellian-averaged cross-sections of the reaction rates as a function of temperature.  The bottom panel shows the reaction rates relative to the ST rate.  As indicated by Fig. \ref{fig:C12rates}, the peak of the CU and CI rates is at $\sim 0.5$ GK and is a factor of approximately $50,000$ and $250$ times the ST rate at that temperature respectively.  The choice of branching ratio for the $\alpha$- and p-exit channels is 13:7, which is valid within the energy range $4.42 < E_{\rm com} < 6.48$ MeV \citep{2006PhRvC..73f4601A}.  It is assumed in this work that the branching ratio is preserved to lower centre of mass energies.  For the n-exit channel, we use the branching ratio from \citet{1977NuPhA.279...70D}.

\subsection{Stellar models}\label{sec:models}

Non-rotating stellar models at solar metallicity (Z=0.02) were generated using the Geneva Stellar Evolution Code (GENEC), with a small nuclear reaction network that takes into account the reactions important for energy generation.  Five masses were considered for each carbon-burning rate, which are 15, 20, 25, 32 and 60 $M_{\sun}$, for a total of 15 stellar models.  These will be referred to as XXYY where XX is the initial mass of the star in solar masses and YY denotes the rate and is `ST', `CI' or `CU' for the standard, intermediate and upper limit rates respectively.  The reason for this choice of initial masses is to provide yields data over a range of masses with approximately even spacing in log-space.

GENEC is described in detail in \citet{2008ApSS.316...43E}, but some important features are recalled here for convenience.  The Schwarzschild criterion for convection is used and convective mixing is treated as a diffusive process from oxygen burning onwards.  No overshooting is included except for hydrogen- and helium-burning cores, where an overshooting parameter of $\alpha = 0.2 H_P$ is used.  Neutrino loss rates are calculated using fitting formulae from \citet{1989ApJ...339..354I}, which are the same as those of the more recent evaluation from \citet{1996ApJS..102..411I} for pair and photoneutrino processes.  The initial abundances used were those of \citet{1993oee..conf...15G}, which correspond directly to the OPAL opacity tables used \citep{1996ApJ...456..902R}.  For lower temperatures, opacities from \citet{2005ApJ...623..585F} are used.

Several mass loss rates are used depending on the effective temperature, $T_{\rm eff}$, and the evolutionary stage of the star.  For main sequence massive stars, where $\log T_{\rm eff} > 3.9$, mass loss rates are taken from \citet{2001A&A...369..574V}.  Otherwise the rates are taken from \citet{1988A&AS...72..259D}.  However, for lower temperatures ($\log T_{\rm eff} < 3.7$), a scaling law of the form

\begin{equation}
\dot{M} = - 1.479 \times 10^{-14} \times \left(\frac{L}{L_{\sun}}\right)^{1.7}
\end{equation}
is used, where $\dot{M}$ is the mass loss rate in solar masses per year, $L$ is the total luminosity and $L_{\sun}$ is the solar luminosity.  For a recent discussion on mass loss rates in the red-supergiant phase, see \citet{2011A&A...526A.156M}.  During the Wolf-Rayet (WR) phase, mass loss rates by \citet{2000A&A...360..227N} are used.  

In GENEC the reaction rates are chosen to be those of the NACRE compilation; \citet{1999NuPhA.656....3A} for the experimental rates and from their website\footnote{http://pntpm3.ulb.ac.be/Nacre/nacre.htm} for theoretical rates.  However, there are a few exceptions.  The rate of \citet{2003PhRvC..67f5804M} was used for $^{14}$N(p, $\gamma$)$^{15}$O below $0.1$ GK and the lower limit NACRE rate was used for temperatures above $0.1$ GK.  This combined rate is very similar to the more recent LUNA rate \citep{2005EPJA...25..455I} at relevant temperatures.  The \citet{2005Natur.433..136F} rate was used for the $3\alpha$ reaction and the \citet{2002ApJ...567..643K} rate was used for $^{12}$C($\alpha,\gamma$)$^{16}$O.  The $^{22}$Ne($\alpha$,n)$^{25}$Mg rate was taken from \citet{2001PhRvL..87t2501J} and used for the available temperature range ($T \leq 1$ GK).  Above this range, the NACRE rate was used.  The $^{22}$Ne($\alpha$,n)$^{25}$Mg rate competes with $^{22}$Ne($\alpha,\gamma$)$^{26}$Mg for $\alpha-$particles.  For this rate, the NACRE rate was used.  The $^{16}$O neutron poison is effective at capturing neutrons, forming $^{17}$O, which can either resupply the `recycled' neutrons via the $^{17}$O($\alpha$,n)$^{20}$Ne reaction or undergo the competing reaction $^{17}$O($\alpha,\gamma$)$^{21}$Ne.  For $^{17}$O($\alpha$, n)$^{20}$Ne the NACRE reaction is used and for the $^{17}$O($\alpha,\gamma$)$^{21}$Ne reaction the correction of the \citet{1988ADNDT..40..283C} rate by \citet{1993PhRvC..48.2746D} is applied.

The models were calculated for as far into the evolution as possible, which for most models is after or during the silicon-burning stage.  The models that ceased before silicon burning were the 15CI, 15CU, 60CI and 60CU models, which proceeded to oxygen-shell burning, and the 20CI and 20CU models, which proceeded to just after the oxygen-shell burning stage.  The s-process yields are not significantly affected by hydrostatic burning stages following oxygen burning because most of the isotopes produced via the s process will be destroyed by photodisintegration and the choice of remnant mass for the supernova explosion, which defines the boundary between matter that falls back onto the remnant and matter that forms supernova ejecta, reduces the impact of nucleosynthesis that neon, oxygen and silicon burning stages would have on the total yields (see also \S \ref{sec:calcs}).  However, it must be noted that there will be explosive burning processes during the supernova explosion and photodisintegration occurring at the bottom of the convective carbon, neon and oxygen shells during the advanced stages, which will affect the abundances \citep[see for example][]{2002ApJ...576..323R, 2009ApJ...702.1068T}.  In this work the contribution of explosive burning and photodisintegration to the total yields is not considered.

Since the $^{12}$C + $^{12}$C reactions do not become efficient until after helium-core burning, the CU and CI models for a particular choice of initial mass were started just before the end of helium-core burning using the ST model data as initial conditions, reducing some of the computational expense.

\subsection{Post-processing}

The NuGrid\footnote{http://forum.astro.keele.ac.uk:8080/nugrid} Multi-Zone Post-Processing tool (the parallel variant; MPPNP) is described in \citet{2008nuco.confE..23H} and Pignatari et al. (2011, in prep.). See also appendix \ref{sec:parallel} for details of the parallel implementation.  The system of equations for the rate of change of abundances of isotopes is solved using an implicit finite differencing method combined with the Newton-Raphson scheme, with the output temperature, density and the distribution of convection (and radiation) zones from GENEC as input.   Additional features have been included to enhance the calculations or save on unnecessary computations.  Sub-timesteps are inserted where appropriate to improve convergence in the case where the timescale of reactions is smaller than the stellar evolution timestep.  Also, the nuclear network is dynamic, adding or removing isotopes from the network depending on the stellar conditions (up to the maximal network defined in Table \ref{tab:MPPNPnet}).  This is useful in reducing the number of computations associated with nuclear reactions where the change in abundance is zero or negligible.  The same (adaptive) mesh used in GENEC was used for the post-processing calculations.

\begin{figure*}
\includegraphics[width=\textwidth]{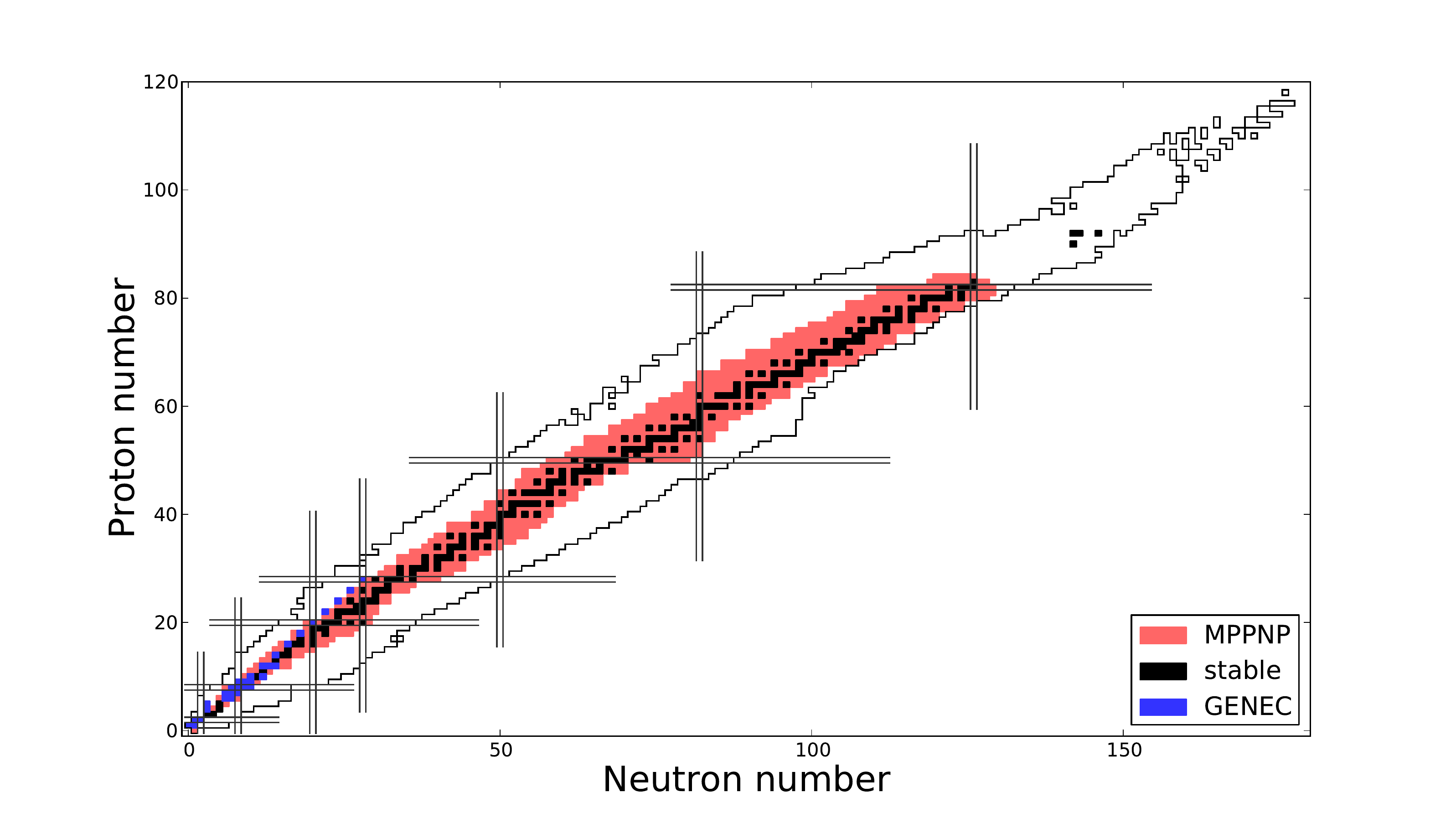}
\caption{Chart of isotopes indicating the nuclear reaction networks used in this work: GENEC (blue squares) and MPPNP (pale red squares).  The network used by MPPNP includes all stable isotopes, which are indicated by black squares.  The outer boundary to each side of the valley of stability indicates the position of all currently known isotopes, including heavy transuranic isotopes.  Parallel grid lines indicate values of $Z$ or $N$ that are magic as specified in the nuclear shell model (2,8,20,28,50,82,126).}
\label{fig:segrechart}
\end{figure*}

The nuclear networks used are shown in Fig. \ref{fig:segrechart}.  The isotopes used in each network are discriminated depending on whether they are involved in reactions important for energy generation (featured in both the stellar model and the post-processing tool) or not (featured only in the post-processing tool).  GENEC uses a skeleton network of 31 isotopes, which is the same network used in previous GENEC models \citep[see for example][]{2004A&A...425..649H, 2005A&A...433.1013H}.  This network is a combination of fundamental isotopes relevant for pp-chain reactions, the CNO tricycle and helium burning and a network similar to the $\alpha 7$ network of \citet{1998ApJ...503..332H}, enacted during the advanced burning stages, which reduces the computational expense associated with a larger network without causing significant errors in energy generation rates.  The isotopes included in the network for MPPNP are specified in Table \ref{tab:MPPNPnet} and are shown in Fig. \ref{fig:segrechart}.  Five isomeric states are also included, which are treated as separate nuclei from their ground state equivalents.  These are $^{26}$Al$^m$, $^{85}$Kr$^m$, $^{115}$Cd$^m$, $^{176}$Lu$^m$ and $^{180}$Ta$^m$

\begin{table}
\centering
\caption{Nuclides included in the nuclear reaction network used for the post-processing calculations}
\begin{tabular}{lrrlrr}
\hline
\hline
Element & $A_{min}$ & $A_{max}$ & Element & $A_{min}$ & $A_{max}$\\
\hline
n  & 1  & 1   & Tc & 93  & 105 \\
H  & 1  & 2   & Ru & 94  & 106 \\
He & 3  & 4   & Rh & 98  & 108 \\
Li & 7  & 7   & Pd & 99  & 112 \\
Be & 7  & 8   & Ag & 101 & 113 \\
B* & 8  & 11  & Cd & 102 & 118 \\
C  & 11 & 14  & In & 106 & 119 \\
N  & 13 & 15  & Sn & 108 & 130 \\
O  & 14 & 18  & Sb & 112 & 133 \\
F  & 17 & 20  & Te & 114 & 134 \\
Ne & 19 & 22  & I  & 117 & 135 \\
Na & 21 & 24  & Xe & 118 & 138 \\
Mg & 23 & 28  & Cs & 123 & 139 \\
Al & 25 & 29  & Ba & 124 & 142 \\
Si & 27 & 32  & La & 127 & 143 \\
P  & 29 & 35  & Ce & 130 & 146 \\
S  & 31 & 38  & Pr & 133 & 149 \\
Cl & 34 & 40  & Nd & 134 & 152 \\
Ar & 35 & 44  & Pm & 137 & 154 \\
K  & 38 & 46  & Sm & 140 & 158 \\
Ca & 39 & 49  & Eu & 143 & 159 \\
Sc & 43 & 50  & Gd & 144 & 162 \\
Ti & 44 & 52  & Tb & 147 & 165 \\
V  & 47 & 53  & Dy & 148 & 168 \\
Cr & 48 & 56  & Ho & 153 & 169 \\
Mn & 51 & 57  & Er & 154 & 175 \\
Fe & 52 & 61  & Tm & 159 & 176 \\
Co & 55 & 63  & Yb & 160 & 180 \\
Ni & 56 & 68  & Lu & 165 & 182 \\
Cu & 60 & 71  & Hf & 166 & 185 \\
Zn & 62 & 74  & Ta & 169 & 186 \\
Ga & 65 & 75  & W  & 172 & 190 \\
Ge & 66 & 78  & Re & 175 & 191 \\
As & 69 & 81  & Os & 179 & 196 \\
Se & 72 & 84  & Ir & 181 & 197 \\
Br & 74 & 87  & Pt & 184 & 202 \\
Kr & 76 & 90  & Au & 185 & 203 \\
Rb & 79 & 91  & Hg & 189 & 208 \\
Sr & 80 & 94  & Tl & 192 & 210 \\
Y  & 85 & 96  & Pb & 193 & 211 \\
Zr & 86 & 98  & Bi & 202 & 211 \\
Nb & 89 & 99  & Po & 204 & 210 \\
Mo & 90 & 102 &    &     &     \\
\hline
\end{tabular}
\medskip

$^{*}$ $^{9}$B is not included.
\label{tab:MPPNPnet}
\end{table}

The reaction rates in MPPNP were set to those used in the skeleton network of GENEC, as specified in \S \ref{sec:models}, for the same reactions.  Additional reactions are taken from the default setup of MPPNP and are specified as follows: charged particle reactions are from \citet{1999NuPhA.656....3A} and \citet{2001ApJS..134..151I}.  $\beta$-decays and electron captures are from \citet{1994ADNDT..56..231O}, \citet{1985ApJ...293....1F} and \citet{2005A&A...441.1195A}.  Neutron captures are from the Karlsruhe astrophysical database of nucleosynthesis in stars (KADoNiS) \citep{2006AIPC..819..123D}.  For reactions not found in these references, reaction rates from the Reaclib database \footnote{http://nucastro.org/reaclib.html} were used, which incorporates a compilation of experimental rates and theoretical rates from NON-SMOKER \citep{2000ADNDT..75....1R,2001ADNDT..79...47R}.


\section{Stellar structure and evolution}\label{sec:Stellarstructure}

\subsection{Hydrogen and helium burning}\label{sec:hydroheliumevol}

\begin{figure*}
\includegraphics[width=\textwidth]{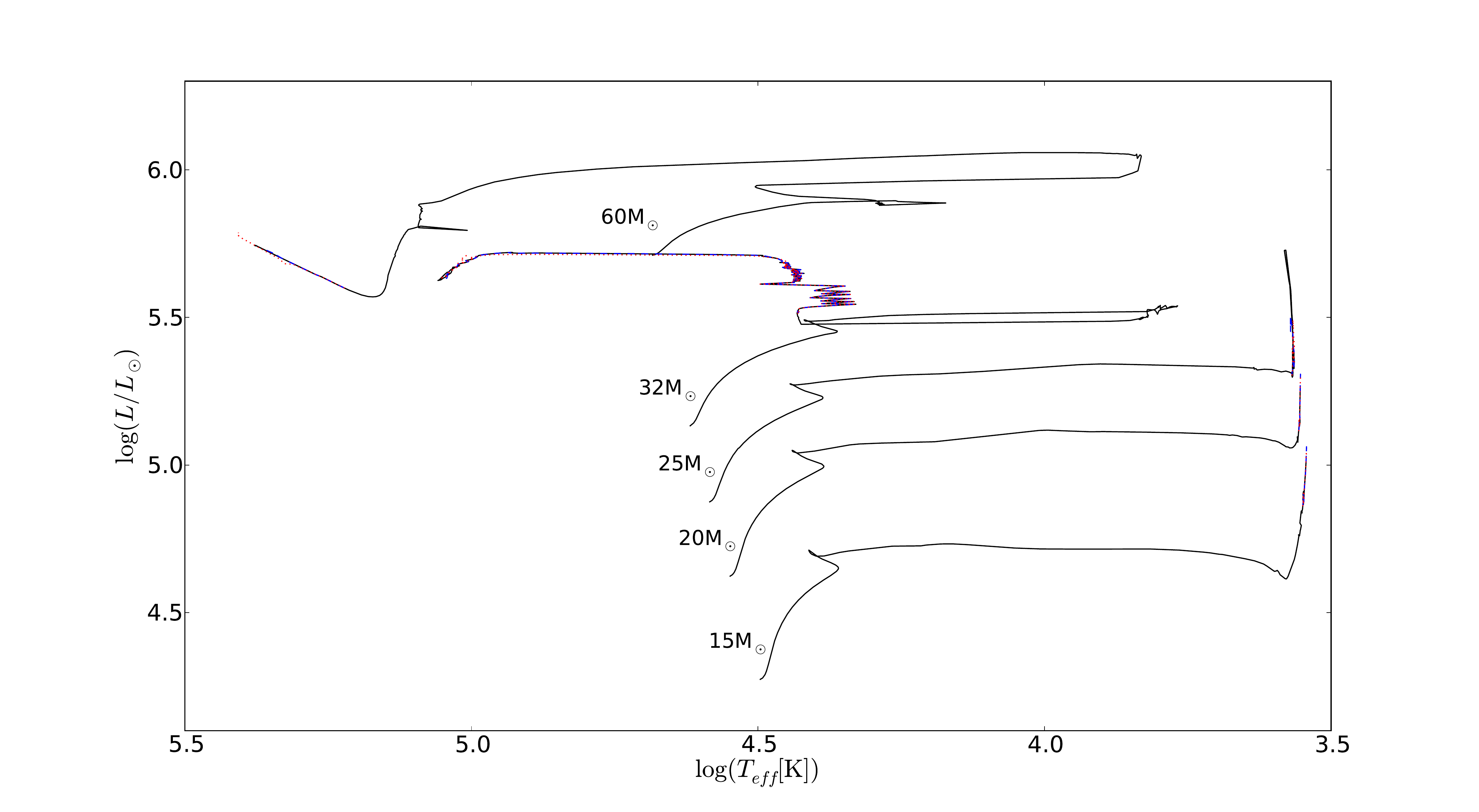}
\caption{Hertzsprung-Russell diagram for all models.  Solid black lines refer to ST model tracks, dashed blue lines refer to CI model tracks and the dotted red lines refer to the CU model tracks.  The tracks indicate that the enhanced rates do not affect the surface evolution, since changes in the carbon-burning rate do not affect the surface properties.  The tracks exhibited by the 32 $M_{\sun}$ and 60 $M_{\sun}$ models show evolution into the WR phase, which is explained by mass loss.}
\label{fig:HR}
\end{figure*}

The evolution of each stellar model during hydrogen- and helium-burning is given entirely by the ST models, as the CI and CU models were started using the profile just before the end of helium burning.  Figure \ref{fig:HR} shows the Hertzsprung-Russell (HR) diagram for all models, which shows that the evolutionary tracks for all models follow their course in the HR diagram primarily during the hydrogen- and helium-burning phases and are not modified by enhanced rates.  The reason for this is that the surface evolution of the stellar models is unaffected by changes in the carbon-burning rate, which is a consequence of the small timescale for burning associated with advanced burning stages in massive stars; the envelope has insufficient time to react significantly to changes in core properties.

Overall, the ST models are very similar to those previously published by the Geneva group, such as the non-rotating stars of \citet{2003A&A...404..975M} and \citet{2004A&A...425..649H}.  The 15, 20 and 25 $M_{\sun}$ model stars evolve towards the red and remain as red supergiants (RSGs) during the advanced stages of evolution.  The 32 and 60 $M_{\sun}$ model stars evolve towards the Humphreys-Davidson limit at $\log T_{\rm eff} \sim 3.8$ before becoming WR stars

The 32 $M_{\sun}$ proceeds to the WR phase during helium-burning.  This is because the mass loss is strong enough for the star to expel the entire hydrogen envelope during helium-burning, with the composition of the remaining envelope rich in helium.  The lower opacity of the helium-rich envelope lowers the radius and favours evolution towards the blue \citep[\S 27.3.2]{2009pfer.book.....M}.  The deviations from the ST track for the CI or CU tracks for this mass are slightly larger than for other masses.  These deviations are generally of the order of $0.1$ per cent with a maximum deviation of $0.01$ in $\log T_{\mathrm{eff}}$ ($\simeq 2$ per cent), which occurs during the rapid transit to the blue after helium burning.

The 60 $M_{\sun}$ star becomes a WR star just after hydrogen-burning.  At the end of the hydrogen-burning phase, the star enters the first `loop' towards the blue (at $\log T_{\mathrm{eff}} \simeq 4.4$), which occurs because of mass loss being high enough to expose the helium-rich outer layer.  Following the first loop to the blue, helium-burning is ignited.  During this phase the core shrinks, lowering the core fraction, $q$, favouring evolution to the red \citep[\S 27.3.2]{2009pfer.book.....M}.  However, the star approaches the Humphreys-Davidson limit in the HR diagram during the evolution and the mass loss becomes high enough to, eventually, peel away the envelope, exposing the helium-burning core ($q \simeq 75$ per cent during helium-burning).  The star consequently evolves towards the blue (at $\log T_{\mathrm{eff}} \simeq 5.0$).

\subsection{Carbon burning}\label{sec:carbonevol}

Unlike the surface evolution, the interior evolution of the star is modified significantly by the enhanced carbon burning rates and changes to the central evolution of the star are important in order to assess changes to the main burning regimes. 

\begin{figure}
\includegraphics[width=0.5\textwidth]{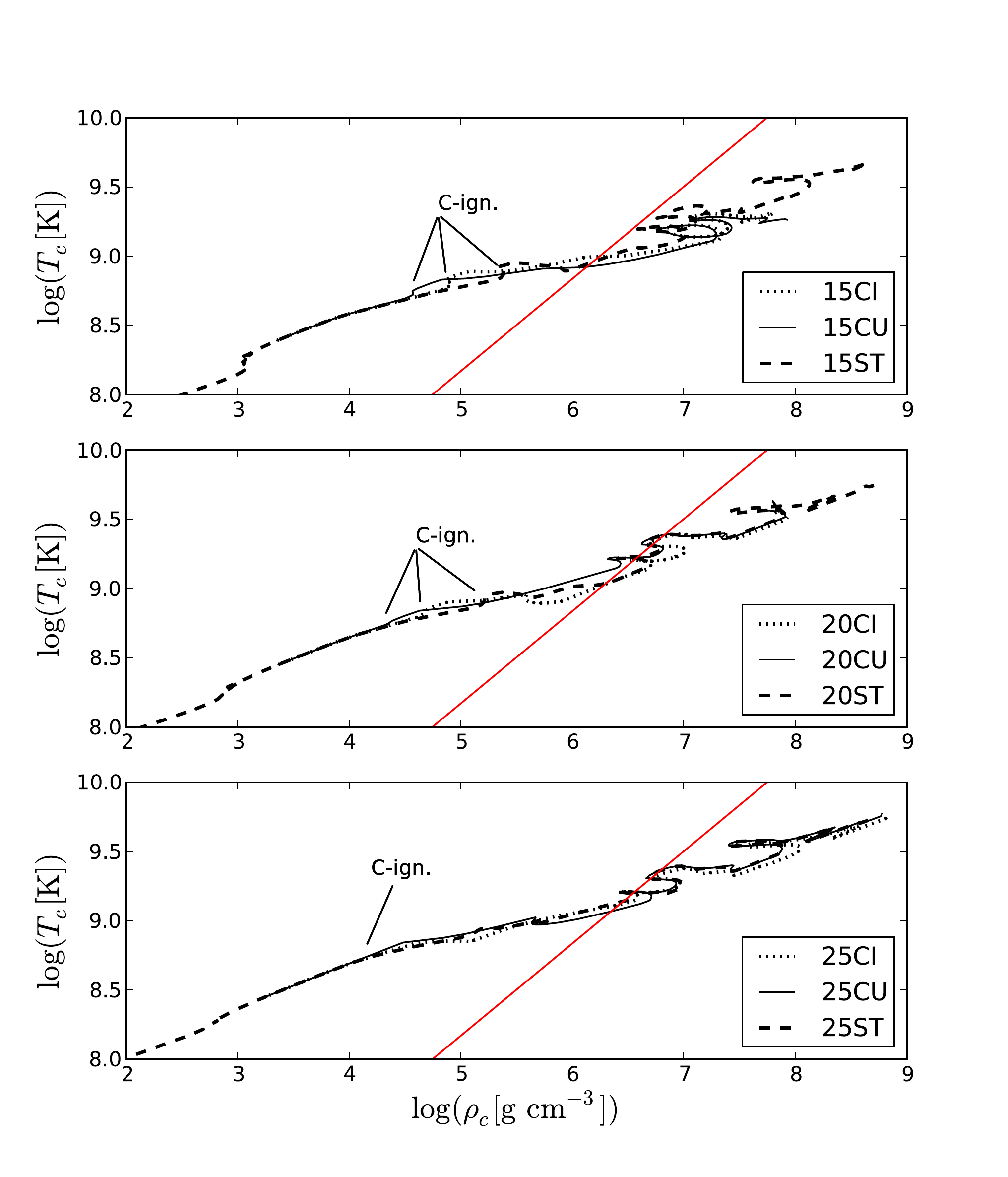}
\caption{$T_c-\rho_c$ diagram for all 15 (top panel), 20 (middle panel) and 25 $M_{\sun}$ (bottom panel) models.  The straight line in each panel indicates the location in the diagram where the ideal gas pressure is equal to the electron degeneracy pressure; $P_{gas} = P_{e,deg}$.  Ignition points for convective core carbon burning are indicated by the annotation.}
\label{fig:tcrc152025}
\end{figure}

Figure \ref{fig:tcrc152025} shows $T_c-\rho_c$ diagrams for the 15, 20 and 25 $M_{\sun}$ models, separated into panels by initial mass.  The enhanced rate models in all cases (including the 32 and 60 $M_{\sun}$ models) ignite carbon burning at lower temperatures and densities, which consequently affects the evolution of the central properties of the star.  This is seen, for example, in the top and middle panels of Fig. \ref{fig:tcrc152025},  where the curves for the CI and CU cases deviate away from that of the ST case towards the higher temperature (at a given density) side of the curve (see also column 7. in Table \ref{tab:Cprops}).  The tendency to deviate in this direction is caused by the presence of a convective core.  This is verified in the bottom panel for the case of the 25CU model whereby the `kink' at carbon ignition is larger than that of the 25ST and 25CI models, since the CU model is the only 25 $M_{\sun}$ model to have a convective core (see also Fig. \ref{fig:kip2532}).

\begin{figure}
\includegraphics[width=0.5\textwidth]{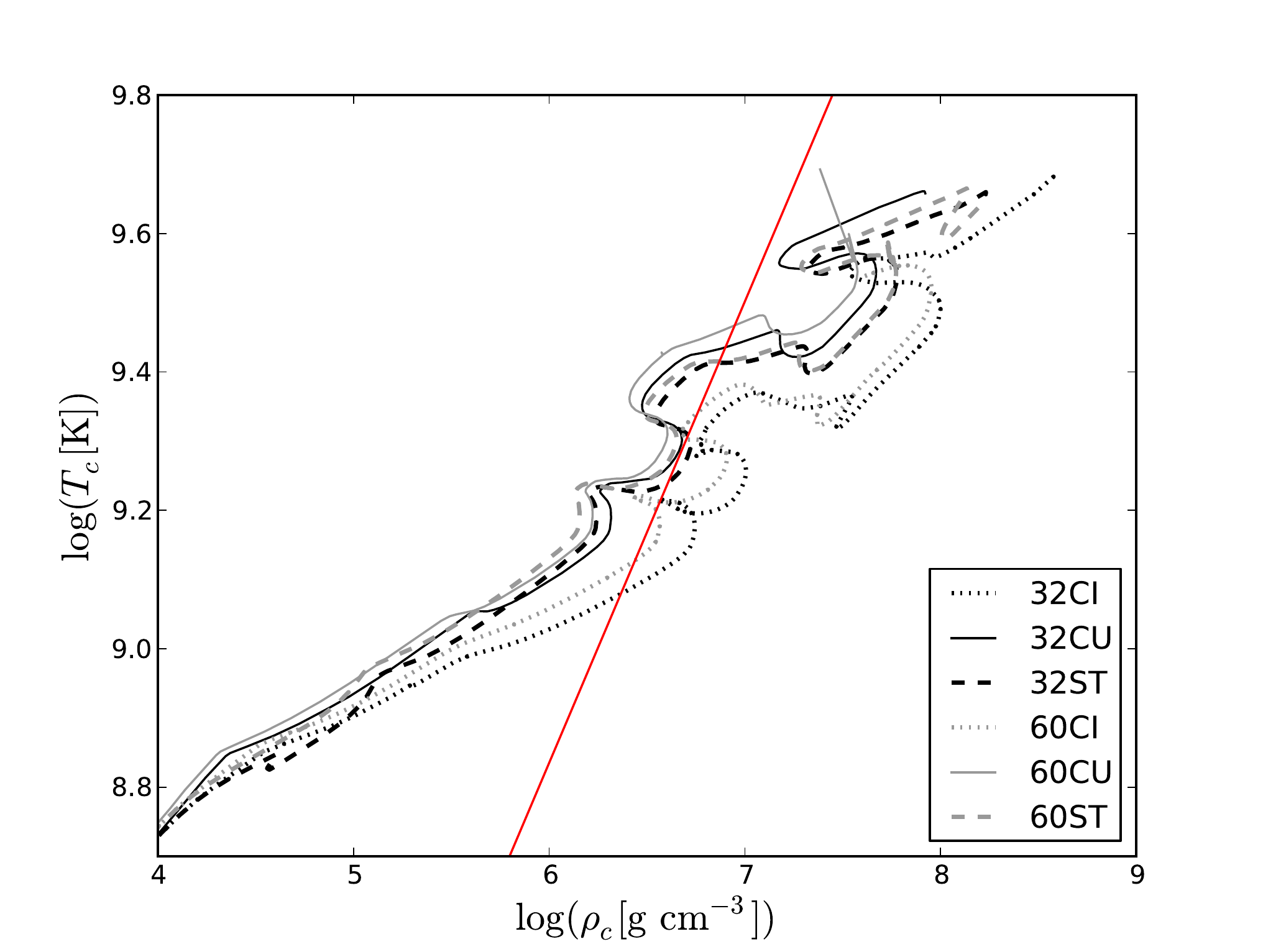}
\caption{$T_c-\rho_c$ diagram for all 32 and 60 $M_{\sun}$ models.  The straight line indicates the location in the diagram where the ideal gas pressure is equal to the electron degeneracy pressure; $P_{gas} = P_{e,deg}$.}
\label{fig:tcrc3260}
\end{figure}

Figure \ref{fig:tcrc152025} shows the impact that the enhanced carbon burning rates have on the central evolution during carbon burning.  However, despite the deviations, many of the models at a particular mass are similar, especially the 25 $M_{\sun}$ models.  Figure \ref{fig:tcrc3260} shows $T_c-\rho_c$ diagrams for the 32 and 60 $M_{\sun}$, which are also quite similar.  In the case of Fig. \ref{fig:tcrc3260}, the 32 and 60 $M_{\sun}$ models exhibit significant mass loss during the hydrogen- and helium-burning stages such that the total mass during the advanced burning stages is very similar ($\sim 13 M_{\sun}$).  Combined with the fact that the helium cores at this stage are qualitatively similar, the models from this point onwards evolve similarly, with the 32CI and 60CI models entering the more degenerate region of the diagram.  Consequently, the tracks follow similar paths dependent on the choice of $^{12}$C + $^{12}$C reaction rate.

\begin{figure*}
\includegraphics[width=\textwidth]{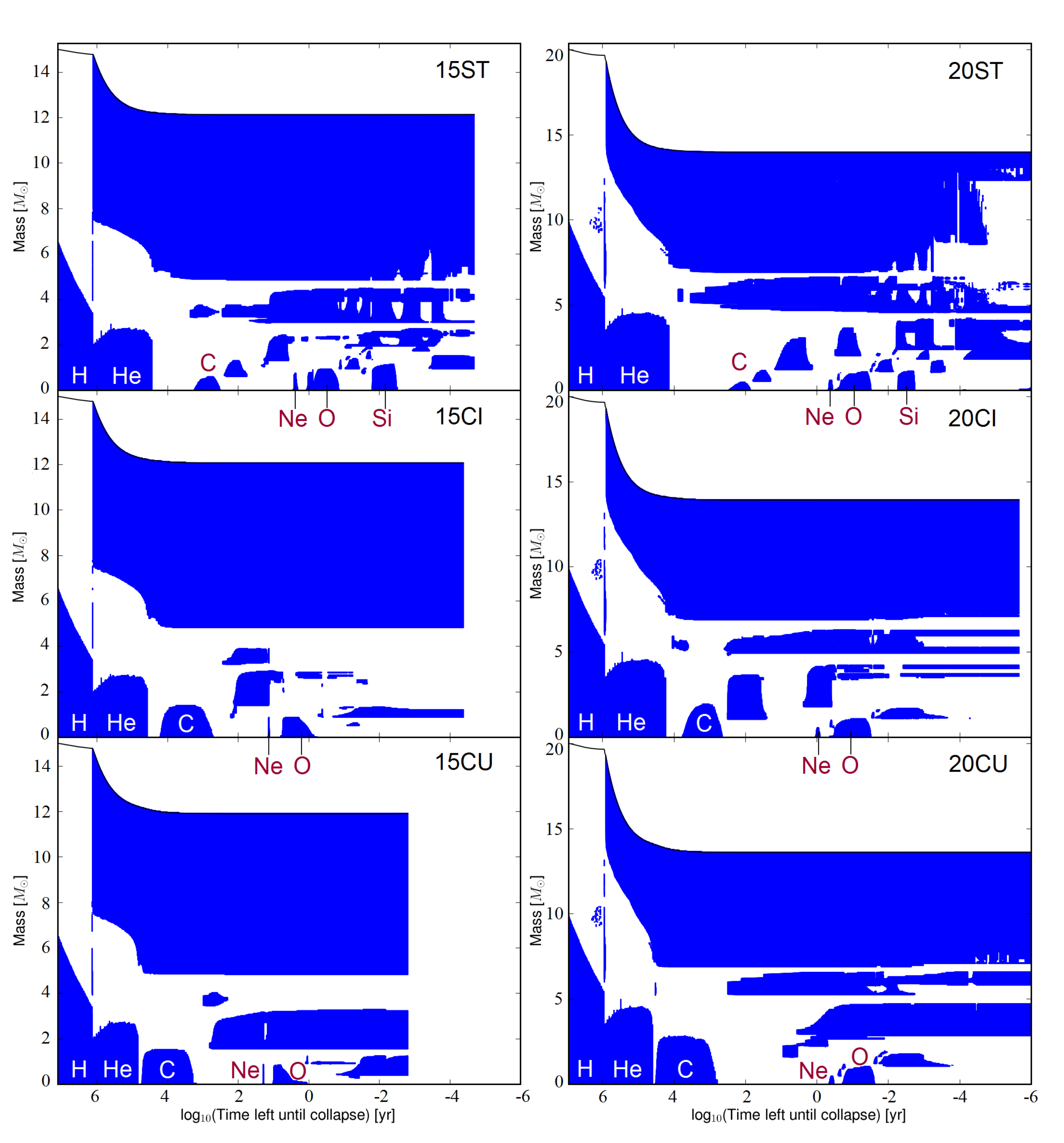}
\caption{Kippenhahn diagrams for the ST, CI and CU models for initial masses of 15 and 20 $M_{\sun}$.  Shaded regions correspond to convection zones.  The major central burning regimes are indicated by the text.}
\label{fig:kip1520}
\end{figure*}

\begin{figure*}
\includegraphics[width=\textwidth]{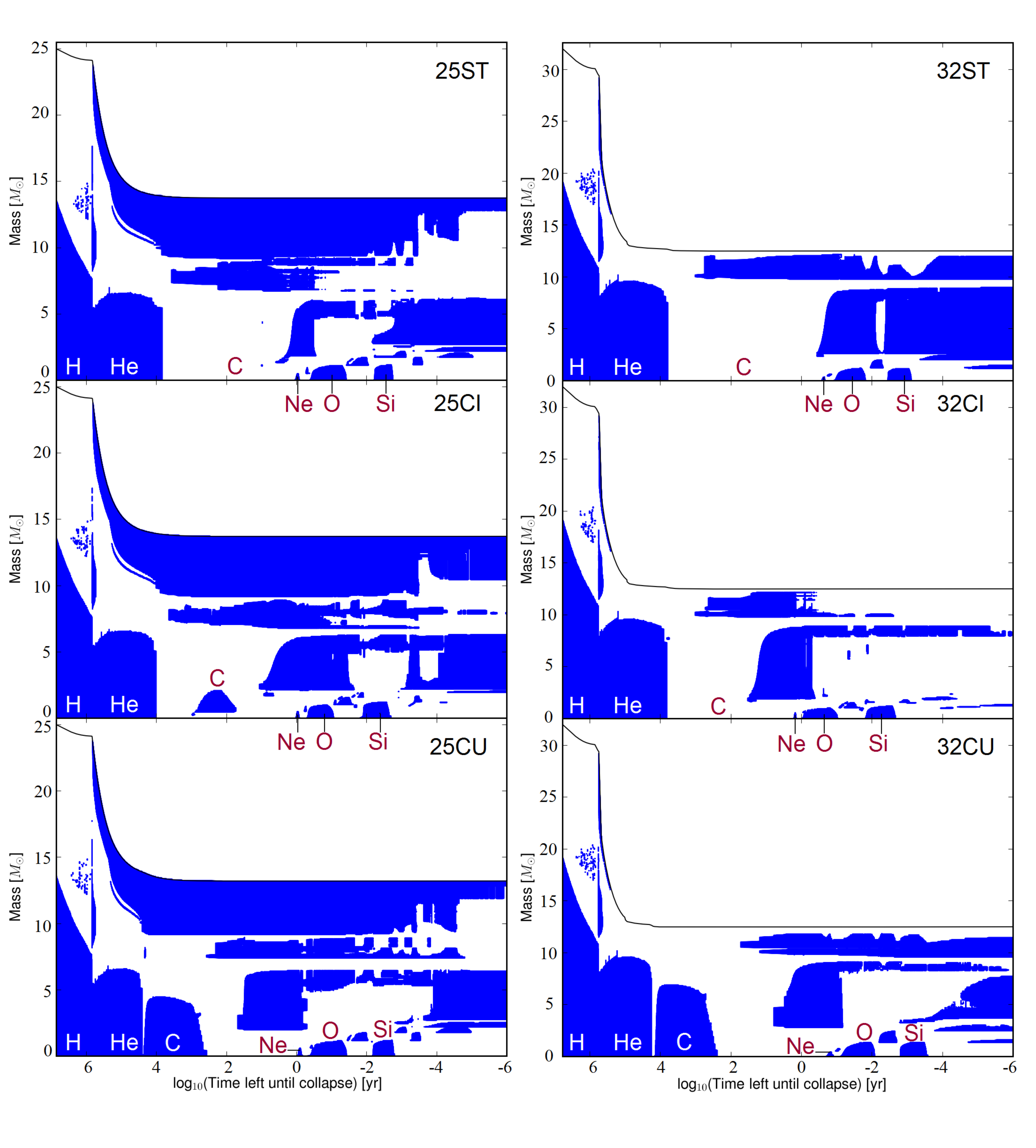}
\caption{Kippenhahn diagrams for the ST, CI and CU models for initial masses of 25 and 32 $M_{\sun}$.  Shaded regions correspond to convection zones.  The major central burning regimes are indicated by the text.}
\label{fig:kip2532}
\end{figure*}

\begin{figure}
\includegraphics[width=0.5\textwidth]{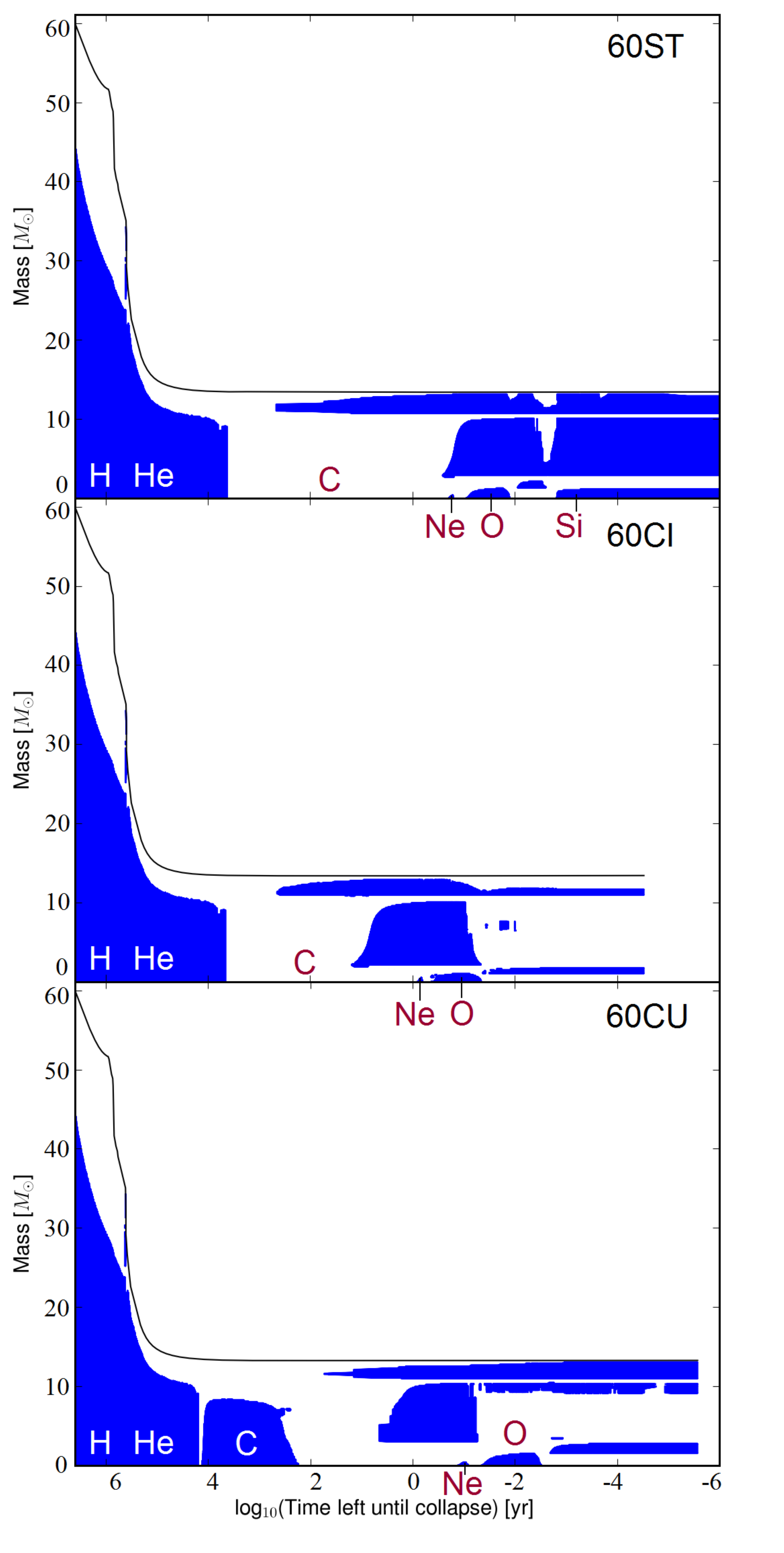}
\caption{Kippenhahn diagrams for the ST, CI and CU models for initial masses of 15 and 20 $M_{\sun}$.  Shaded regions correspond to convection zones.  The major central burning regimes are indicated by the text.}
\label{fig:kip60}
\end{figure}

Kippenhahn diagrams for all models are presented in Fig. \ref{fig:kip1520}, \ref{fig:kip2532} and \ref{fig:kip60}, with the shaded regions corresponding to convection zones and the intermediate regions corresponding to radiative zones.  The total mass is given by the thin black line at the top of each diagram.  Overall, Fig. \ref{fig:kip1520}, \ref{fig:kip2532} and \ref{fig:kip60} show that the convection zone structure of the carbon-burning stage is heavily modified by the increased rates, particularly for the CU cases where a convective carbon-core is present over the entire mass range considered.  The presence of a convective carbon-core is important for nucleosynthesis as the convective mixing provides more fuel for carbon-burning and the carbon-core s process.  The mass loss increases significantly with initial mass, but does not change much with the $^{12}$C + $^{12}$C rate.  Small deviations in the mass loss, which are less than 1 per cent, are due to the increased lifetime of the core carbon burning stage in the CI and CU models (see Table \ref{tab:lifetimes}).

Model data complementary to Figures \ref{fig:kip1520}, \ref{fig:kip2532} and \ref{fig:kip60} are presented in Table \ref{tab:Cprops}, which specify properties pertaining to convection zones during carbon burning.  Column 2. (`Core/Shell') identifies the presence, or not, of a convective core or shell and labels the shells in chronological order during the evolution.  The other columns specify the lifetime of the convection zone\footnote{Many of the convective shells persist until the presupernova stage.  In models 15CI, 20CI, 25ST, 25CI, 25CU, 32CI, 32CU and 60CU however, the carbon shell shrinks because of the influence of another burning stage (such as neon or oxygen burning).  The convective carbon shell can therefore feature a rather complicated structure through the following advanced stages.  In these cases, the lifetime is calculated from the onset of convection to the point where the convective shell shrinks significantly in size.} ($\tau_{\mathrm{C}}$) in years, the lower and upper limits in mass coordinate of the convection zone ($M_{\mathrm{low}}$ and $M_{\mathrm{upp}}$ respectively, in $M_{\sun}$), the size of the convection zone in mass ($\Delta M$, in $M_{\sun}$) and the temperature ($T$, in GK), density ($\rho$, in g cm$^{-3}$) and the mass fraction abundances of $^{12}$C and $^{16}$O ($X_{^{12}\mathrm{C}}$ and $X_{^{16}\mathrm{O}}$ respectively) at the onset of convection at position $M_{\mathrm{low}}$.

The ST models indicate an upper mass limit for the presence of a convective carbon core with a value between 20 and 25 $M_{\sun}$, which is consistent with previous models \citep{2000ApJ...528..368H, 2004A&A...425..649H}.  For model 25CI a strong convective shell is ignited slightly off-centre (at a mass coordinate of $0.436 M_{\sun}$) and model 25CU exhibits a large convective carbon core.  In all CU models the carbon-core burning stage is convective, which, in models 25CU, 32CU and 60CU, replaces the radiative cores.  In Model 25CI the first carbon shell ignites close to the centre and models 20CI and 15CI have larger convective cores.  Considering these facts and the presence of a convective core in every CU model, one can hypothesise that the limiting mass for the presence of a convective carbon core increases with the carbon burning rate, which will consequently represent a source of uncertainty for the presence of a convective core near to the limiting mass of $\sim 22 M_{\sun}$.  A firm verification of the limiting mass for the CI case would however require a finer grid of stellar models between 20 and 25 $M_{\sun}$.

The sizes, in mass, of the carbon-burning zones (column 6 in Table \ref{tab:Cprops}) are generally larger in CI and CU models.  This affects the $^{12}$C abundance profile within the star and consequently the number of carbon-burning shells during the evolution.  The Kippenhahn diagrams for the 15 and 20 $M_{\sun}$ models (Fig. \ref{fig:kip1520}) demonstrate this effect well; the 15ST and 20ST models have many carbon burning shells where the ignition of a successive shell lies at a position that corresponds to the maximum coordinate reached by the previous convection zone.

As the rate is increased, the tendency for convective shells to `overlap' (where the lower bound in mass of the convective region extends below the upper bound of the previous convection zone) is increased.  All CU models, except the 15CU model, show this overlap, which occurs between a convective carbon core and the first convective carbon shell.  The amount of overlap between the carbon core and the first carbon shell, and the first and second carbon shells, in the 20CI model (in Fig. \ref{fig:kip1520}) is also much larger than that in the 20ST model.  This overlap effect occurs because successive carbon-shell burning episodes, caused by ignition of residual $^{12}$C fuel left over from previous burning stages, can occur at a lower temperature and density or with a lower abundance of $^{12}$C fuel (see column 9 of Table \ref{tab:Cprops}).  This effect has been noted previously by \citet{1998ApJ...502..737C} and in the preliminary studies (\citealt{2010JPhCS.202a2023B}a,b).  

\begin{figure}
\includegraphics[width=0.5\textwidth]{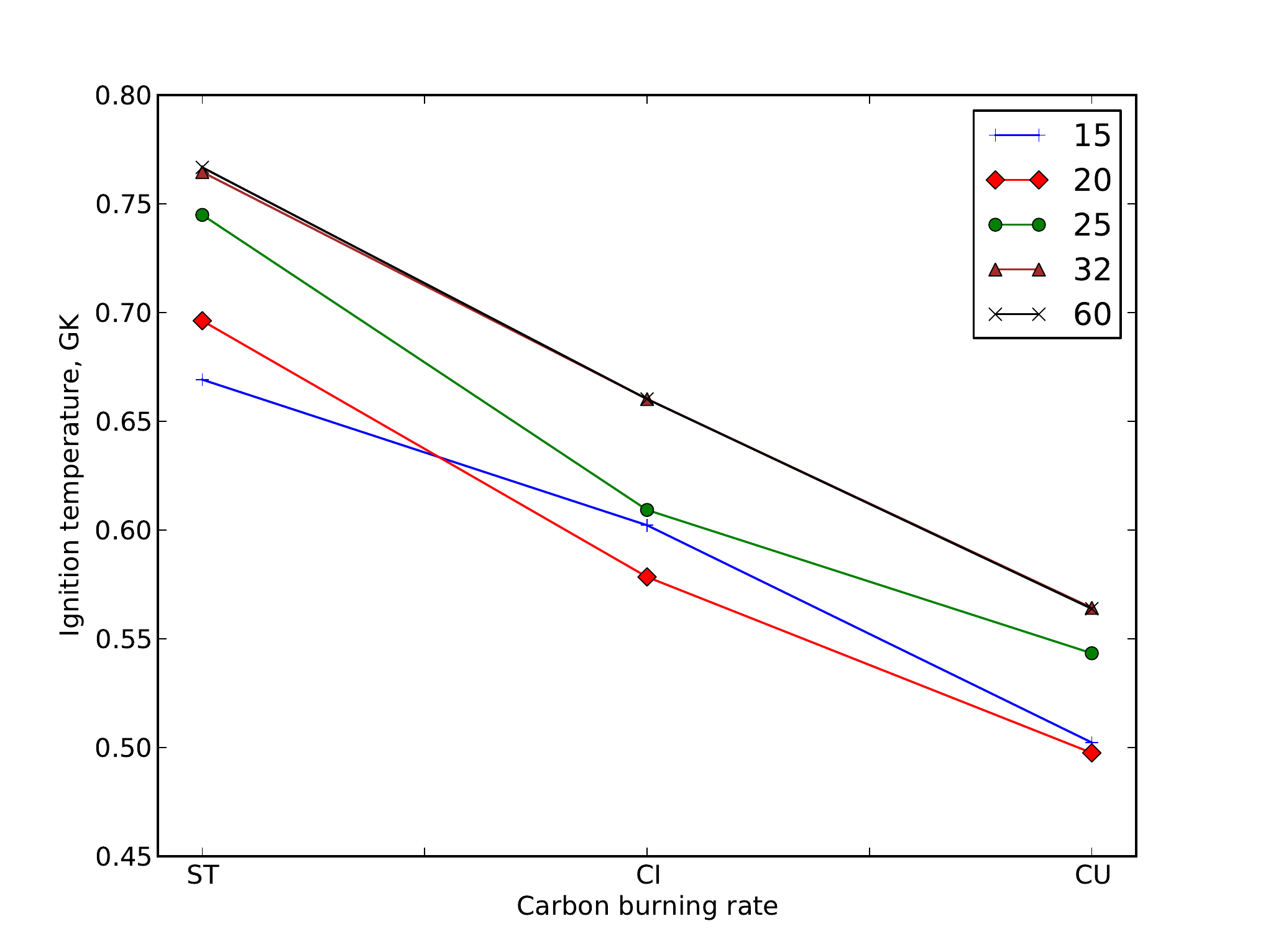}
\caption{Ignition temperatures for core carbon burning for all models.}
\label{fig:igntemp}
\end{figure}

\begin{figure}
\includegraphics[width=0.5\textwidth]{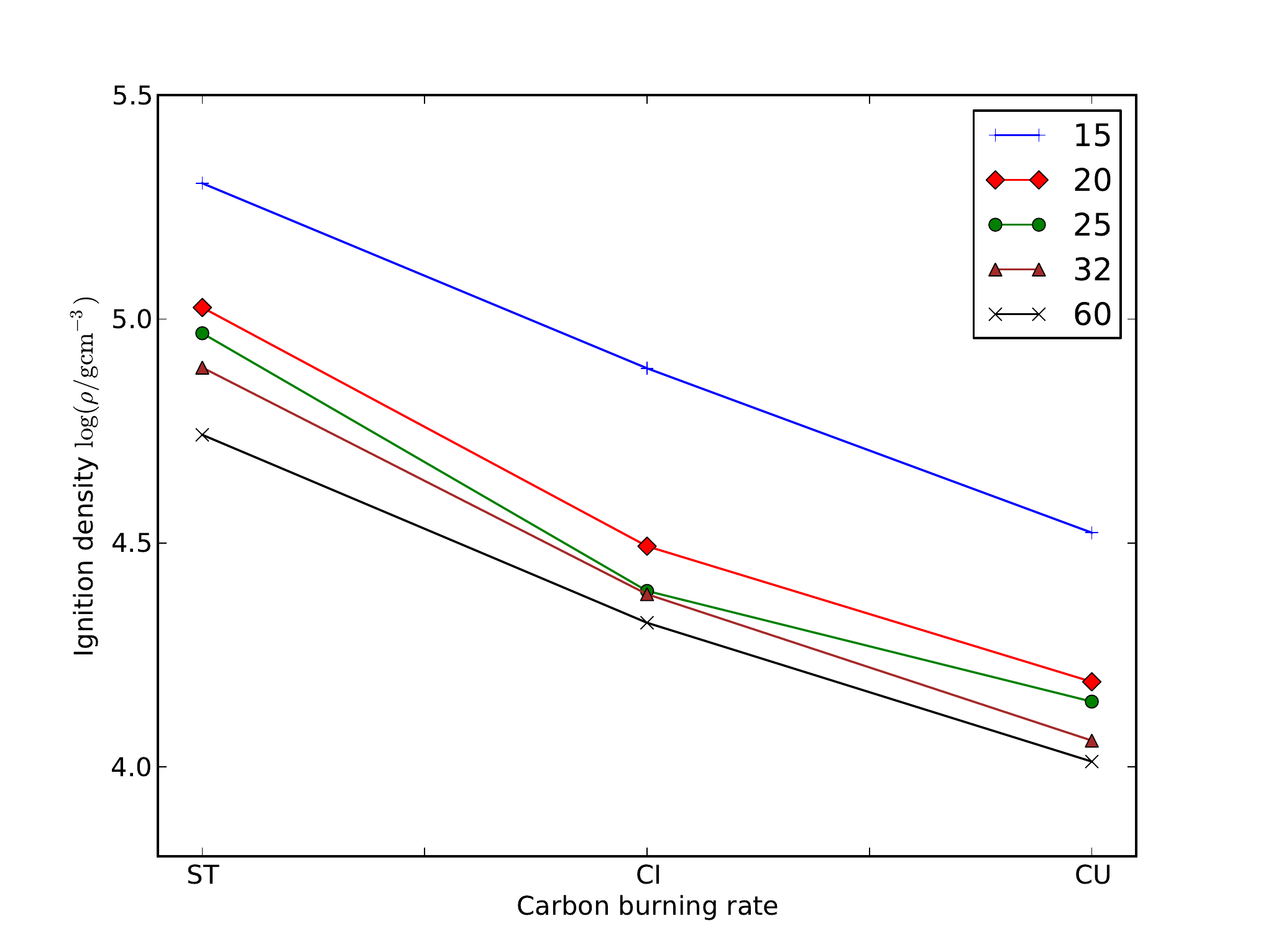}
\caption{Ignition densities for core carbon burning for all models.}
\label{fig:igndens}
\end{figure}

The total energy generation of the $^{12}$C + $^{12}$C reaction is given by \citep{2002RvMP...74.1015W}:

\begin{equation}
\epsilon_{\rm nuc}(^{12}\mathrm{C}) \approx 4.8 \times 10^{18} Y^2(^{12}\mathrm{C}) \rho \lambda_{12,12} \, {\rm erg \, g}^{-1} {\rm s}^{-1}
\label{eqn:energyrate}
\end{equation}

where $Y^2(^{12}$C), is the number abundance of $^{12}$C ($Y = X/A$), $\rho$ is the density and $\lambda_{\rm 12,12}$ is the nuclear reaction rate, which is dependent on temperature.  For a given density and abundance, an increased $^{12}$C + $^{12}$C rate increases the energy generation rate from nuclear reactions.  The effect this has on the ignition conditions (temperature and density) for core carbon burning are displayed in Fig. \ref{fig:igntemp} and \ref{fig:igndens} (the ignition point is defined as the point in time when the central mass fraction abundance of $^{12}$C is $0.3$ per cent lower than its maximum value).  An increased rate allows a star to reach the required energy output to support the star against gravitational contraction at a lower temperature (and also lower density).  Note also the dependence on initial mass, with ignition conditions favouring higher temperatures and lower densities with increasing initial mass.  In the case of lower ignition temperatures and densities, the convective core ignites more promptly in the CI and CU models.  Changes to the ignition conditions and the $^{12}$C abundance at the start of core carbon burning are responsible for the increased likelihood of having overlapping convection zones.

\begin{table*}
\caption{Stellar structure properties for carbon-burning cores and shells at the onset of convection.  Shells are labelled in chronological order.  $\tau_{\mathrm{conv}}$ is the lifetime of the convection zone, $M_{\mathrm{low}}$ and $M_{\mathrm{upp}}$ are lower and upper mass coordinates for the location of the zone.  $\Delta M$ is the size of the zone in mass, $T$ and $\rho$ are the temperature and density of the zone at $M_{\mathrm{low}}$ and $X_{^{12}\mathrm{C}}$ and $X_{^{16}\mathrm{O}}$ are the $^{12}$C and $^{16}$O mass fraction abundances within the convection zone.}
\begin{tabular}{lcccrccccc}
\hline
\hline
Model & Core/Shell & $\tau_{\mathrm{conv}}$ & $M_{\mathrm{low}}$ & $M_{\mathrm{upp}}$ & $\Delta M$ & $T$ & $\rho$ & $X_{^{12}\mathrm{C}}$ & $X_{^{16}\mathrm{O}}$\\
& & (yr) & $(M_{\sun})$ & $(M_{\sun})$ & $(M_{\sun})$ & (GK) & (g cm$^{-3}$) & & \\
\hline
15ST & Core & 1458  & 0     & 0.588 & 0.588 & 0.717 & $2.367 \times 10^5$ & 0.2947 & 0.6296 \\
     & 1    & 187.2 & 0.604 & 1.293 & 0.689 & 0.773 & $1.816 \times 10^5$ & 0.3002 & 0.6332 \\
     & 2    & 17.92 & 1.302 & 2.435 & 1.134 & 0.904 & $1.936 \times 10^5$ & 0.0862 & 0.5041 \\
15CI & Core & 15720 & 0     & 1.381 & 1.381 & 0.589 & $7.409 \times 10^4$ & 0.3104 & 0.6400 \\
     & 1    & 150.1 & 1.396 & 2.907 & 1.511 & 0.758 & $1.139 \times 10^5$ & 0.0472 & 0.4883 \\
15CU & Core & 51890 & 0     & 1.517 & 1.517 & 0.486 & $3.011 \times 10^4$ & 0.3192 & 0.6458 \\
     & 1    & 594.2 & 1.536 & 3.270 & 1.734 & 0.531 & $3.557 \times 10^4$ & 0.3185 & 0.6453 \\
20ST & Core &   219 & 0     & 0.466 & 0.466 & 0.783 & $1.587 \times 10^5$ & 0.2320 & 0.6441 \\
     & 1    & 41.55 & 0.507 & 1.157 & 0.650 & 0.843 & $1.390 \times 10^5$ & 0.2150 & 0.6332 \\
     & 2    & 13.40 & 1.024 & 3.088 & 1.884 & 0.873 & $1.109 \times 10^5$ & 0.2438 & 0.6516 \\
     & 3    & 0.228 & 2.021 & 3.319 & 1.298 & 1.132 & $1.447 \times 10^5$ & 0.0469 & 0.5350 \\
20CI & Core &  5418 & 0     & 1.921 & 1.921 & 0.626 & $4.155 \times 10^4$ & 0.2636 & 0.6647 \\
     & 1    & 290.9 & 1.047 & 3.631 & 2.584 & 0.781 & $7.203 \times 10^4$ & 0.0675 & 0.5481 \\
     & 2    & 1.985 & 1.784 & 4.137 & 2.354 & 0.872 & $6.615 \times 10^4$ & 0.0488 & 0.5380 \\
20CU & Core & 32280 & 0     & 2.771 & 2.771 & 0.498 & $1.553 \times 10^4$ & 0.2861 & 0.6794 \\
     & 1    & 10.05 & 2.158 & 2.609 & 0.450 & 0.712 & $4.792 \times 10^4$ & 0.0147 & 0.5275 \\
     & 2    & 3.714 & 2.815 & 4.696 & 1.880 & 0.592 & $2.706 \times 10^4$ & 0.2861 & 0.6794 \\
25ST & 1    & 3.734 & 1.819 & 5.928 & 4.109 & 0.946 & $1.017 \times 10^5$ & 0.1449 & 0.6306 \\
25CI & 1    & 925.4 & 0.436 & 2.075 & 1.640 & 0.718 & $3.656 \times 10^4$ & 0.1830 & 0.6554 \\
     & 2    & 12.69 & 2.111 & 6.208 & 4.097 & 0.516 & $3.893 \times 10^4$ & 0.2492 & 0.6975 \\
25CU & Core & 22520 & 0     & 4.452 & 4.452 & 0.510 & $1.191 \times 10^4$ & 0.2586 & 0.7038 \\
     & 1    & 34.77 & 1.954 & 6.429 & 4.475 & 0.735 & $3.622 \times 10^4$ & 0.0191 & 0.5656 \\
32ST & 1    & 0.373 & 2.586 & 8.948 & 6.361 & 1.059 & $7.925 \times 10^4$ & 0.1346 & 0.6869 \\
32CI & 1    & 33.06 & 1.869 & 8.789 & 6.920 & 0.773 & $3.290 \times 10^4$ & 0.1507 & 0.6973 \\
32CU & Core & 13780 & 0     & 6.897 & 6.897 & 0.539 & $1.001 \times 10^4$ & 0.2164 & 0.7399 \\
     & 1    & 5.679 & 2.774 & 9.077 & 6.303 & 0.710 & $2.390 \times 10^4$ & 0.0269 & 0.6265 \\
60ST & 1    & 0.260 & 2.900 & 10.12 & 7.221 & 1.073 & $7.159 \times 10^4$ & 0.1360 & 0.6794 \\
60CI & 1    & 15.04 & 2.171 & 10.04 & 7.866 & 0.793 & $3.080 \times 10^4$ & 0.1541 & 0.6911 \\
60CU & Core & 12900 & 0     & 8.326 & 8.326 & 0.542 & $9.210 \times 10^3$ & 0.2205 & 0.7341 \\
     & 1    & 4.276 & 2.975 & 10.39 & 7.412 & 0.721 & $2.207 \times 10^4$ & 0.0309 & 0.6207 \\
\hline
\end{tabular}
\label{tab:Cprops}
\end{table*}

The lifetime of convection zones is generally longer in the CI and CU models.  This could be perceived as counter-intuitive, since with an enhanced rate one would expect that the $^{12}$C fuel would be expended more rapidly.  However, the burning takes place in lower temperature and density conditions, which affect the neutrino losses.  Table \ref{tab:energy_neut} shows the energy generation terms for nuclear reactions ($\epsilon_{\rm nuc}$) and neutrino losses ($\epsilon_{\nu}$) at the centre of the star when the mass fraction of $^{12}$C is half the amount available just prior to carbon-core burning.  The proportion of neutrinos formed by various neutrino processes are also specified in Table \ref{tab:energy_neut}, which are given as fractions, $f$, of the total neutrino losses (in per cent).  These processes are pair production ($f_{\rm pair}$), photoneutrino interactions ($f_{\rm phot}$) and the rest ($f_{\rm rest}$), which are bremsstrahlung, recombination and plasmon decay processes \citep{1996ApJS..102..411I}.  Neutrino formation through these last three processes is negligibly small at carbon burning temperatures.

\begin{figure}
\includegraphics[width=0.5\textwidth]{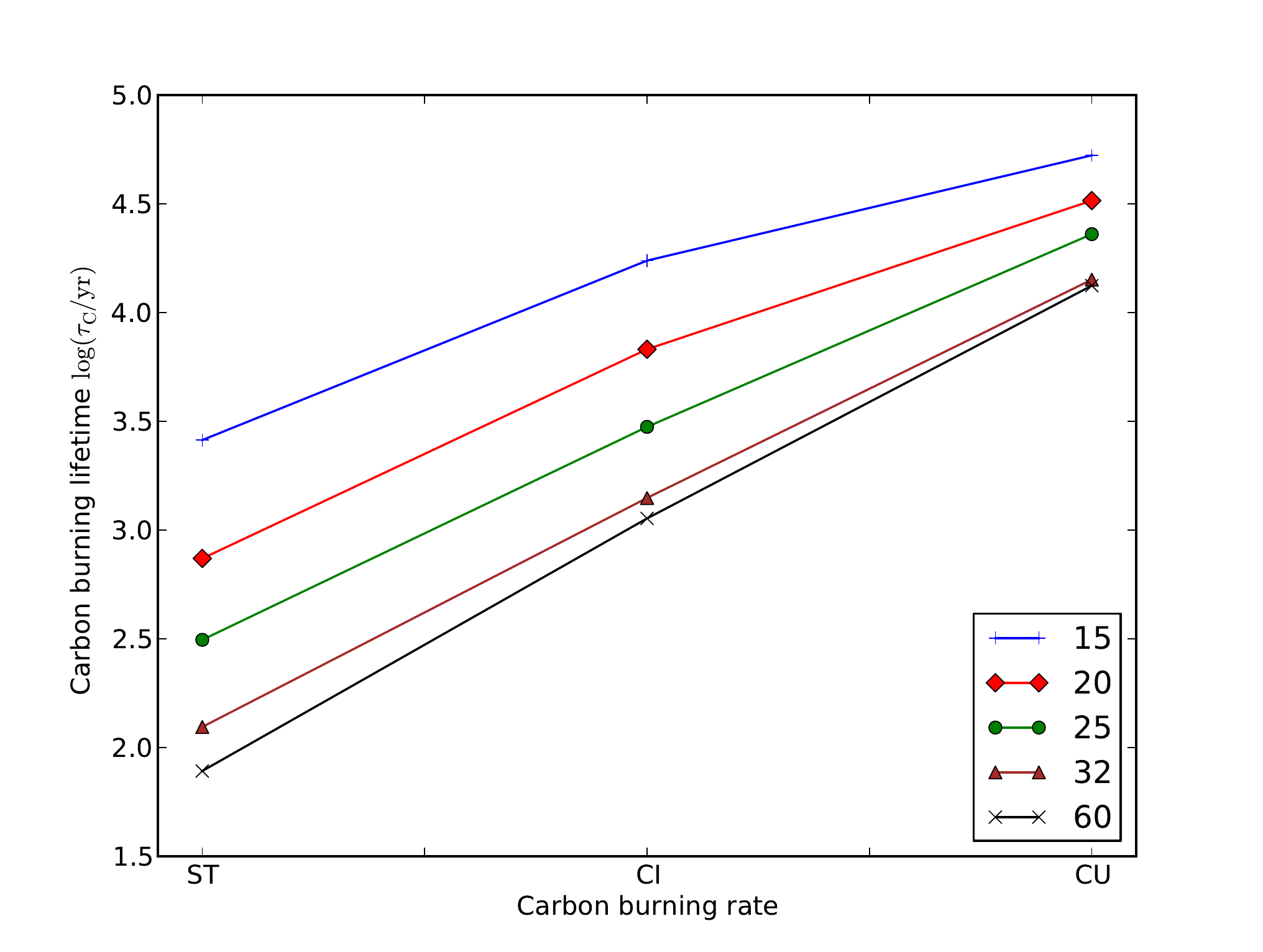}
\caption{Carbon core burning lifetimes for all models.  Note that for some models, the core carbon burning is radiative rather than convective.  The carbon burning lifetime is defined as the time for the mass fraction abundance of $^{12}$C to reduce from 0.3 per cent of its maximum value to a value of $10^{-3}$.}
\label{fig:clifetimes}
\end{figure}

As shown by Table \ref{tab:energy_neut}, the energy generation rate from nuclear reactions and the neutrino losses are reduced in the CI and CU models, although an increase in energy generation rate is seen in models 25CU, 32CU and 60CU from their CI counterparts.  This increase is due to the presence of the convective carbon core, where there is an increased availability of $^{12}$C fuel from mixing.  During carbon burning, the timescale for burning is governed primarily by the neutrino losses (as is true for all advanced burning stages) and these losses generally increase monotonically with increasing temperature.  In fact, massive star evolution during the advanced stages of evolution can be described as a  neutrino-mediated Kelvin-Helmholtz contraction of a carbon-oxygen core \citep{2002RvMP...74.1015W, 2009SSRv..147....1E}.  Therefore, a reduction in the neutrino-losses has the consequence of increasing the lifetime of carbon-burning stages.  Only the carbon shells in models 32CU and 60CU do not show this behaviour (see Fig. \ref{fig:kip2532} and \ref{fig:kip60}).  This can be explained by the presence of a previous convective carbon core in those models, which reduces the abundance of carbon fuel available for burning in these shells.  Systematic trends during shell burning are less clear because of the rather complicated evolution of the shell structure, but convective shells often form at lower temperatures in CI and CU models (see column 7 in Table \ref{tab:Cprops}), similar to the situation in the core.  For carbon core burning, on the other hand, there is a clear increase in the lifetime with increasing rate, which is shown in Fig. \ref{fig:clifetimes}.

The main neutrino processes during carbon burning are those caused by pair production and photoneutrino interactions \citep{2002RvMP...74.1015W,1996ApJS..102..411I}.  It is worth noting that the decrease in temperature in the CI and CU models is responsible for a larger proportion of neutrinos formed by the photoneutrino process rather than pair production.  This trend at larger carbon-burning rates is opposite to the trend with initial mass, which favours higher temperatures and production of neutrinos by pair production with increasing initial mass.

\begin{table*}
\caption{Energy generation and neutrino parameters during core carbon burning.  For each model the central values of temperature, $T$, density, $\rho$, energy generation rates for nuclear burning, $\epsilon_{\rm nuc}$, and neutrino losses, $\epsilon_{\nu}$, and percentage fractions of the total neutrinos formed by pair production ($f_{\rm pair}$), photoneutrino interactions ($f_{\rm phot}$) and other processes ($f_{\rm rest}$) are specified.  These parameters are determined at the time when the mass fraction of $^{12}$C is half of the value just prior to carbon burning.}
\begin{tabular}{lccccrrr}
\hline
\hline
Model & $T$ & $\rho$ & $\epsilon_{\rm nuc}$ & $\epsilon_{\nu}$ & $f_{\rm pair}$ & $f_{\rm phot}$ & $f_{\rm rest}$ \\
     & (GK) & (g cm$^{-3}$) & (erg g$^{-1}$ s$^{-1}$) & (erg g$^{-1}$ s$^{-1}$) & & & \\
\hline
15ST & 0.830 & $2.141 \times 10^{5}$ & $4.762 \times 10^{7}$ & $-1.542 \times 10^{7}$ & 89.665 & 10.253 & 0.082 \\
15CI & 0.686 & $7.659 \times 10^{4}$ & $6.822 \times 10^{6}$ & $-1.454 \times 10^{6}$ & 70.007 & 29.861 & 0.132 \\
15CU & 0.566 & $3.772 \times 10^{4}$ & $2.277 \times 10^{6}$ & $-1.448 \times 10^{5}$ & 19.800 & 79.902 & 0.298 \\
20ST & 0.883 & $1.679 \times 10^{5}$ & $1.663 \times 10^{8}$ & $-5.910 \times 10^{7}$ & 95.651 & 4.327  & 0.022 \\
20CI & 0.723 & $5.356 \times 10^{4}$ & $1.529 \times 10^{7}$ & $-5.260 \times 10^{6}$ & 87.461 & 12.508 & 0.031 \\
20CU & 0.588 & $2.477 \times 10^{4}$ & $3.727 \times 10^{6}$ & $-2.643 \times 10^{5}$ & 41.935 & 57.943 & 0.122 \\
25ST & 0.859 & $1.439 \times 10^{5}$ & $5.176 \times 10^{7}$ & $-4.435 \times 10^{7}$ & 95.061 & 4.917  & 0.022 \\
25CI & 0.690 & $3.942 \times 10^{4}$ & $2.603 \times 10^{6}$ & $-2.975 \times 10^{6}$ & 83.475 & 16.490 & 0.035 \\
25CU & 0.603 & $1.889 \times 10^{4}$ & $4.975 \times 10^{6}$ & $-4.533 \times 10^{5}$ & 58.913 & 41.026 & 0.061 \\
32ST & 0.904 & $1.313 \times 10^{5}$ & $1.360 \times 10^{8}$ & $-1.234 \times 10^{8}$ & 97.310 & 2.680  & 0.010 \\
32CI & 0.711 & $3.532 \times 10^{4}$ & $3.682 \times 10^{6}$ & $-5.995 \times 10^{6}$ & 89.439 & 10.543 & 0.018 \\
32CU & 0.621 & $1.510 \times 10^{4}$ & $5.725 \times 10^{6}$ & $-9.148 \times 10^{5}$ & 74.347 & 25.625 & 0.028 \\
60ST & 0.919 & $1.106 \times 10^{5}$ & $1.900 \times 10^{8}$ & $-1.954 \times 10^{8}$ & 98.053 & 1.941  & 0.006 \\
60CI & 0.725 & $3.260 \times 10^{4}$ & $5.863 \times 10^{6}$ & $-9.442 \times 10^{6}$ & 92.247 & 7.741  & 0.012 \\
60CU & 0.625 & $1.375 \times 10^{4}$ & $6.244 \times 10^{6}$ & $-1.096 \times 10^{6}$ & 77.670 & 22.309 & 0.021 \\
\hline
\end{tabular}
\label{tab:energy_neut}
\end{table*}

These effects on the central evolution are responsible for the different tracks exhibited by the CI and CU models with respect to the ST models in Figures \ref{fig:tcrc152025} and \ref{fig:tcrc3260}.  For the 15 and 20$M_{\sun}$ models the larger cores cause the CI and CU tracks to tend towards the higher temperature, lower density side of the ST track, but only for the duration the convective core is present.  When the star moves onto carbon shell burning, the core cools and the track returns to the standard curve.

As explained above, the overlap exhibited by convective shells over the ashes of convective carbon cores is due to the ignition of carbon that represents the unburnt remainder from carbon-core burning.  The presence of this remainder is caused by the gradual shrinking of the carbon-core near the end of the burning stage.  This occurs in model 20CI and all CU models, except model 15CU where the shell is located at the top of the previous convective carbon core.  The convective carbon shell in the 20CU model (see Fig. \ref{fig:kip1520}), however, shows an interesting structure.  In this case a carbon shell is ignited at a position that overlaps with the core and then shortly after an additional shell is ignited at the point corresponding to the top of the previous core.  Because of the unusual structure, the lifetime given in Table \ref{tab:Cprops} for the 20CU model, shell 1, is defined from the onset of convection to the time it shrinks back up into the second shell.  

The presence of overlap with a carbon core has a significant impact on the composition of the shell at the onset of convection.  Indeed, carbon-core burning ashes, including s-process nuclides, will mix out to a position above the remnant mass and be present in the supernova ejecta.  As mentioned above, overlapping shells have previously been noted in the literature, but the consequences of overlapping shells of this nature are not well studied.  The nucleosynthetic consequences of overlap will be discussed in \S \ref{sec:Nucleosynthesis}. 

\subsection{Advanced stages beyond carbon burning}\label{sec:lateevol}

Despite the changes to the stellar structure during carbon burning, the evolution of the advanced burning stages in the core following carbon-burning seems only slightly affected in terms of the convection zone structure, as seen in Fig. \ref{fig:kip1520}, \ref{fig:kip2532} and \ref{fig:kip60}, but exhibit burning stages with different lifetimes.  The burning lifetimes for the hydrostatic burning stages are presented in Table \ref{tab:lifetimes}, which are defined for each stage as the difference in age from the point where the principal fuel for that stage ($^1$H for hydrogen burning, $^4$He for helium burning, etc.) is depleted by 0.3 per cent from its maximum value to the age where the mass fraction of that fuel depletes below a value of $10^{-5}$, except for carbon burning and neon burning, where this value is $10^{-3}$, and oxygen burning, where this value is $10^{-2}$.  These criteria are necessary to ensure a lifetime is calculated in those cases where residual fuel is unburnt (such as during oxygen burning in the 15CU model, where the $^{16}$O mass fraction abundance that remains unburnt following the end of core oxygen burning is $\sim 3.177 \times 10^{-3}$) and to ensure that the burning stages are correctly separated.  The lifetime of the advanced stages is relatively sensitive to the mass fractions of isotopes defining the lifetime.

Carbon burning lifetimes are longer for the CI and CU rates, as explained in \S \ref{sec:carbonevol}, but lifetimes for the other advanced stages do not show a general trend with the carbon burning rate.  This lack of trend also applies to the central properties, as seen in Fig. \ref{fig:tcrc152025}, where the tracks are modified by the enhanced rate models but the modifications do not follow a general pattern.  In fact, there are examples of $T_c-\rho_c$ tracks, e.g. the 25CI and 25CU models in Fig. \ref{fig:tcrc152025}, where following the deviation caused by carbon ignition the track returns to that of the ST rate (especially for the 15, 20 and 25 $M_{\sun}$ models).  The main property determining the variations in the lifetime is the central temperature, which is linked with the neutrino loss rates.

\begin{table*}
\caption{Lifetimes for all core burning stages in all models (in yrs).  Lifetimes are provided for hydrogen burning ($\tau_{\mathrm{H}}$), helium burning ($\tau_{\mathrm{He}}$), carbon burning ($\tau_{\mathrm{C}}$), neon burning ($\tau_{\mathrm{Ne}}$), oxygen burning ($\tau_{\mathrm{O}}$) and silicon burning ($\tau_{\mathrm{Si}}$).  The total lifetime is given by ($\tau_{\mathrm{Total}}$).}
\begin{tabular}{lcccrrcc}
\hline
\hline
Model & $\tau_{\mathrm{H}}$ & $\tau_{\mathrm{He}}$ & $\tau_{\mathrm{C}}$ & $\tau_{\mathrm{Ne}}$ & $\tau_{\mathrm{O}}$ & $\tau_{\mathrm{Si}}$ & $\tau_{\mathrm{Total}}$\\
\hline
15ST & $1.137 \times 10^{7}$ & $1.255 \times 10^{6}$ & $2.595 \times 10^{3}$ & $1.253$ & $1.233$ & $1.685 \times 10^{-2}$ & $1.268 \times 10^{7}$ \\
15CI & $1.137 \times 10^{7}$ & $1.255 \times 10^{6}$ & $1.735 \times 10^{4}$ & $14.296$ & $4.745$ & $         -         $ & $1.269 \times 10^{7}$ \\
15CU & $1.137 \times 10^{7}$ & $1.255 \times 10^{6}$ & $5.288 \times 10^{4}$ & $12.918$ & $8.815$ & $         -         $ & $1.272 \times 10^{7}$ \\
20ST & $7.926 \times 10^{6}$ & $8.396 \times 10^{5}$ & $7.409 \times 10^{2}$ & $0.193$ & $0.293$ & $1.302 \times 10^{-2}$ & $8.799 \times 10^{6}$ \\
20CI & $7.926 \times 10^{6}$ & $8.396 \times 10^{5}$ & $6.786 \times 10^{3}$ & $0.655$ & $0.542$ & $         -         $ & $8.803 \times 10^{6}$ \\
20CU & $7.926 \times 10^{6}$ & $8.396 \times 10^{5}$ & $3.275 \times 10^{4}$ & $0.265$ & $0.253$ & $         -         $ & $8.825 \times 10^{6}$ \\
25ST & $6.492 \times 10^{6}$ & $6.519 \times 10^{5}$ & $3.131 \times 10^{2}$ & $0.634$ & $0.603$ & $4.322 \times 10^{-3}$ & $7.168 \times 10^{6}$ \\
25CI & $6.492 \times 10^{6}$ & $6.519 \times 10^{5}$ & $2.984 \times 10^{3}$ & $0.539$ & $0.597$ & $1.097 \times 10^{-2}$ & $7.169 \times 10^{6}$ \\
25CU & $6.492 \times 10^{6}$ & $6.519 \times 10^{5}$ & $2.296 \times 10^{4}$ & $0.505$ & $0.515$ & $1.746 \times 10^{-2}$ & $7.186 \times 10^{6}$ \\
32ST & $5.287 \times 10^{6}$ & $5.346 \times 10^{5}$ & $1.245 \times 10^{2}$ & $0.111$ & $0.167$ & $8.997 \times 10^{-3}$ & $5.840 \times 10^{6}$ \\
32CI & $5.287 \times 10^{6}$ & $5.346 \times 10^{5}$ & $1.406 \times 10^{3}$ & $0.726$ & $1.123$ & $1.173 \times 10^{-2}$ & $5.840 \times 10^{6}$ \\
32CU & $5.287 \times 10^{6}$ & $5.346 \times 10^{5}$ & $1.419 \times 10^{4}$ & $0.148$ & $0.111$ & $5.458 \times 10^{-3}$ & $5.852 \times 10^{6}$ \\
60ST & $3.549 \times 10^{6}$ & $3.935 \times 10^{5}$ & $7.808 \times 10^{1}$ & $0.090$ & $0.119$ & $8.624 \times 10^{-3}$ & $3.955 \times 10^{6}$ \\
60CI & $3.549 \times 10^{6}$ & $3.935 \times 10^{5}$ & $1.132 \times 10^{3}$ & $0.425$ & $0.505$ & $         -         $ & $3.955 \times 10^{6}$ \\
60CU & $3.549 \times 10^{6}$ & $3.935 \times 10^{5}$ & $1.331 \times 10^{4}$ & $0.112$ & $0.071$ & $         -         $ & $3.966 \times 10^{6}$ \\
\hline
\end{tabular}
\label{tab:lifetimes}
\end{table*}

The last column of Table \ref{tab:lifetimes} shows that the total lifetime of the star increases slightly with an enhanced carbon burning rate, because of the longer carbon burning lifetime.  Since the total lifetime increases by $\approx 1-5 \times 10^4$ years, the strong mass loss (characteristic of massive stars), which can increase by up to $\sim 10^{-5} M_{\sun}$ yr$^{-1}$, increases the mass lost by up to $0.5 M_{\sun}$.  This is demonstrated in column 2 of Table \ref{tab:cores}, which shows the core masses at the end of oxygen burning for all models.  In column 3 of Table \ref{tab:cores}, we see that the carbon burning rate does not affect the helium core mass (the helium core mass is defined as the mass coordinate where the mass fraction abundance of $^{4}$He is $0.75$ at the interface between the hydrogen and helium-rich layers). There is only a tiny difference for the 25 $M_{\sun}$ case because of the small structure re-arrangement of the hydrogen burning shell.  In column 4, we see that with an increasing carbon burning rate, the CO core mass is larger (the CO core mass is defined as the mass coordinate where the $^{4}$He mass fraction abundance is $10^{-3}$).  The reason is the following.  With an increased rate, carbon burning occurs at lower temperatures where the energy production dominates over neutrino cooling and this leads to a stronger carbon core burning in a larger convective zone.  Thus the carbon burning core produces more energy and this leads to a less energetic helium-burning shell that is radiative rather than convective, which is the case for the ST models.  When the He-shell is radiative the burning front depletes completely the helium available at one mass coordinate and then moves upwards leading to a more massive CO core whereas with a convective He-shell, the bottom of the shell stays at the same mass coordinate since the helium in the convective shell is never completely exhausted due to mixing.  Note also that the 32 and 60 $M_{\sun}$ models do not exhibit a value for $M^{75\%}_{\mathrm{\alpha}}$.  This is because the mass loss is strong enough in these WR stars to expel the majority of their helium-rich envelopes and the $^{4}$He abundance is not high enough to satisfy the criterion for $M^{75\%}_{\mathrm{\alpha}}$.  In these cases, the helium core mass is taken as the final mass, $M_{\mathrm{Final}}$ (see column 2 of Table \ref{tab:cores}).

As mentioned above, the size of the convective cores during neon, oxygen and silicon burnings is only slightly affected by the changes in carbon burning rate, as can be seen in the last column of Table \ref{tab:cores} for the oxygen-free core, $M_{\mathrm{O-free}}$, calculated at the end of core oxygen burning.  The changes in $M_{\mathrm{O-free}}$ with carbon burning rate are because of changes in the position of the lower boundary of the last convective carbon shell.  Generally, the magnitude of the changes in $M_{\mathrm{O-free}}$ are small and do not present a clear pattern.

\begin{table}
\centering
\caption{Core masses at the end of oxygen burning, in solar masses.  For each model, the final total mass ($M_{\mathrm{Final}}$), helium core mass ($M^{75\%}_{\mathrm{\alpha}}$), CO core mass ($M_{\mathrm{CO}}$) and the oxygen-free core mass ($M_{\mathrm{O-free}}$) are specified.  Note that the 32 and 60 $M_{\sun}$ models expel most of their helium-rich envelopes, consequently becoming WR stars.}
\begin{tabular}{lrrrr}
\hline
\hline
Model & $M_{\mathrm{Final}}$ & $M^{75\%}_{\mathrm{\alpha}}$ & $M_{\mathrm{CO}}$ & $M_{\mathrm{O-free}}$ \\
\hline
15ST & $12.132$ & $4.791$ & $2.805$ & $0.921$ \\
15CI & $12.069$ & $4.791$ & $2.923$ & $0.867$ \\
15CU & $11.907$ & $4.791$ & $3.239$ & $0.849$ \\
20ST & $13.974$ & $6.826$ & $4.494$ & $1.083$ \\
20CI & $13.916$ & $6.826$ & $4.491$ & $1.099$ \\
20CU & $13.602$ & $6.826$ & $4.696$ & $1.040$ \\
25ST & $13.738$ & $9.199$ & $6.301$ & $1.081$ \\
25CI & $13.710$ & $9.092$ & $6.384$ & $0.980$ \\
25CU & $13.202$ & $9.092$ & $6.544$ & $1.124$ \\
32ST & $12.495$ & $12.495$ & $9.146$ & $1.187$ \\
32CI & $12.495$ & $12.495$ & $9.146$ & $0.984$ \\
32CU & $12.493$ & $12.493$ & $9.425$ & $1.334$ \\
60ST & $13.428$ & $13.428$ & $10.701$ & $1.242$ \\
60CI & $13.423$ & $13.423$ & $10.446$ & $0.990$ \\
60CU & $13.278$ & $13.278$ & $10.929$ & $1.519$ \\
\hline
\end{tabular}
\label{tab:cores}
\end{table}


\section{Nucleosynthesis}\label{sec:Nucleosynthesis}

\subsection{Neutron sources}

The main effects on the nucleosynthesis in the stellar models are due to the lower central temperature of the star and the increased lifetime.  In particular, the lower central temperature will affect the efficiency of neutron source reactions.  We recall that the main neutron sources for the s process are $^{13}$C, which is important during carbon core burning, and $^{22}$Ne, which is important during helium core burning and carbon shell burning.  The $^{13}$C neutron source is mainly produced during carbon core burning by the $^{12}$C(p,$\gamma$)$^{13}$N($\beta^+$)$^{13}$C reaction chain.  Neutrons are then produced by $^{13}$C($\alpha$,n)$^{16}$O reactions.  The protons and $\alpha$-particles originate directly from the $^{12}$C + $^{12}$C fusion reactions. There is competition between the $^{13}$N($\beta^+$)$^{13}$C and $^{13}$N($\gamma$,p)$^{12}$C, where at temperatures above $0.8$ GK, the ($\gamma$,p) reaction dominates over the $\beta$-decay.  The $^{13}$C neutron source is  thus an efficient neutron producer only at lower temperatures.  During carbon shell burning, where the temperatures are higher, the $^{22}$Ne source is the dominant neutron source.  One can therefore expect that as the carbon burning rate is increased and the interior temperature is lowered, the efficiency of the $^{13}$C neutron source will increase.  This efficiency will also be higher given the increased lifetimes.

A non-negligible fraction of neutrons are also present from the $^{17}$O and $^{21}$Ne neutron sources, but these nuclei are mainly produced by neutron captures on $^{16}$O and $^{20}$Ne (and $^{17}$O($\alpha,\gamma$)$^{21}$Ne) and therefore only act as mediators of the neutron irradiance. The $^{25}$Mg($\alpha$,n)$^{28}$Si and $^{12}$C($^{12}$C,n)$^{23}$Mg neutron sources are marginal for all models considered here, despite the increases to the carbon burning rate.  We refer to Pignatari et al. (2011) for a more detailed discussion about the $^{12}$C($^{12}$C,n)$^{23}$Mg reaction.

\subsection{S-process parameters}

Several indicators for the neutron capture nucleosynthesis are considered.  The s process is typically characterised by the neutron density, $n_n$, the neutron captures per iron seed, $n_c$, and the neutron exposure, $\tau_n$.  $n_c$ is defined as follows:

\begin{equation}
n_c = \frac{\sum_i^n (A_i-56) (X_i - X^0_i)}{X_{^{56}\rm Fe}}, \label{eqn:ncap}
\end{equation}

where $X^0_i$ is the initial mass fraction abundance of isotope $X_i$ with atomic mass $A_i$ and $X_{^{56}\rm Fe}$ is the intial mass fraction abundance of $^{56}$Fe, which is the dominant seed isotope for s-process nucleosynthesis.  $\tau_n$ is defined as $\tau_n = \int v_T n_n dt$ \citep[][]{1968psen.book.....C}. However, these definitions are of limited use in the multi-zone calculations used here.  The reason for this is that in the multi-zone stellar models, convective mixing affects the neutron irradiance experienced by a given mass element \citep{2007ApJ...655.1058T}.  Stellar matter, including the neutron sources, seeds and poisons, is mixed into and out of the bottom of the convection zone, where the temperature is highest and where the majority of the s process occurs.  Consequently, an evaluation of $n_c$ or $\tau_n$ at a particular mass coordinate will be different to that experienced by a given mass element.

Therefore, in order to evaluate relevant parameters to describe the neutron irradiance, convective mixing needs to be taken into account in the evaluation of the parameter.  This can be achieved for the neutron exposure by considering the initial and final abundances of $^{54}$Fe, an isotope that is slowly destroyed by neutron captures in the s-process sites considered here.  It cannot be used during or after oxygen burning where temperatures are high enough to photodisintegrate heavy elements \citep{1995ApJS..101..181W}.  An estimate of the neutron exposure using $^{54}$Fe can be made using the following formula \citep{1995ApJS..101..181W, 2000ApJ...533..998T},

\begin{equation}
\tau_{54} = -\frac{1}{\sigma} [\ln X_{\rm i}(^{54}{\rm Fe}) - \ln X_{\rm f}(^{54}{\rm Fe})], \label{eqn:fe54taun}
\end{equation}
where $\sigma$ is the $^{54}$Fe(n,$\gamma$)$^{55}$Fe reaction rate \citep[$\sigma = 29.6 \pm 1.3$ mb,][]{2006AIPC..819..123D} and $X_{\rm i}(^{54}{\rm Fe})$ and $X_{\rm f}(^{54}{\rm Fe})$ are the mass fraction abundances of $^{54}$Fe before and after the neutron exposure respectively.  A better estimate of $n_c$ can be obtained by using mass-averaged abundances for $X_i$, $X^0_i$ and $X_{^{56}\rm Fe}$ over the maximum size of the convective region,

\begin{equation}
n_{c,{\rm av}} = \frac{\sum_i^n (A_i-56) (\langle X_i\rangle - \langle X^0_i\rangle)}{\langle X_{^{56}\rm Fe}\rangle}.\label{eqn:ncapav}
\end{equation}
This takes into account any changes to the size of the convective region during the burning stage where the s-process nucleosynthesis occurs.

Table \ref{tab:neut} lists, for all models, the neutron exposure, $\tau_{54}$, the neutron captures per iron seed, $n_{c, {\rm av}}$, the mass fraction abundances of the isotopes $^{54}$Fe and $^{88}$Sr and the isobaric ratios $^{70}$Ge/$^{70}$Zn, $^{80}$Kr/$^{80}$Se and $^{86}$Sr/$^{86}$Kr.  $^{88}$Sr, like $^{54}$Fe, is also a useful s-process indicator as it has a neutron-magic nucleus ($N=50$) and is slowly built-up over the course of the s process.  The isobaric ratios are also specified, because changes to the ratios demonstrate deviations to the s-process path at branching point nuclides ($^{69}$Zn, $^{79}$Se and $^{85}$Kr for $^{70}$Ge/$^{70}$Zn, $^{80}$Kr/$^{80}$Se and $^{86}$Sr/$^{86}$Kr respectively).  Indeed, if the neutron density increases, the s-process path opens to allow the production of more neutron-rich isotopes, lowering these ratios.

\begin{table*}
\caption{S-process tracers, neutron capture parameters and isotopic ratios at the end of helium-core burning, carbon-core burning and convective carbon-shell burning.  $n_{c,{\rm av}}$ is the neutron captures per iron seed averaged over the convective region and $\tau_{54}$ is the neutron exposure calculated using Eq. \ref{eqn:fe54taun}.  The $^{88}$Sr and $^{54}$Fe abundances are specified as average mass fraction abundances, $X_{^{88}{\rm Sr}}$ and $X_{^{54}{\rm Fe}}$ respectively, at the end of the burning stage over the convective region, except for radiative burning where the central values are taken.  The s-process parameters for a shell that persists to the presupernova stage use final abundances that are evaluated at start of oxygen burning, which removes the effects of photodisintegration occurring during the late evolutionary stages from the evaluation of the s-process parameters.}
\begin{tabular}{lcccrrrrr}
\hline
\hline
Model & Shell & $^{88}$Sr & $^{54}$Fe & $n_{c,{\rm av}}$ & $\tau_{54}$ (mb$^{-1}$) & $^{70}$Ge/$^{70}$Zn & $^{80}$Kr/$^{80}$Se & $^{86}$Sr/$^{86}$Kr \\
\hline
 15ST & He-core & $2.005 \times 10^{-7}$ & $5.750 \times 10^{-6}$  & $1.641$  & $0.088$ & 115.913  & 2.690  & 4.247  \\
 15ST & C-core  & $1.556 \times 10^{-6}$ & $7.721 \times 10^{-7}$  & $6.601$  & $0.062$ & 1165.633 & 5.107  & 46.001 \\
 15ST &     1   & $1.000 \times 10^{-6}$ & $1.089 \times 10^{-6}$  & $4.740$  & $0.048$ & 1036.915 & 3.668  & 20.178 \\
 15ST &     2   & $6.629 \times 10^{-7}$ & $1.266 \times 10^{-6}$  & $3.903$  & $0.042$ & 335.818  & 0.701  & 2.708  \\
 15CI & C-core  & $1.009 \times 10^{-4}$ & $9.137 \times 10^{-8}$  & $29.270$ & $0.134$ & 901.882  & 4.284  & 45.048 \\
 15CI &     1   & $2.803 \times 10^{-5}$ & $6.958 \times 10^{-7}$  & $6.005$  & $0.059$ & 862.687  & 3.172  & 23.268 \\
 15CU & C-core  & $2.182 \times 10^{-4}$ & $3.716 \times 10^{-8}$  & $46.293$ & $0.165$ & 743.822  & 4.080  & 44.065 \\
 15CU &     1   & $5.046 \times 10^{-5}$ & $2.163 \times 10^{-6}$  & $19.423$ & $0.055$ & 638.189  & 0.765  & 1.726  \\
 20ST & He-core & $3.817 \times 10^{-7}$ & $1.070 \times 10^{-6}$  & $3.069$  & $0.143$ & 928.859  & 3.588  & 7.503  \\
 20ST & C-core  & $1.286 \times 10^{-6}$ & $1.615 \times 10^{-7}$  & $8.080$  & $0.062$ & 1315.250 & 4.012  & 30.741 \\
 20ST &     1   & $1.064 \times 10^{-6}$ & $2.303 \times 10^{-7}$  & $6.605$  & $0.043$ & 1245.114 & 2.605  & 17.583 \\
 20ST &     2   & $9.403 \times 10^{-7}$ & $3.382 \times 10^{-7}$  & $4.934$  & $0.033$ & 518.314  & 0.774  & 4.205  \\
 20ST &     3   & $8.762 \times 10^{-7}$ & $4.292 \times 10^{-7}$  & $0.119$  & $0.001$ & 487.403  & 0.696  & 3.802  \\
 20CI & C-core  & $5.197 \times 10^{-5}$ & $8.818 \times 10^{-8}$  & $27.796$ & $0.084$ & 970.039  & 4.200  & 41.853 \\
 20CI &     1   & $2.424 \times 10^{-5}$ & $3.828 \times 10^{-7}$  & $5.920$  & $0.023$ & 975.182  & 2.873  & 20.450 \\
 20CI &     2   & $2.160 \times 10^{-5}$ & $3.869 \times 10^{-7}$  & $2.737$  & $0.012$ & 347.183  & 0.366  & 3.352  \\
 20CU & C-core  & $1.727 \times 10^{-4}$ & $4.802 \times 10^{-9}$  & $60.722$ & $0.182$ & 779.749  & 4.104  & 36.648 \\
 20CU &     1   & $7.074 \times 10^{-5}$ & $5.484 \times 10^{-7}$  & $4.073$  & $0.019$ & 494.139  & 2.019  & 22.567 \\
 20CU &     2   & $1.194 \times 10^{-5}$ & $6.573 \times 10^{-7}$  & $4.651$  & $0.027$ & 151.579  & 0.348  & 4.048  \\
 25ST & He-core & $6.153 \times 10^{-7}$ & $3.539 \times 10^{-7}$  & $4.280$  & $0.180$ & 2220.036 & 3.755  & 11.329 \\
 25ST & C-core  & $1.472 \times 10^{-6}$ & $7.918 \times 10^{-8}$  & $8.271$  & $0.045$ & 1432.597 & 4.385  & 35.554 \\
 25ST &     1   & $9.499 \times 10^{-7}$ & $1.482 \times 10^{-7}$  & $5.632$  & $0.028$ & 87.609   & 0.109  & 0.515  \\
 25CI & C-core  & $4.092 \times 10^{-5}$ & $1.411 \times 10^{-9}$  & $48.421$ & $0.179$ & 970.416  & 4.576  & 59.426 \\
 25CI &     1   & $1.772 \times 10^{-5}$ & $6.313 \times 10^{-8}$  & $23.538$ & $0.045$ & 1063.729 & 4.066  & 38.990 \\
 25CI &     2   & $1.111 \times 10^{-6}$ & $1.564 \times 10^{-7}$  & $5.543$  & $0.028$ & 315.357  & 0.280  & 1.401  \\
 25CU & C-core  & $1.475 \times 10^{-4}$ & $1.509 \times 10^{-9}$  & $73.339$ & $0.184$ & 804.018  & 4.072  & 36.419 \\
 25CU &     1   & $9.824 \times 10^{-5}$ & $1.347 \times 10^{-7}$  & $15.755$ & $0.015$ & 698.157  & 1.283  & 10.094 \\
 32ST & He-core & $1.097 \times 10^{-6}$ & $1.192 \times 10^{-7}$  & $5.623$  & $0.217$ & 3380.614 & 3.900  & 16.340 \\
 32ST & C-core  & $1.788 \times 10^{-6}$ & $5.333 \times 10^{-8}$  & $6.239$  & $0.024$ & 1640.445 & 3.640  & 28.449 \\
 32ST &     1   & $1.315 \times 10^{-6}$ & $8.625 \times 10^{-8}$  & $3.016$  & $0.010$ & 75.996   & 0.130  & 1.014  \\
 32CI & C-core  & $1.825 \times 10^{-5}$ & $3.955 \times 10^{-9}$  & $38.296$ & $0.110$ & 1042.993 & 4.740  & 60.126 \\
 32CI &     1   & $2.045 \times 10^{-6}$ & $6.562 \times 10^{-8}$  & $5.220$  & $0.017$ & 1021.836 & 1.646  & 9.944  \\
 32CU & C-core  & $1.007 \times 10^{-4}$ & $8.498 \times 10^{-10}$ & $77.718$ & $0.167$ & 837.791  & 3.949  & 39.032 \\
 32CU &     1   & $7.633 \times 10^{-5}$ & $3.346 \times 10^{-8}$  & $16.738$ & $0.011$ & 509.651  & 0.428  & 4.911  \\
 60ST & He-core & $1.524 \times 10^{-6}$ & $6.404 \times 10^{-8}$  & $6.489$  & $0.238$ & 1741.270 & 1.125  & 12.267 \\
 60ST & C-core  & $1.701 \times 10^{-6}$ & $5.297 \times 10^{-8}$  & $5.862$  & $0.023$ & 1743.568 & 3.246  & 25.865 \\
 60ST &     1   & $1.335 \times 10^{-6}$ & $7.814 \times 10^{-8}$  & $2.779$  & $0.009$ & 69.670   & 0.146  & 1.136  \\
 60CI & C-core  & $1.491 \times 10^{-5}$ & $4.808 \times 10^{-9}$  & $33.897$ & $0.104$ & 1072.384 & 4.619  & 52.637 \\
 60CI &     1   & $1.622 \times 10^{-6}$ & $5.837 \times 10^{-7}$  & $3.800$  & $0.029$ & 871.777  & 0.921  & 5.676  \\
 60CU & C-core  & $1.076 \times 10^{-4}$ & $6.551 \times 10^{-10}$ & $81.743$ & $0.172$ & 837.512  & 3.877  & 36.865 \\
 60CU &     1   & $8.908 \times 10^{-5}$ & $2.512 \times 10^{-8}$  & $17.940$ & $0.010$ & 455.999  & 0.370  & 4.862  \\
\hline
\end{tabular}
\label{tab:neut}
\end{table*}

\subsection{Core carbon burning}

\begin{figure*}
\includegraphics[width=\textwidth]{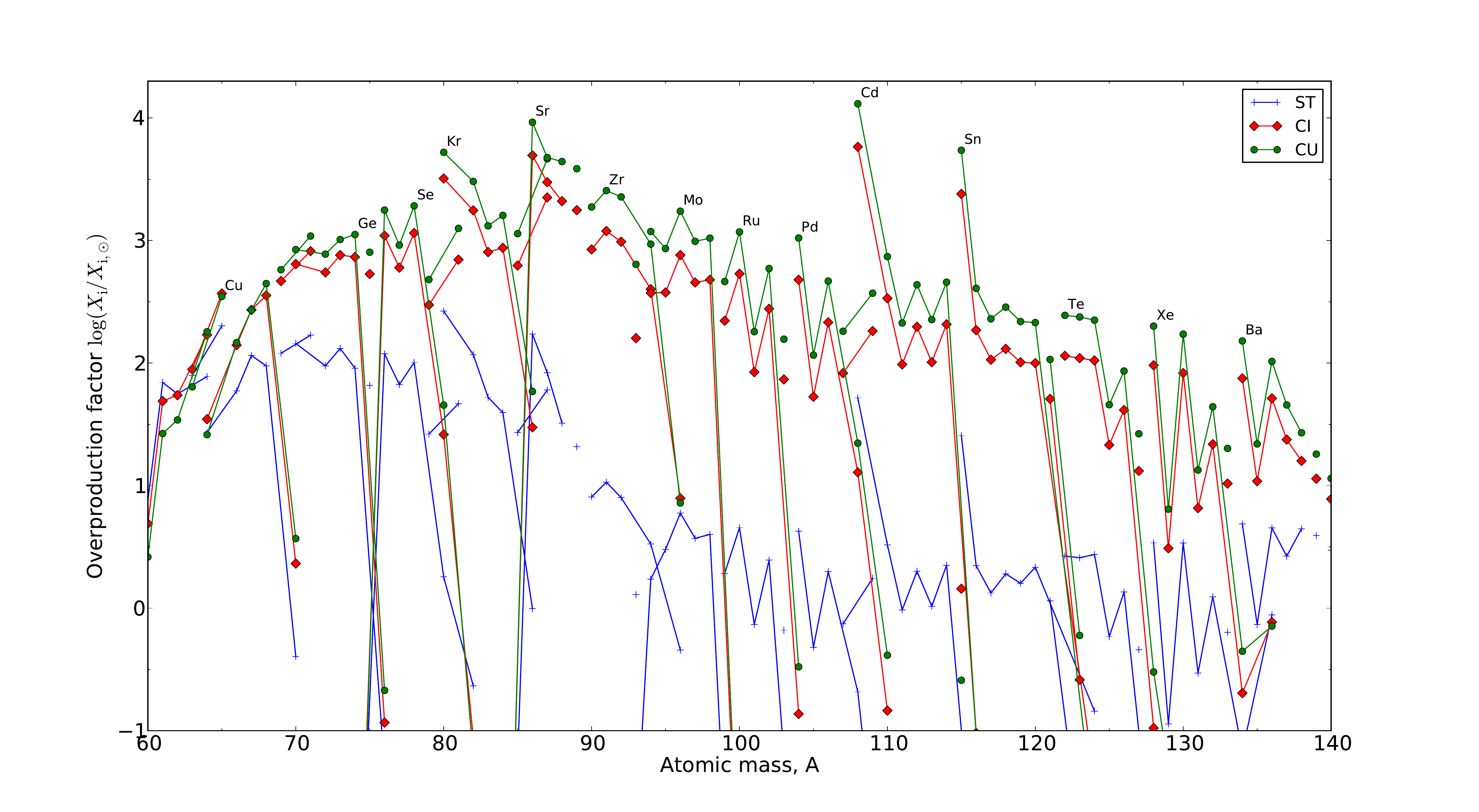}
\caption{Central overproduction factors for most stable isotopes at the end of central carbon burning for the 15$M_{\odot}$ models.  The plot shows a significant increase in nucleosynthesis of isotopes between $60 < A < 140$ in the CI and CU models, which is beyond the Sr-Y-Zr peak at an atomic mass $\approx 90$.}
\label{fig:15Ccore}
\end{figure*}

According to Table \ref{tab:neut}, all CI and CU models show a depletion of $^{54}$Fe and production of $^{88}$Sr relative to the ST case, indicating that a higher neutron exposure is present in the convective carbon core.  For all CI and CU models, irrespective of mass, the neutron exposure is high enough to allow an increasing production of isotopes beyond the Sr-Y-Zr peak, which is quantified in a higher neutron captures per iron seed.  An example of this nucleosynthesis for the 15$M_{\odot}$ model is seen in Fig. \ref{fig:15Ccore}, which shows the central overproduction factors for heavy, stable isotopes in the star at the end of carbon burning.  The distribution of synthesised isotopes is extended, with increasing rate, beyond the Sr-Y-Zr peak to include isotopes up to the Ba-La peak at $A \approx 140$.  This is an anomalous distribution compared to the weak s-process component.

The neutron density in the carbon core decreases from a typical value of $\sim 10^8$ cm$^{-3}$, which is maintained throughout the burning, to $\sim 10^7$ cm$^{-3}$ in the models with an increasing carbon burning rate.  In the 25CU, 32CU and 60CU models the neutron density is enhanced over the CI cases because of the presence of the convective core; the mixing into and out of the centre acts to maintain a supply of neutron sources at the centre.  Concerning the ST case, the neutron exposures for the cores are similar in magnitude to that of the helium burning core ($\sim 0.06$ mb$^{-1}$), but are lower for the most massive stars considered here ($\sim 0.02$ mb$^{-1}$ for the 32ST and 60ST models).  For the CI and CU rates the neutron exposures are significantly enhanced, typically exceeding $0.1$ mb$^{-1}$.  This is mainly due to the rising efficiency of the $^{13}$C neutron source at lower temperatures, coupled with the increased lifetime of the core carbon burning stage.

\subsection{Carbon shell burning} 

Nucleosynthesis in the carbon shells is characterised by the s process with a high neutron density but lower neutron exposure compared to carbon core, with $^{22}$Ne being the dominant neutron source.  In the ST models, the neutron densities vary from $\sim 10^8$ cm$^{-3}$ for early convective shells (models 15ST and 20ST), and increase to a typical value of $\sim 10^{10}$ cm$^{-3}$ in the final carbon burning shell.  In the CI and CU models, the neutron density is $\sim 10^{7}$ cm$^{-3}$ in early shells, similar to the values obtained during core carbon burning, and then rises to $\sim 10^{8-9}$ cm$^{-3}$.  The lifetimes for the carbon shell burning stages vary quite differently from model to model, but are generally increasing with increasing rate.  For example, in the 15CU case, the lifetimes of the last carbon shell in Table \ref{tab:Cprops} for the 15ST, CI and CU models are $17.92$, $150.1$ and $594.2$ years respectively.  The carbon shell in model 15CU consequently exhibits a strong neutron exposure of similar magnitude to the carbon core (see Table \ref{tab:neut}).  It should be noted however that in almost every instance of a carbon burning shell, the neutron exposure is smaller than that of the carbon core in the same model.  This asserts the fact that carbon shells are characterised by a lower neutron exposure and higher neutron density (with $^{22}$Ne as the main neutron source), although the degree with which this is true is reduced with an increasing carbon burning rate.  That is, the general trend with increasing rate is a decrease in the neutron density and an increase in the neutron exposure in the carbon shells.

The above can be verified by considering the ratios of isotopes involved at branching points, since the lower neutron density will close the s-process path to the synthesis of more neutron-rich isotopes at branching points.  The last three columns of Table \ref{tab:neut} show the isobaric ratios at the end of the core and shell carbon burning stages for $^{70}$Ge/$^{70}$Zn, $^{80}$Kr/$^{80}$Se and $^{86}$Sr/$^{86}$Kr, with values for the end of helium core burning specified for reference.  For most models, the ratios increase in the last carbon shell with increasing carbon burning rate, favouring production of the s-only isotopes $^{70}$Ge, $^{80}$Kr and $^{86}$Sr, due to the lower neutron density in the carbon-shells in the CI and CU models.  However, the ratios are sensitive to convection, since shell overlap causes the shells to be polluted with carbon core s-process ashes.  Consequently, the 25CU, 32CU and 60CU models instead show a decrease in the ratios.  Considering that the ratios in the initial composition are 3.271, 6.124 and 0.036 for $^{70}$Ge/$^{70}$Zn, $^{80}$Kr/$^{80}$Se and $^{86}$Sr/$^{86}$Kr, the presence of lower isobaric ratios than these in the shells indicates that the branching is indeed affected during the carbon shell s-process and that the decrease is not associated purely with the mixing of carbon core matter with helium burning ashes.


\section{Yields}\label{sec:Yields}

\subsection{Calculations}\label{sec:calcs}

The yields calculations were made in the same manner as that of \citet{2005A&A...433.1013H}, which considers two contributions to the yields: the stellar wind and the supernova explosion.  The wind yield for nuclide $i$ for a star with initial mass $m$ is calculated using: 

\begin{equation}
mp^{\rm wind}_{im} = \int^{\tau(m)}_0 \dot{M}(m,t)[X^S_i(m,t) - X^0_i] dt \label{eqn:windyield1}
\end{equation}
where $\tau(m)$ is the final age of the star, $\dot{M}(m,t)$ is the mass loss rate, $X^S_i$ is the surface mass-fraction abundance, $X^0_i$ is the initial mass-fraction abundance.  The majority of the matter lost through the stellar wind occurs during hydrogen and helium burning.  The composition of the wind is similar to that of the initial composition, except for the 32$M_{\sun}$ and 60$M_{\sun}$ models where the mass loss is significant enough to include some of the hydrogen burning ashes.  Table \ref{tab:cores} shows that the total mass lost over the stellar evolution due to the stellar wind increases significantly with initial mass ($\approx 20$ per cent lost for the 15 $M_{\sun}$ models to $\approx 80$ per cent lost for the 60 $M_{\sun}$ models).

The presupernova yields are calculated using:

\begin{equation}
mp^{\rm preSN}_{im} = \int^{m_{\tau}}_{M_{{\rm rem},m}} [X_i(m_r) - X^0_i] dm_r \label{eqn:presn1}
\end{equation}
where $m_{\tau}$ is the total mass of the star at $\tau(m)$, $M_{{\rm rem},m}$ is the remnant mass, $X^0_i$ is the initial mass fraction abundance of element $i$ and $X_i(m_r)$ is the mass fraction abundance at mass coordinate $m_r$.  The total yields are then just the sum of the wind and the presupernova yields.  The calculated yields of selected isotopes for model 15ST are shown in Table \ref{tab:yields15ST} (full yield tables for all models are provided with the electronic edition of this paper).

\begin{table*}
\caption{Yields for model 15ST.  For each isotope, $i$, the atomic mass (A), atomic number (Z), initial mass fraction abundance ($X^0_i$), wind yield ($mp^{\rm wind}$, in $M_{\sun}$), presupernova yield ($mp^{\rm preSN}$, in $M_{\sun}$), total yield ($mp^{\rm total}$, in $M_{\sun}$), total ejected mass ($E_{im}$, in $M_{\sun}$) and average overproduction factor ($\langle OP\rangle$) are specified.  The decays of unstable species to their stable isobars are taken into account.}
\begin{tabular}{lrrrrrrrr}
\hline
\hline
Isotope & A & Z & $X^0_i \quad$ & $mp^{\rm wind}$ \quad & $mp^{\rm preSN}$ \quad & $mp^{\rm total}$ \quad & $E_{im}$ \quad \quad & $\langle OP \rangle$ \\
\hline
$^{1}$H & 1 & 1 & 7.064E-01 & -4.366E-02 & -2.933E+00 & -2.977E+00 & 6.485E+00 & 0.685 \\
$^{4}$He & 4 & 2 & 2.735E-01 & 4.345E-02 & 1.435E+00 & 1.479E+00 & 5.142E+00 & 1.404 \\
$^{12}$C & 12 & 6 & 3.425E-03 & -2.639E-03 & 3.101E-01 & 3.074E-01 & 3.533E-01 & 7.703 \\
$^{13}$C & 13 & 6 & 4.156E-05 & 2.302E-04 & 2.276E-04 & 4.577E-04 & 1.014E-03 & 1.822 \\
$^{14}$N & 14 & 7 & 1.059E-03 & 4.132E-03 & 3.401E-02 & 3.814E-02 & 5.232E-02 & 3.689 \\
$^{16}$O & 16 & 8 & 9.624E-03 & -1.474E-03 & 7.579E-01 & 7.564E-01 & 8.853E-01 & 6.868 \\
$^{19}$F & 19 & 9 & 5.611E-07 & -9.796E-08 & -2.190E-06 & -2.288E-06 & 5.227E-06 & 0.696 \\
$^{20}$Ne & 20 & 10 & 1.818E-03 & -2.514E-06 & 3.238E-01 & 3.238E-01 & 3.482E-01 & 14.302 \\
$^{23}$Na & 23 & 11 & 4.000E-05 & 3.023E-05 & 1.337E-02 & 1.340E-02 & 1.394E-02 & 26.021 \\
$^{24}$Mg & 24 & 12 & 5.862E-04 & -1.079E-08 & 2.747E-02 & 2.747E-02 & 3.532E-02 & 4.498 \\
$^{27}$Al & 27 & 13 & 6.481E-05 & 4.579E-08 & 3.142E-03 & 3.142E-03 & 4.010E-03 & 4.620 \\
$^{28}$Si & 28 & 14 & 7.453E-04 & -1.752E-08 & 1.844E-03 & 1.844E-03 & 1.183E-02 & 1.185 \\
$^{31}$P & 31 & 15 & 7.106E-06 & 1.394E-09 & 7.106E-05 & 7.106E-05 & 1.662E-04 & 1.747 \\
$^{32}$S & 32 & 16 & 4.011E-04 & -9.512E-09 & -1.897E-04 & -1.897E-04 & 5.182E-03 & 0.965 \\
$^{36}$Ar & 36 & 18 & 8.202E-05 & -1.944E-09 & -7.472E-05 & -7.472E-05 & 1.024E-03 & 0.932 \\
$^{39}$K & 39 & 19 & 3.900E-06 & -9.244E-11 & 7.466E-06 & 7.466E-06 & 5.970E-05 & 1.143 \\
$^{40}$Ca & 40 & 20 & 7.225E-05 & -1.706E-09 & -5.212E-05 & -5.212E-05 & 9.156E-04 & 0.946 \\
$^{45}$Sc & 45 & 21 & 5.414E-08 & -1.283E-12 & 8.303E-07 & 8.303E-07 & 1.555E-06 & 2.145 \\
$^{50}$Ti & 50 & 22 & 2.208E-07 & -5.234E-12 & 3.801E-06 & 3.801E-06 & 6.758E-06 & 2.285 \\
$^{51}$V & 51 & 23 & 4.138E-07 & -9.808E-12 & -6.535E-08 & -6.536E-08 & 5.476E-06 & 0.988 \\
$^{52}$Cr & 52 & 24 & 1.658E-05 & -3.929E-10 & -1.282E-05 & -1.282E-05 & 2.092E-04 & 0.942 \\
$^{55}$Mn & 55 & 25 & 1.098E-05 & -2.603E-10 & 3.666E-06 & 3.666E-06 & 1.507E-04 & 1.025 \\
$^{54}$Fe & 54 & 26 & 8.118E-05 & -1.924E-09 & -1.208E-04 & -1.208E-04 & 9.665E-04 & 0.889 \\
$^{56}$Fe & 56 & 26 & 1.322E-03 & -3.133E-08 & -1.213E-03 & -1.213E-03 & 1.649E-02 & 0.931 \\
$^{59}$Co & 59 & 27 & 3.991E-06 & -9.461E-11 & 2.580E-04 & 2.580E-04 & 3.114E-04 & 5.825 \\
$^{60}$Ni & 60 & 28 & 2.276E-05 & -5.394E-10 & 1.437E-04 & 1.437E-04 & 4.485E-04 & 1.472 \\
$^{63}$Cu & 63 & 29 & 6.600E-07 & -1.564E-11 & 5.493E-05 & 5.493E-05 & 6.376E-05 & 7.213 \\
$^{65}$Cu & 65 & 29 & 3.035E-07 & -7.193E-12 & 3.249E-05 & 3.249E-05 & 3.655E-05 & 8.993 \\
$^{64}$Zn & 64 & 30 & 1.131E-06 & -2.680E-11 & 1.792E-05 & 1.792E-05 & 3.306E-05 & 2.183 \\
$^{66}$Zn & 66 & 30 & 6.690E-07 & -1.586E-11 & 1.856E-05 & 1.856E-05 & 2.752E-05 & 3.072 \\
$^{70}$Zn & 70 & 30 & 1.577E-08 & -3.737E-13 & -1.160E-08 & -1.160E-08 & 1.996E-07 & 0.945 \\
$^{69}$Ga & 69 & 31 & 4.551E-08 & -1.079E-12 & 2.367E-06 & 2.367E-06 & 2.977E-06 & 4.884 \\
$^{71}$Ga & 71 & 31 & 3.108E-08 & -7.366E-13 & 2.012E-06 & 2.012E-06 & 2.428E-06 & 5.834 \\
$^{70}$Ge & 70 & 32 & 5.157E-08 & -1.222E-12 & 3.185E-06 & 3.185E-06 & 3.876E-06 & 5.611 \\
$^{72}$Ge & 72 & 32 & 6.910E-08 & -1.638E-12 & 2.614E-06 & 2.614E-06 & 3.539E-06 & 3.824 \\
$^{75}$As & 75 & 33 & 1.430E-08 & -3.390E-13 & 4.113E-07 & 4.113E-07 & 6.028E-07 & 3.147 \\
$^{76}$Se & 76 & 34 & 1.296E-08 & -3.072E-13 & 6.260E-07 & 6.260E-07 & 7.995E-07 & 4.606 \\
$^{78}$Se & 78 & 34 & 3.376E-08 & -8.003E-13 & 1.441E-06 & 1.441E-06 & 1.894E-06 & 4.188 \\
$^{80}$Se & 80 & 34 & 7.226E-08 & -1.713E-12 & 2.985E-07 & 2.985E-07 & 1.266E-06 & 1.308 \\
$^{79}$Br & 79 & 35 & 1.389E-08 & -3.293E-13 & 1.867E-07 & 1.867E-07 & 3.728E-07 & 2.003 \\
$^{81}$Br & 81 & 35 & 1.386E-08 & -3.285E-13 & 2.041E-07 & 2.041E-07 & 3.897E-07 & 2.100 \\
$^{80}$Kr & 80 & 36 & 2.575E-09 & -6.103E-14 & 2.610E-07 & 2.610E-07 & 2.955E-07 & 8.569 \\
$^{82}$Kr & 82 & 36 & 1.320E-08 & -3.128E-13 & 7.028E-07 & 7.028E-07 & 8.795E-07 & 4.977 \\
$^{84}$Kr & 84 & 36 & 6.602E-08 & -1.565E-12 & 1.031E-06 & 1.031E-06 & 1.915E-06 & 2.166 \\
$^{86}$Kr & 86 & 36 & 2.044E-08 & -4.846E-13 & 1.289E-07 & 1.289E-07 & 4.027E-07 & 1.471 \\
$^{85}$Rb & 85 & 37 & 1.282E-08 & -3.040E-13 & 1.721E-07 & 1.721E-07 & 3.438E-07 & 2.002 \\
$^{87}$Rb & 87 & 37 & 5.063E-09 & -2.025E-12 & 6.776E-08 & 6.776E-08 & 1.356E-07 & 1.999 \\
$^{84}$Sr & 84 & 38 & 3.228E-10 & -7.651E-15 & -6.777E-10 & -6.777E-10 & 3.646E-09 & 0.843 \\
$^{86}$Sr & 86 & 38 & 5.845E-09 & -1.385E-13 & 3.642E-07 & 3.642E-07 & 4.424E-07 & 5.652 \\
$^{87}$Sr & 87 & 38 & 4.443E-09 & 1.800E-12 & 1.858E-07 & 1.858E-07 & 2.453E-07 & 4.123 \\
$^{88}$Sr & 88 & 38 & 5.011E-08 & -1.188E-12 & 5.602E-07 & 5.602E-07 & 1.231E-06 & 1.835 \\
$^{89}$Y & 89 & 39 & 1.229E-08 & -2.914E-13 & 9.875E-08 & 9.875E-08 & 2.634E-07 & 1.600 \\
$^{90}$Zr & 90 & 40 & 1.534E-08 & -3.637E-13 & 4.445E-08 & 4.445E-08 & 2.500E-07 & 1.216 \\
$^{92}$Zr & 92 & 40 & 5.227E-09 & -1.239E-13 & 1.871E-08 & 1.871E-08 & 8.872E-08 & 1.267 \\
$^{94}$Zr & 94 & 40 & 5.413E-09 & -1.283E-13 & 6.178E-09 & 6.178E-09 & 7.868E-08 & 1.085 \\
$^{93}$Nb & 93 & 41 & 1.900E-09 & -4.504E-14 & 7.083E-09 & 7.082E-09 & 3.253E-08 & 1.278 \\
$^{92}$Mo & 92 & 42 & 1.012E-09 & -2.400E-14 & -1.687E-09 & -1.687E-09 & 1.187E-08 & 0.876 \\
$^{94}$Mo & 94 & 42 & 6.448E-10 & -1.528E-14 & 2.073E-11 & 2.072E-11 & 8.656E-09 & 1.002 \\
$^{96}$Mo & 96 & 42 & 1.188E-09 & -2.815E-14 & 3.811E-09 & 3.811E-09 & 1.972E-08 & 1.240 \\
$^{98}$Mo & 98 & 42 & 1.754E-09 & -4.158E-14 & 3.213E-09 & 3.213E-09 & 2.671E-08 & 1.137 \\
$^{100}$Mo & 100 & 42 & 7.146E-10 & -1.694E-14 & -1.219E-09 & -1.219E-09 & 8.352E-09 & 0.873 \\
\hline
\end{tabular}
\label{tab:yields15ST}
\end{table*} 

The point in the evolution in which the yields are taken in this work is at the end of central oxygen burning, as explained in \S \ref{sec:models}.  This choice was made since not all the models were post-processed until the end of silicon burning.  Notice that, as mentioned in \S \ref{sec:models}, after central oxygen burning, the material outside the remnant mass is not affected much by the pre-explosive evolution.  The only potential contributions that may affect the s-process abundances are during the early collapse, when the neutron density may increase significantly \citep[e.g. in the carbon shell, see][]{2010ApJ...710.1557P} or partial or complete photodisintegration at the bottom of the carbon, neon and oxygen shells.  The effects of photodisintegration will be discussed in a forthcoming paper (Pignatari et al. 2011, in prep.).

With regards to explosive burning, the supernova explosion is responsible for destroying and recreating a portion of the ejecta, which includes p-process rich and, to a smaller extent, s-process rich layers, possibly having a relevant impact on the total yields of s-process nuclides \citep[see for instance][]{2002ApJ...576..323R,2009ApJ...702.1068T}.  However, the explosive burning process is sensitive to uncertainties in the supernova explosion mechanism for the range of initial masses considered here \citep{2009APS..APR.B4001F}.  The uncertainties associated with the supernova explosion, namely the explosion energy, the ignition mechanism and the amount of fall-back, are important especially for the 15, 20 and 25 $M_{\sun}$ models.  These uncertainties would also affect the amount of matter locked up in the remnants.  In this work, the remnant mass takes into account the additional matter that falls back onto the remnant following the initial explosion.  The choice of remnant masses for the models is taken from the analytical fits of Fryer et al. (2011, in prep.) for solar metallicity stars, which derive from energy-driven explosions \citep[see for instance][]{2009APS..APR.B4001F}.  The remnant masses, $M_{{\rm rem},m}$, are given by

\begin{equation}
M_{{\rm rem},m} = \left\{ \begin{array}{ll}
		  1.1 + 0.2 e^{(m-11)/4} - 3 e^{0.4(m-26)}, & 11 < m \leq 30 \\
		  18.35 - 0.3m,                             & 30 < m < 50 \end{array} \right.
\end{equation}
which gives remnant masses of 1.61, 2.73, 5.71 and 8.75 $M_{\sun}$ for initial masses, $m$, of 15, 20, 25 and 32 $M_{\sun}$ respectively.  For the 60 $M_{\sun}$ models a remnant mass was calculated by scaling with the CO core mass ratio for the ST models,

\begin{equation}
M_{{\rm rem},60M_{\sun}} = M_{{\rm rem},32M_{\sun}} \left(\frac{M_{{\rm CO},60M_{\sun}}}{M_{{\rm CO},32M_{\sun}}}\right),
\end{equation}
giving a remnant mass of 10.24 $M_{\sun}$.  The resultant remnant masses are such that for the 15 $M_{\sun}$ models, the oxygen shell is partially included in the supernova ejecta.  For the other models however, the remnants are large and the ejecta includes the upper portion of the carbon shell and the overlying layers only.  The remnant masses here are larger in comparison with those used in previous studies of explosive nucleosynthesis \citep{2000ApJS..129..625L,2002ApJ...576..323R}.  This is due to the use, in those studies, of piston-driven models that are known to underestimate the amount of fall-back onto the supernova remnant \citep{2007ApJ...664.1033Y}.  The large remnant masses may cause the explosive nucleosynthesis to occur predominantly in the layers that fall back onto the remnant.

In addition to the yields, the ejected masses, $E_{im}$ can be calculated, which are the exact analogues of Eq. \ref{eqn:windyield1} and \ref{eqn:presn1}, but without the inclusion of the $X^0_i$ term.  If the total mass of matter ejected is $M_{{\rm ej},m} = m_{\tau} - M_{{\rm rem},m}$, the overproduction factors averaged over the ejecta are calculated using

\begin{equation}
\langle{\rm OP}\rangle_{im} = \frac{E_{im}}{M_{{\rm ej},m} X^0_i}
\end{equation}

\begin{figure*}
\includegraphics[width=\textwidth]{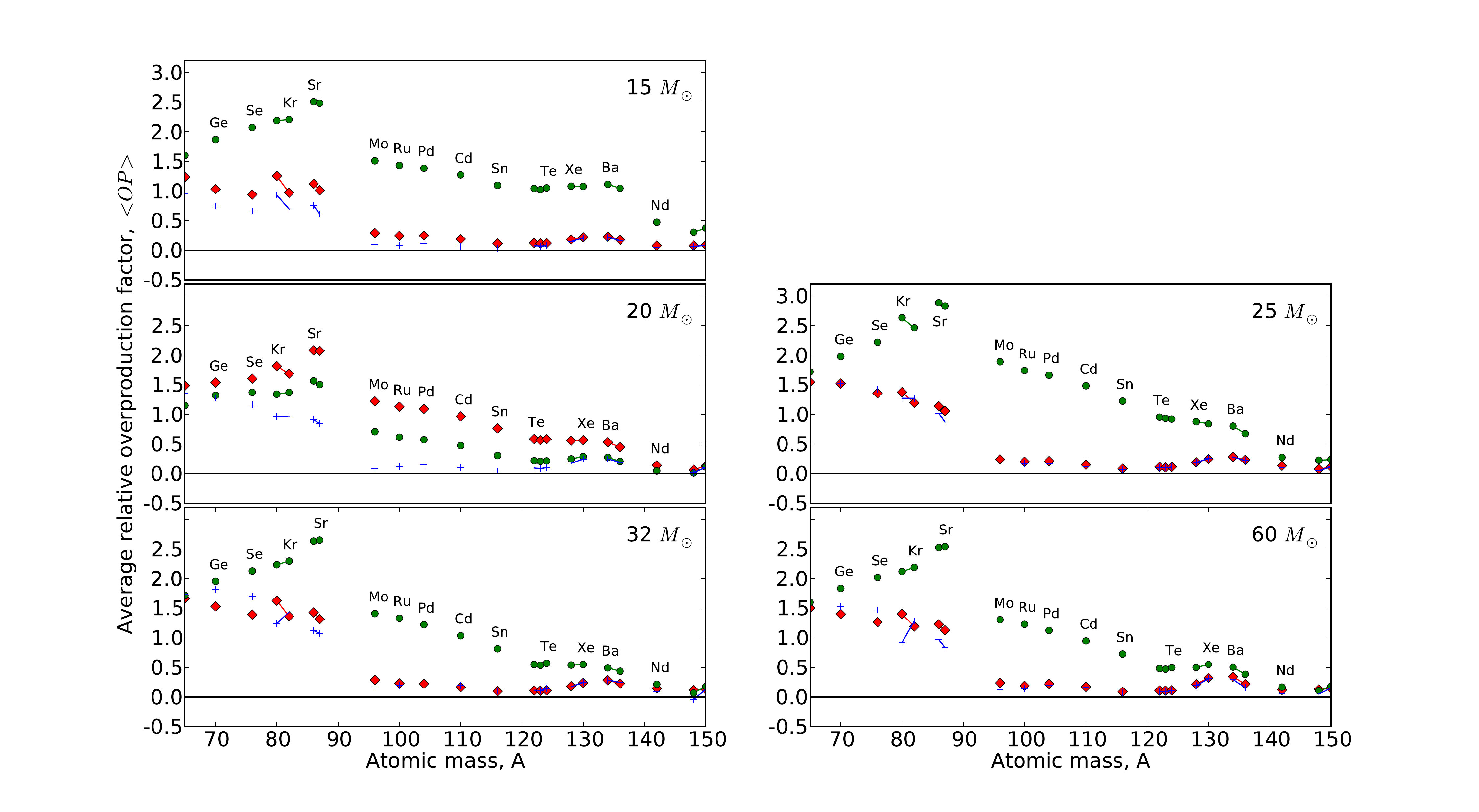}
\caption{The overproduction factors averaged over the total ejected mass for s-only nuclides as a function of atomic mass.  The ST, CI and CU rates are indicated by the blue crosses, red diamonds and green circles respectively.  Isotopes of the same element are connected by adjoining lines.}
\label{fig:sonly}
\end{figure*}

The overproduction factors averaged over the ejecta for the s-only isotopes are shown in Fig. \ref{fig:sonly}, which represents well the general abundance distribution for stable isotopes created by the models.  A considerable amount of s-process nucleosynthesis occurs for all CU models by up to 3 dex, which is either because of overlap between the carbon shells and the carbon core (for models 20CI, 25CU, 32CU and 60CU) or because of strong neutron exposures in the carbon shells (models 15CU and 20CU).  The 20CI model features a strong overlap between the convective carbon core and the successive carbon shells, which is not seen in model 20CU and therefore has more significant production than model 20CU.  In fact, for the CI rate, only the 20 $M_{\sun}$ model shows a significantly enhanced production over the ST rate.  The 15CI model also shows some production, but the distribution of isotopes is very similar to that of model 15ST.  This is in contrast to the 20CI model, which shows an extended distribution of production featuring heavier nuclides.

A first order approximation of the weak s-process component can be made by taking the sum of the yields for each stellar model, taking into account the number of stars with that initial mass formed,

\begin{equation}
y_{{\rm weak},i} = \frac{\sum_m r_m E_{im}}{\sum_m M_{{\rm ej},m} r_m},
\end{equation}
where $r_m$ is a weighting factor determined by the integration of the Salpeter initial mass function (IMF), $dN/dm = \xi_0 m^{-2.35}$, over a certain range.  Yields from the 15, 20, 25, 32 and 60 $M_{\sun}$ models were applied to stars within the initial mass ranges of 12.5-17.5, 17.5-22.5, 22.5-28.5, 28.5-46 and 46-80 $M_{\sun}$ respectively, giving values of $r_m$ equal to 39.75, 19.89, 13.45, 14.59, 12.32 per cent respectively (with $\xi_0 = 0.304$).  Consequently, the 15 and 20 $M_{\sun}$ models dominate as the main contributors to the evaluation of the weak component ($\approx 60$ per cent of all stars in the total massive star mass range considered here).  Stars with initial masses less than 12.5 $M_{\sun}$ or greater than 80 $M_{\sun}$ are assumed to have a zero contibution to the weak s-process component.

The $^{13}$C neutron source during carbon core burning is mainly primary whereas the $^{22}$Ne source is secondary\footnote{The products of nucleosynthesis processes in stars, to first order, can be described as being primary or secondary depending on whether the processes responsible for the production depend on the initial metallicity.  The production of primary nuclides does not vary with metallicity whereas secondary nuclides will be produced in proportion to their initial seed nuclei.}, since it depends on the initial $^{14}$N abundance from the CNO cycle.  If a solar metallicity star of a given mass is the dominant site for the production of particular primary and secondary nuclides, A and B, respectively, the overproduction factor for B is expected to be approximately twice that of A \citep{1971Ap&SS..14..179T}.  Although this is a rather crude approximation regarding the detailed nature of chemical evolution within galaxies and/or star clusters and the nucleosynthesis processes themselves \citep{1979ApJ...229.1046T}, the weak s process in massive stars is expected to hold reasonably to this approximation because the dominant neutron sources, seeds and poisons of the weak s-process are secondary.  It can be expected therefore that the overproduction factors for the weak s-process nuclides reproduce the solar system abundances when the overproduction factor is approximately twice that of $^{16}$O \citep{2009ApJ...702.1068T}.  In any case, this rule of thumb can be used as a rough guide to indicate the typical solar production of s-process nuclides \citep{2002ApJ...576..323R,2010ApJ...710.1557P}.

\begin{figure*}
\includegraphics[width=\textwidth]{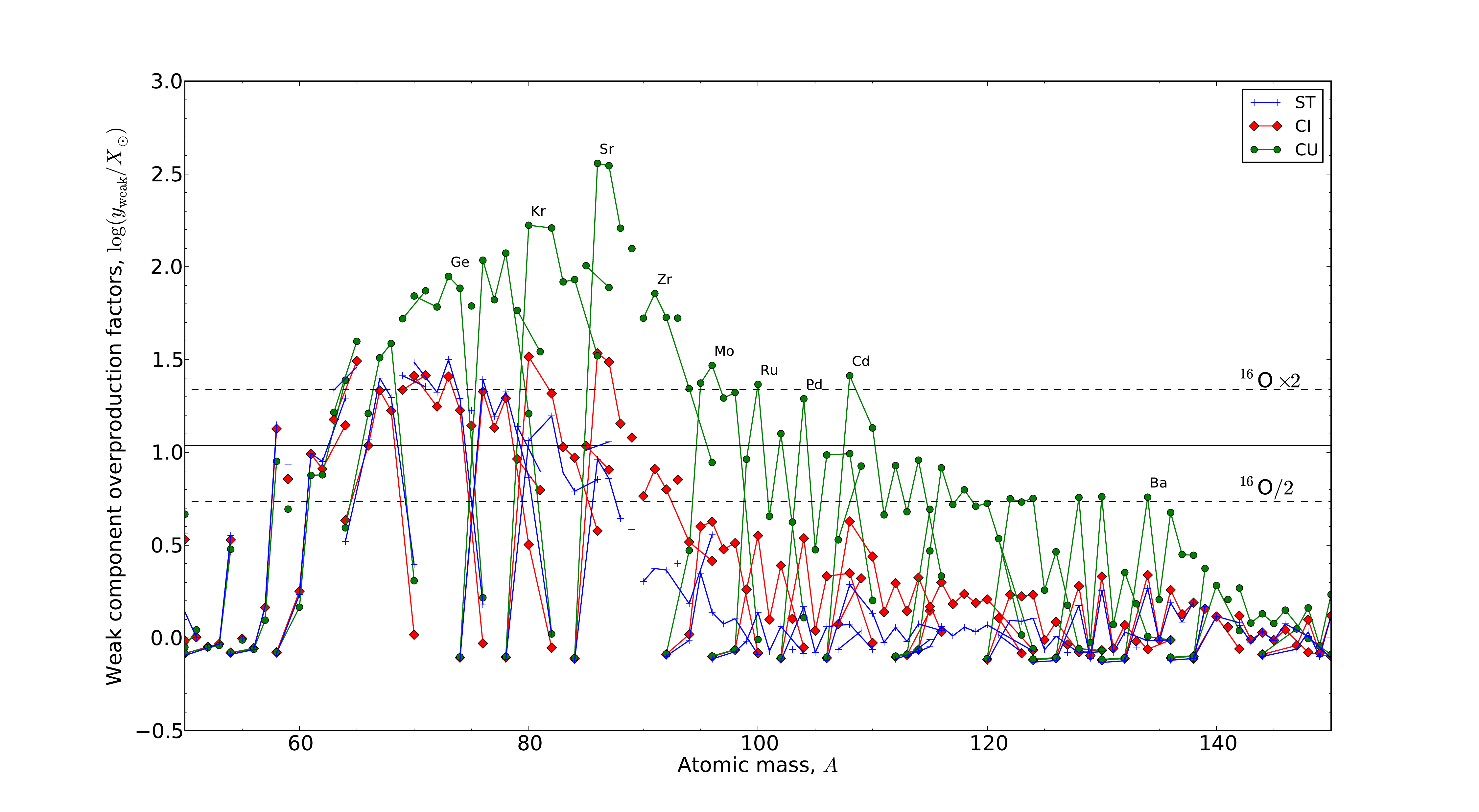}
\caption{The overproduction factors of the predicted weak component for each rate, focussing on isotopes with atomic mass $50 < A < 150$.  Isotopes of the same element are connected by adjoining lines.  The solid black line indicates the overproduction factor $^{16}$O and the two dashed lines corresponds to the overproduction factors of $^{16}$O multiplied and divided by two.  Changes to the overproduction factor of $^{16}$O are negligibly small with changes to the carbon burning rate.  The isotopic chains for Ge, Kr, Sr, Zr, Mo, Ru, Pd, Cd and Ba in the CU model are labelled for brevity.}
\label{fig:weakOP}
\end{figure*}

The overproduction factors of the weak component, $y_{{\rm weak},i}/X^0_i$, for nuclides with atomic masses $50 < A < 150$, are displayed in Fig. \ref{fig:weakOP}.  Concerning the CU rate, the overproduction factors are very large (up to $2.56$ dex for $^{86}$Sr) with respect to the ST model, with significant s-process production of nuclides up to the Ba-La peak at $A \approx 140$.  The resulting s-process distribution, peaked at the Sr-Y-Zr, is not characteristic of the weak s-process component, stopping at $A \approx 90$.  The s-process nuclides with $90 < A < 110$ have overproduction factors that are comparable to $^{16}$O multiplied by two.  Such differences for the CU case compared to the classical weak s-process component occur because of the $^{13}$C neutron source.

For the CI case, the overabundances of many nuclides are similar to the ST case, except for nuclides that are close to the Sr-Y-Zr peak or with higher atomic mass (Mo, Ru, Cd, and Pd for example).  S-process isotopes of Kr and Sr have overproduction factors that are higher than $^{16}$O multiplied by two.  The abundances of the heavier nuclides Y, Zr, Mo, Ru, Cd and Pd show an enhanced production, which is $0.5$ to $1.0$ dex lower than the Kr-Sr peak.  Overall, the resulting s-process distribution is approximately flat from Ni to Sr.

\begin{figure*}
\includegraphics[width=\textwidth]{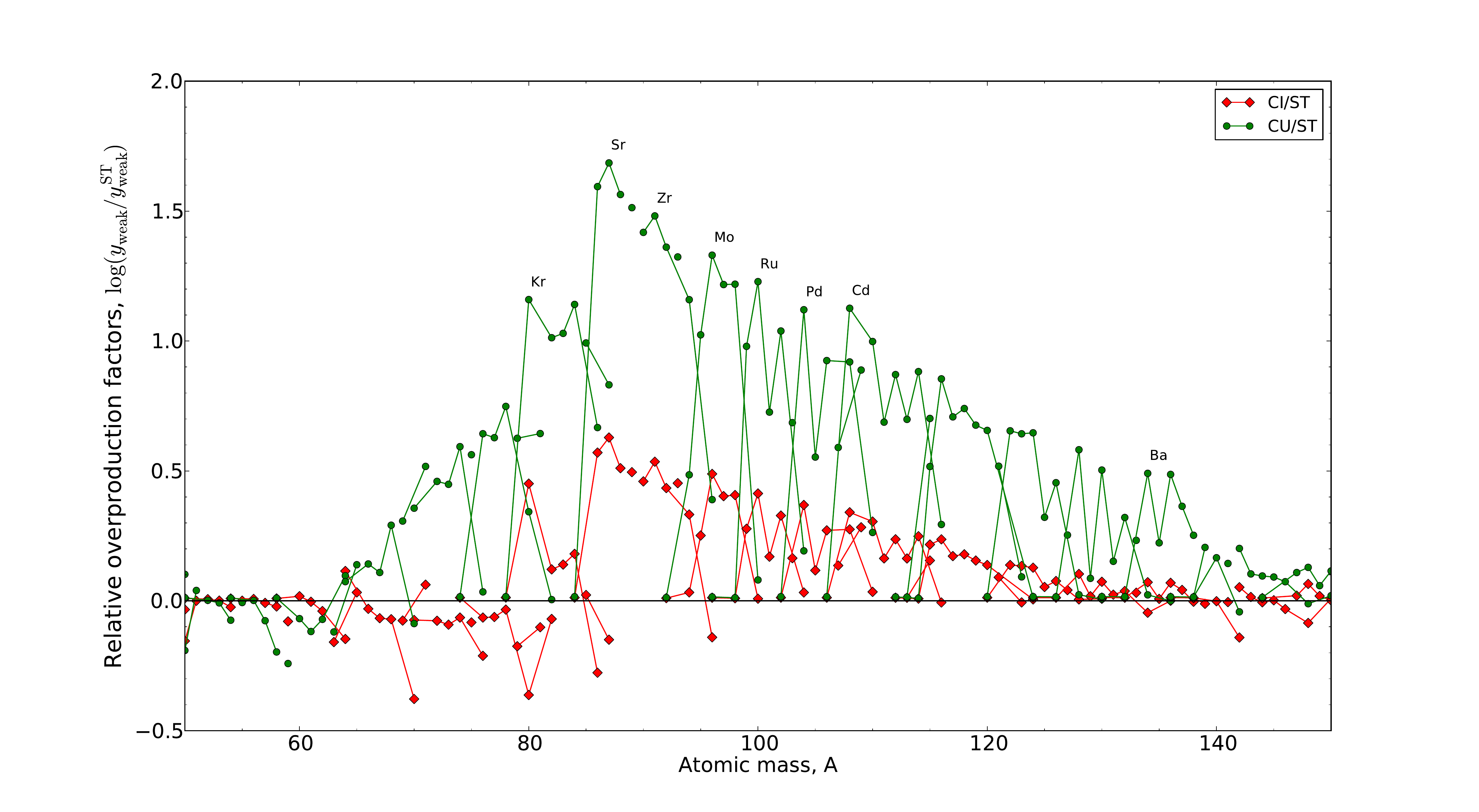}
\caption{The overproduction factors of the predicted weak component for the CI and CU rates relative to the ST rate.  Isotopes are connected by adjoining lines.  The isotopic chains for Kr, Sr, Zr, Mo, Ru, Pd, Cd and Ba in the CU model are labelled for brevity.}
\label{fig:weakrel}
\end{figure*}

Fig. \ref{fig:weakrel} shows the overproduction factors for the weak components of the CI and CU cases plotted relative to the ST case.  The peak of the relative production of s-process nuclides lies at $^{87}$Sr in both cases and declines smoothly with increasing mass number, although the overproduction factor for $^{86}$Sr is slightly larger than the $^{87}$Sr for all cases (see Fig. \ref{fig:weakOP}).  For the CU case, the overabundance of $^{87}$Sr is $1.7$ dex larger than for the ST case.  The enhancement stops at Ba, with $0.5$ dex more production and declines steeply, with a production of heavier nuclides similar to that of the ST case.  For the CI case however, the peak production at $^{87}$Sr is $0.6$ dex larger than the ST case and tends to $0.0$ at Ba.

\begin{figure*}
\includegraphics[width=\textwidth]{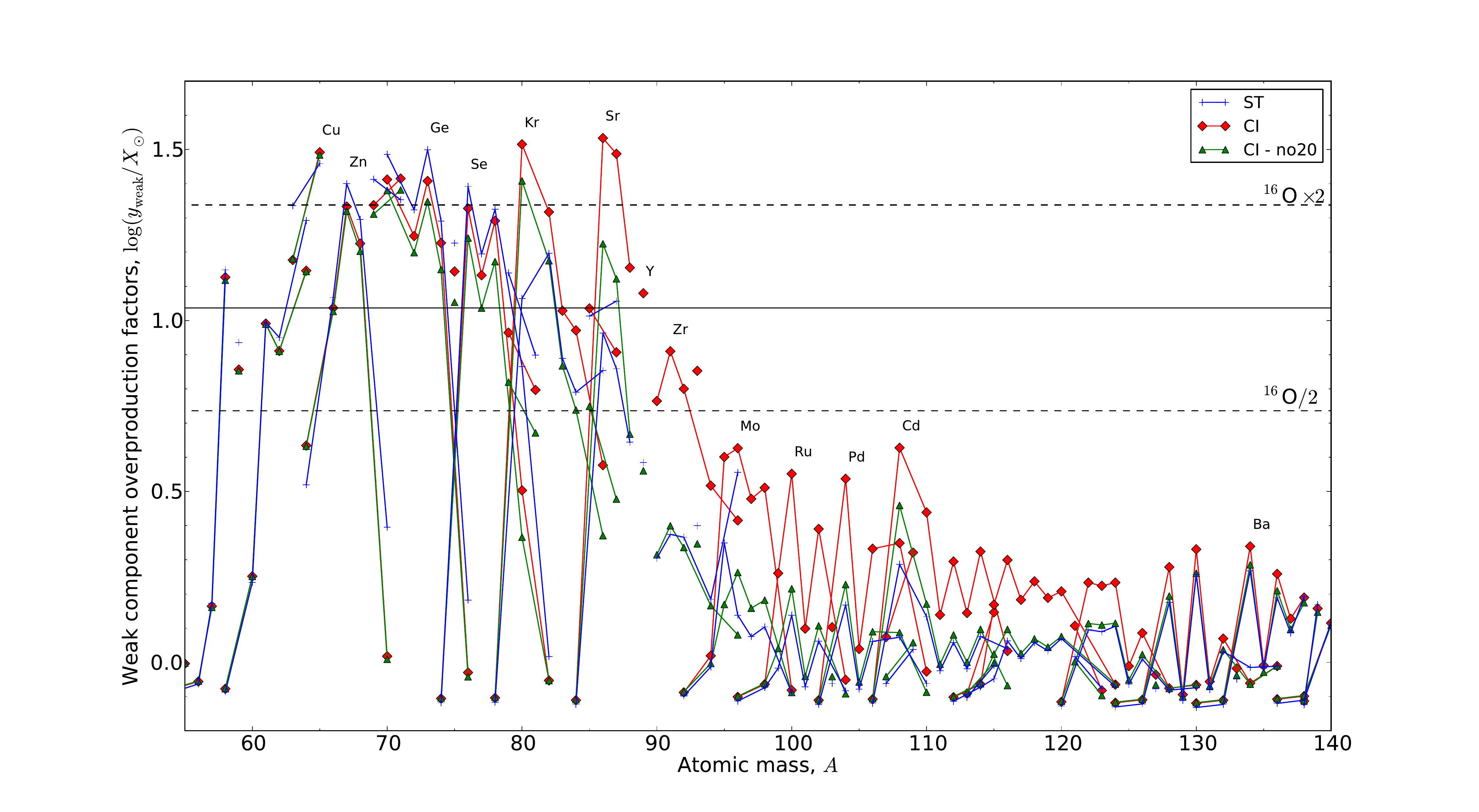}
\caption{The overproduction factors of the predicted weak component relative to the solar system abundances with the $20 M_{\sun}$ models removed from the calculation (CI - no20).  The weak components for the ST and CI case including the 20 $M_{\sun}$ models are included for comparison.  Isotopes of a given element are connected by adjoining lines.  The isotopic chains for Cu, Zn, Ge, Se, Kr, Sr, Y, Zr, Mo, Ru, Pd, Cd and Ba in the CU model are labelled for brevity.}
\label{fig:weakno20}
\end{figure*}

The overproduction factors of Sr, Y, Zr, Mo, Ru, Pd and Cd are enhanced in the carbon-core s-process (for example, see Fig. \ref{fig:15Ccore}).  In the CU case, this occurs for all models other than model 15CU.  In the CI case, the overlap between the convective carbon core and the carbon shell only occurs for model 20CI.  Removing the 20 $M_{\sun}$ models from the evaluation of the weak component allows for a comparison between the predicted weak component with and without the occurrence of an overlap.  Fig. \ref{fig:weakno20} shows the predicted weak component (CI - no20) using the 15, 25, 32 and 60 $M_{\sun}$ models using initial mass ranges of 12.5-20.0, 20.0-28.5, 28.5-46 and 46-80 $M_{\sun}$ in the IMF calculation.  The overproduction factors for the CI - no20 case show a reduction in Sr isotopes to values just less than the $^{16}$O$\times 2$ line and a significant reduction in Y, Zr, Mo, Ru, Pd and Cd isotopes to values similar to the ST case and a reduction in Br and Rb isotopes to values close to the $^{16}$O$/2$ line.  The branching at $^{95}$Zr is also affected, which mainly affects the relative overproduction factors of $^{96}$Zr and $^{95}$Mo.

\section{Discussion}\label{sec:Discussion}

The results in the previous section show that with an increased carbon burning rate, the contribution of the neutron-capture processes during hydrostatic burning stages to the yields of massive stars is modified significantly.

The CU case exhibits a strong production of isotopes between the iron-group nuclides and the Ba-peak nuclides with regards to current massive star models (see Fig. \ref{fig:weakOP}).  This production originates from the s-process production in a convective carbon core in which mixing has caused the ashes of carbon burning to be transported out from the centre of the star where it will be present in the supernova ejecta.  This overlap was found in all but one of the CU models (15CU).  Fig. \ref{fig:weakOP} shows that the yields of the CU case are inconsistent with the weak s-process contribution to the solar system abundances (see for example the anomolously high abundance of Sr-Y-Zr peak and Ba-La peak nuclides compared to those with $60 < A < 90$).  Therefore, a strong resonance with $(\omega \gamma) \simeq 6.8 \times 10^{-5}$ eV at a centre-of-mass energy $E_{\rm com} = 1.5$ MeV in the $^{12}$C + $^{12}$C reaction rate is unlikely to be present in the reaction rate, according to the models used in the present analysis.

For the CI case, an extended distribution is found but the overproduction factors are not as high as the CU case (see Fig. \ref{fig:weakrel}).  The main nucleosynthesis differences occur at the Sr-Y-Zr peak and beyond, which is a signature dominated by the presence of overlap of a carbon shell with the convective carbon core.  The large overproduction of Kr and Sr could suggest that the CI carbon burning rate is too high.  In any case, it is unlikely that a solar metallicity model should demonstrate a strong overlap between the convective carbon core and the carbon-shell of the kind experienced in model 20CI.  However, considering the present uncertainties in the stellar models  such as the reaction rates (for example, the critical reactions $^{12}$C($\alpha,\gamma$)$^{16}$O and $^{22}$Ne($\alpha$,n)$^{25}$Mg), the initial composition and the treatment of convective-radiative boundaries, the abundance of Sr is not a significant enough constraint to assert that the CI rate would be inconsistent with the solar system abundance distribution.

The production of Sr, Y, Zr and other heavier nuclides has been studied extensively as galactic chemical evolution models and observations have suggested the existence of an additional primary nucleosynthesis process, the lighter element primary process (LEPP) \citep[][]{2004ApJ...601..864T,2007ApJ...671.1685M}.  The carbon core s-process and the mixing of heavy nuclei out from the centre could provide an alternative nucleosynthesis scenario for the LEPP.  It is tempting to underline the similarity between the LEPP signature and the anomalous carbon burning s-process component present in the CU models and partly in the CI models.  However, we recall that the LEPP process should be primary if the solar LEPP and (low metallicity) stellar LEPP are indeed the same process \citep[see for example][]{2007ApJ...671.1685M}.  Although the carbon-core s-process features a primary neutron source, $^{13}$C, the seed nuclei, $^{56}$Fe, are secondary.  Consequently, an s-process component using iron seeds in these conditions cannot reproduce the stellar LEPP abundances at low metallicity.  Therefore, the carbon core s-process component is unlikely to represent the site for the stellar LEPP component at low metallicity.  In addition, when the number of seeds is lowered, the neutron captures per iron seed increases (see Eq. \ref{eqn:ncap}) and the distribution of s-process nuclides extends to higher atomic mass.  However, if the solar and stellar LEPPs differ in origin, the carbon core s-process may provide a solution to the solar LEPP.


\section{Summary and conclusions}\label{sec:Conclusions}

In order to investigate the sensitivity of massive star evolution to the potentially large uncertainties in the carbon burning rate, fifteen stellar models with five initial masses of $15, 20, 25, 32$ and $60 M_{\sun}$ and three different carbon burning rates were generated with GENEC and post-processed with the parallel post-processing code MPPNP.  The yields for each model were then calculated and the consequences of the different rates on stellar evolution and nucleosynthesis were examined.  The main conclusions are summarized as follows.

An enhanced carbon burning rate directly affects the ignition conditions for carbon burning, which move to lower temperatures and densities.  The reduced temperature lowers the neutrino losses, causing the carbon burning stage to occur for a longer lifetime.  An increasing dominance of neutrinos formed through photoneutrino interactions is seen, rather than formation by pair-production.  The change in temperature and the neutrino losses affect the convection zone structure.  In the models using the CI rate, the maximum initial mass for the formation of a convective carbon core increases by a few solar masses from its current value of $\approx 22 M_{\sun}$.  In models using the CU rate, carbon core burning occurs in a convective core in the entire mass range.  The increased carbon burning rates generally reduce the number of carbon burning shells (because they have a larger mass extent) and increase the probability of overlap between different convective zones.  Although the increased carbon burning rates used in this study strongly affect carbon burning, the impact on further burning stages (neon, oxygen and silicon) is small and does not present any clear trend.  Therefore, no constraint can be applied to the $^{12}$C + $^{12}$C rate directly from stellar evolution considerations.

The presence of a significant overlap between the convective carbon core and the convective carbon shell, as seen in most of the CU models and in model 20CI may present a further nucleosynthesis site worthy of investigation.  This is especially true considering the present uncertainties in stellar models with regards to convective-radiative boundaries and the abundance distribution exhibited by the carbon core s-process.  In particular, the carbon core s-process bares similarities to the solar LEPP.  However, because of the secondary nature of the iron seeds, it cannot provide a solution to the stellar LEPP at low metallicity.  Further studies into the uncertainties relevant for low metallicity massive stars are required to confirm this statement.
  
According to the present models, a strongly enhanced rate (the CU rate) due to the presence of a low energy resonance (near to the Gamow peak) causes a large convective carbon core to exist in every stellar model.  The large convective core will mix isotopes a considerable distance away from the centre of the star causing the ejecta to be polluted with matter rich in s-process isotopes.  The overabundance distribution obtained with the CU rate is too high and has a vastly different shape.  The yields are therefore incompatible with the weak s-process contribution to the solar system and the CU rate is therefore ruled out.

A moderately enhanced rate (the CI rate), like the strongly enhanced rate, also affects the interior convection zones and consequently the structure of the star.  With the CI rate, an overlap is only present in the $20 M_{\sun}$ case, which enriches the ejecta with products of the carbon core s process.  This enrichment predominantly involves nuclides at the Sr-Y-Zr peak and the heavier elements Mo, Ru, Pd and Cd.  With this additional nucleosynthesis component, the overproduction factor for Kr and Sr seems to be too high to be consistent with the solar system abundances since it would imply that the majority, if not all, of the solar Kr and Sr comes from massive stars, with only a smaller contribution from AGB stars at the Sr peak.  For all the other masses, the changes in nucleosynthesis occur only from changes to carbon-shell burning, which are more subtle and involve isotopes primarily at branching points.  If the contribution from the 20 $M_{\sun}$ model is not included (CI-no20), the yields obtained are very similar to the standard yields.  Consequently, the CI rate is probably very close to the `upper limit' for the carbon burning rate to lead to a weak s-process production compatible with the solar system composition.

Given that an overlap between the convective carbon core and shells has such a strong impact on the yields and that 1D stellar models use the mixing length theory, which might not exactly represent the complex 3D nature of convective-radiative interfaces, it will be crucial to study such potential shell overlaps as well as overlap between burning shells of different burning stages \citep{2011ApJ...733...78A} in 3D hydrodynamic simulations.  It should also be acknowledged that the present conclusions are built on the assumption that the ratio of the $\alpha-$ and p-exit channels of the $^{12}$C + $^{12}$C (13:7) reaction is preserved to lower energies.  Further studies of this uncertainty, including also an analysis of the p-process in massive stars, will be discussed in a forthcoming paper (Pignatari et al. 2011, in prep.).

The effects of the carbon burning rate on the stellar evolution and nucleosynthesis of massive stars demonstrates that nuclear physics experiments investigating the $^{12}$C + $^{12}$C continue to remain relevant for the understanding of stars and further nuclear physics experiments, particularly at energies close to the Gamow peak for hydrostatic carbon fusion, are highly desirable in order to improve stellar models.

NuGrid acknowledges significant support from NSF grant PHY0922648 (Joint Institute for Nuclear Astrophysics, JINA) and EU MIRG-CT-2006-046520.  K.N. and R.H. acknowledge support from the World Premier International Research Center Initiative (WPI Initiative), MEXT, Japan.  R.H. acknowledges support from the STFC (UK).  M.P. acknowledges support from the Ambizione grant of the SNSF (Switzerland).  R.H. and M.P also acknowledge support from the EUROCORE Eurogenesis programme.  F.H. acknowledges NSERC Discovery Grant funding.  The work of C.F. and G.R. was funded in part under the auspices of the National Nuclear Security Administration of the U.S. Department of Energy at Los Alamos National Laboratory and supported by Contract No. DE-AC52-06NA25396.  Computations were performed at the Arizona State University's Fulton High-performance Computing Center (USA), the high-performance computer KHAOS at EPSAM Institute at Keele University (UK) and the CFI (Canada) funded computing resources at the Department of Physics and Astronomy at the University of Victoria.  This work used the SE library (LA-CC-08-057) developed at Los Alamos National Laboratory as part of the NuGrid collaboration; the SE library makes use of the HDF5 library, which was developed by The HDF Group and by the National Center for Supercomputing Applications at the University of Illinois at Urbana-Champaign.


\appendix
\section{Parallel-programming implementation}\label{sec:parallel}

At a particular timestep, the parameters for a 1D spherical shell (or zone) are loaded into memory and a nuclear reaction network is calculated for that zone.  This requires the inverse of a square matrix to be calculated, which has dimensions equal to the number of isotopes included in the network.  For each timestep there are typically $10^3$ zones, dependent on the stellar model and the evolutionary stage of the model, and there are $\sim 10^6$ timesteps per model.  Therefore, the post-processing of a single stellar model requires $\sim 10^9$ nuclear network calculations.  With the nuclear reaction network specified in Table \ref{tab:MPPNPnet} including $\simeq 1.3 \times 10^4$ reactions, the computational expense involved becomes significant; the typical duration of a single MPPNP run on a uniprocessor is approximately $10-12$ months with current serial technology.  Therefore, the application of parallel programming is an absolute necessity to allow the calculations to complete over a reasonable timescale.

The choice of parallelism is a simple master-slave (or Workqueue) strategy where a single, master, processor allocates work to a number of slave processors, which is implemented using the Message Passing Interface (MPI) library routines in Fortran \citep{mpi_330577}.  This is an implementation of parallelism where processors communicate information by passing `messages' to each other with each processor having access to a local, private memory.  The advantage of message passing is the ability to operate on distributed memory resources (such as cluster networks), as well as shared memory resources, and the ability to control explicitly how communications are handled and the parallel behaviour of the program.  It is an embarrassingly parallel program\footnote{An embarrassingly parallel program is one where slave processors are not required to communicate information to each other during the run; the problem can simply be split and allocated in parts to a large number of processors.}, which allows for an efficient parallelisation and reduces dramatically the potential communication overhead.  This was achieved by distributing `work' over mass zones for each timestep, which are calculated independently from each other during the post-processing calculations.  Here, a single unit of `work' is defined as the nuclear reaction network calculation (in flops) for all involved species for a single zone at a particular timestep.

The operation of the parallel program is as follows.  First, the nuclear reaction rates and other global parameters are broadcasted to each slave so that each processor has the required data available in local memory.  Then a loop over timesteps is entered.  For each iteration of the loop a simple first-in first-out (FIFO) scheduler is invoked, which assigns work (in the form of a message containing the temperature, density and abundances) zone-by-zone (from the centre to the surface), first to all assigned processors and then to idle processors as they become available for further work.

Load balancing is important to reduce the impact of idle processors on the performance.  In MPPNP a simple load balancing scheme is specified, where the zones are allocated in order from the centre to the surface.  This choice is made in lieu with the typical distribution of work over the interior of the star at any particular timestep.  The distribution is set by the dynamic network implemented in MPPNP, which adds or removes isotopes from the network calculation depending on the nucleosynthesis flux limits (negligible changes in abundances are ignored to save on unnecessary computation).  In general, the dynamic network assigns more isotopes to zones that have higher temperatures (since higher temperatures increase the nuclear reaction rates) and are convective (since the resultant mixing can cause an increase in the abundance of fuel).  Therefore the distribution has a maximum in the centre and decreases with mass coordinate towards the surface, affected by the presence of convection zones.  However, this is a general case; it is not unusual to have a non-monotonic distribution of work at particular steps throughout the model evolution, especially at the boundaries of convection zones and where neutron sources are efficient.

The parallel burning step is followed by a serial mixing step.  The change in mass fraction abundance of species $i$, $X_i$, over time, $t$, is calculated using the diffusion equation

\begin{equation}\label{eqn:diffmix}
\frac{\partial X_i}{\partial t} = \frac{\partial}{\partial m_r} \left[D (4\pi r^2 \rho)^2 \frac{\partial X_i}{\partial m_r}\right]
\end{equation} 

where $m_r$ is the mass coordinate (at radius $r$), $\rho$ is the density and $D$ is the diffusion coefficient calculated from mixing-length theory (MLT).  The diffusion coefficient is normally large enough ($\sim 10^{16}$ cm$^2$ s$^{-1}$ for hydrogen and helium burning) so that all convection zones, over a timestep $\Delta t$, act to smooth out immediately any sharp changes in abundance associated with concentrated nuclear burning.

\begin{figure}
\includegraphics[width=0.5\textwidth]{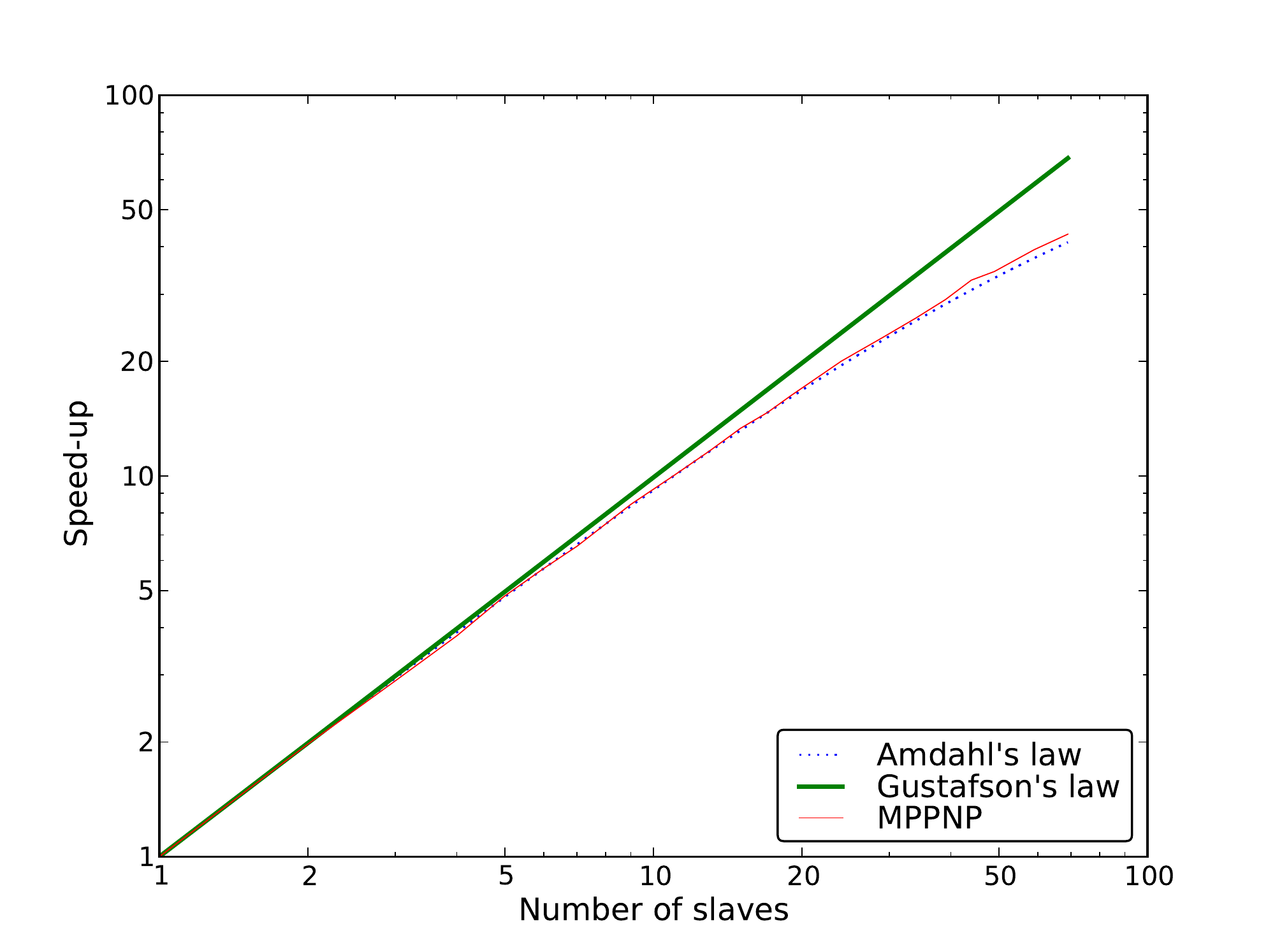}
\caption{Speed-up factor for MPPNP with respect to those of Gustafson's law and Amdahl's law with a serial fraction of 1 per cent.}
\label{fig:speedup}
\end{figure}

Figure \ref{fig:speedup} shows the speed-up factor of MPPNP for a small test run (with $250$ zones and $2000$ timesteps; a typical stellar model uses $\approx 10^3$ zones and $\sim 10^6$ timesteps) compared to the theoretical laws predicted by Amdahl's and Gustafson's law for a program with a serial fraction of 1 per cent.  Amdahl's law,

\begin{equation}\label{eqn:amdahl}
S(p) = \frac{t_s}{t_p} = \frac{t_s}{ft_s + (1-f)t_s/p} = \frac{p}{1+{p-1}f},
\end{equation}

gives the maximum speed-up, $S(p)$, possible for a program with a fixed amount of work, i.e. the time spent running serial computations is constant.  In Eq. \ref{eqn:amdahl}, $t_s$ is the duration of the program with a serial fraction, $f$, on a uniprocessor and $t_p$ is the parallel duration on a system with $p$ processors.  The close fit of this law with MPPNP suggests that the parallelisation is close to the ideal case and is not hampered by communication overhead or excessive initialisation.  However, it would be preferable to achieve a parallelisation comparable to Gustafson's law, 

\begin{equation}\label{eqn:gustafson}
S(p) = \frac{t_s}{t_p} = \frac{ft_p + p(1-f)t_p}{t_p} = p + f(1-p)
\end{equation}

which is the maximum speed-up possible with a constraint on the parallel time, i.e. the time spent running parallel computations is constant.  This could be achieved by including more zones (for example, with the adaptive mesh refinement routine), but the improved scaling would come at the expense of an increased workload.  In any case, only $250$ zones were used in the test case; as the number of slave processors approaches 250, the total number of jobs allocated to each processor approaches unity.  In this regime, the time spent by idle processors is likely to increase significantly and the speed-up factor will plateau.  The post-processing calculations for each model, using 60 slave processors, took approximately $5$ to $10$ days each, depending on the model.

\begin{thebibliography}{}

\bibitem[\protect\citeauthoryear{{Aguilera}}{{Aguilera} et~al.}{2006}]{2006PhRvC..73f%
4601A}
{Aguilera} E.~F. et al., 2006, Phys. Rev. C, 73, 064601

\bibitem[\protect\citeauthoryear{{Aikawa}, {Arnould}, {Goriely}, {Jorissen} \&
  {Takahashi}}{{Aikawa} et~al.}{2005}]{2005A&A...441.1195A}
{Aikawa} M.,  {Arnould} M.,  {Goriely} S.,  {Jorissen} A.,    {Takahashi} K.,
  2005, A\&A, 441, 1195

\bibitem[\protect\citeauthoryear{{Angulo}}{{Angulo} et~al.}{1999}]{1999NuPhA.656....3%
A}
{Angulo} C. e.~a., 1999, Nuc. Phys. A, 656, 3

\bibitem[\protect\citeauthoryear{{Arcoragi}, {Langer} \& {Arnould}}{{Arcoragi}
  et~al.}{1991}]{1991A&A...249..134A}
{Arcoragi} J.,  {Langer} N.,    {Arnould} M., 1991, A\&A, 249, 134

\bibitem[\protect\citeauthoryear{{Arlandini}, {K{\"a}ppeler}, {Wisshak},
  {Gallino}, {Lugaro}, {Busso} \& {Straniero}}{{Arlandini}
  et~al.}{1999}]{1999ApJ...525..886A}
{Arlandini} C.,  {K{\"a}ppeler} F.,  {Wisshak} K.,  {Gallino} R.,  {Lugaro} M.,
   {Busso} M.,    {Straniero} O., 1999, ApJ, 525, 886

\bibitem[\protect\citeauthoryear{{Arnett}}{{Arnett}}{1972}]{1972ApJ...176..699%
A}
{Arnett} W.~D.,  1972, ApJ, 176, 699

\bibitem[\protect\citeauthoryear{{Arnett} \& {Meakin}}{{Arnett} \&
  {Meakin}}{2011}]{2011ApJ...733...78A}
{Arnett} W.~D.,  {Meakin} C.,  2011, ApJ, 733, 78

\bibitem[\protect\citeauthoryear{{Arnett} \& {Thielemann}}{{Arnett} \&
  {Thielemann}}{1985}]{1985ApJ...295..589A}
{Arnett} W.~D.,  {Thielemann} F.,  1985, ApJ, 295, 589

\bibitem[\protect\citeauthoryear{{Arnett} \& {Truran}}{{Arnett} \&
  {Truran}}{1969}]{1969ApJ...157..339A}
{Arnett} W.~D.,  {Truran} J.~W., 1969, ApJ, 157, 339

\bibitem[\protect\citeauthoryear{{Barr{\'o}n-Palos}}{{Barr{\'o}n-Palos}
  et~al.}{2006}]{2006NuPhA.779..318B}
{Barr{\'o}n-Palos} et al., 2006, Nuc. Phys. A, 779, 318

\bibitem[\protect\citeauthoryear{{Bennett}}{{Bennett} et~al.}{2010}]{2010JPhCS.202a2023B}
{Bennett} M.~E. et al., 2010, JPhCS, 202, 012023

\bibitem[\protect\citeauthoryear{{Bennett}}{{Bennett} et~al.}{2010}]{2010arXiv1012.3258B}
{Bennett} M.~E. et al., 2010, Proc. 11th Symp. on Nuclei in the Cosmos. Heidelberg, Germany.

\bibitem[\protect\citeauthoryear{{Betts} \& {Wuosmaa}}{{Betts} \&
  {Wuosmaa}}{1997}]{1997RPPh...60..819B}
{Betts} R.~R.,  {Wuosmaa} A.~H.,  1997, Rep. Prog. Phys., 60, 819

\bibitem[\protect\citeauthoryear{{Busso} \& {Gallino}}{{Busso} \&
  {Gallino}}{1985}]{1985A&A...151..205B}
{Busso} M.,  {Gallino} R.,  1985, A\&A, 151, 205

\bibitem[\protect\citeauthoryear{{Caughlan} \& {Fowler}}{{Caughlan} \&
  {Fowler}}{1988}]{1988ADNDT..40..283C}
{Caughlan} G.~R.,  {Fowler} W.~A.,  1988, ADNDT,
  40, 283

\bibitem[\protect\citeauthoryear{{Chieffi}, {Limongi} \& {Straniero}}{{Chieffi}
  et~al.}{1998}]{1998ApJ...502..737C}
{Chieffi} A.,  {Limongi} M.,    {Straniero} O.,  1998, ApJ,
  502, 737

\bibitem[\protect\citeauthoryear{{Clayton}}{{Clayton}}{1968}]{1968psen.book...%
..C}
{Clayton} D.~D.,  1968, {Principles of stellar evolution and nucleosynthesis}

\bibitem[\protect\citeauthoryear{{Cooper}, {Steiner} \& {Brown}}{{Cooper}
  et~al.}{2009}]{2009ApJ...702..660C}
{Cooper} R.~L.,  {Steiner} A.~W.,    {Brown} E.~F.,  2009, ApJ, 702, 660

\bibitem[\protect\citeauthoryear{{Couch}, {Schmiedekamp} \& {Arnett}}{{Couch}
  et~al.}{1974}]{1974ApJ...190...95C}
{Couch} R.~G.,  {Schmiedekamp} A.~B.,    {Arnett} W.~D.,  1974, ApJ, 190, 95

\bibitem[\protect\citeauthoryear{{Dayras}, {Switkowski} \& {Woosley}}{{Dayras}
  et~al.}{1977}]{1977NuPhA.279...70D}
{Dayras} R.,  {Switkowski} Z.~E.,    {Woosley} S.~E.,  1977, Nuclear Phys. A,
  279, 70

\bibitem[\protect\citeauthoryear{{de Jager}, {Nieuwenhuijzen} \& {van der
  Hucht}}{{de Jager} et~al.}{1988}]{1988A&AS...72..259D}
{de Jager} C.,  {Nieuwenhuijzen} H.,    {van der Hucht} K.~A.,  1988, A\&AS, 72, 259

\bibitem[\protect\citeauthoryear{{Descouvemont}}{{Descouvemont}}{1993}]{1993Ph%
RvC..48.2746D}
{Descouvemont} P.,  1993, Phys. Rev. C, 48, 2746

\bibitem[\protect\citeauthoryear{{Diaz-Torres}}{{Diaz-Torres}}{2008}]{2008PhRv%
L.101l2501D}
{Diaz-Torres} A.,  2008, Phys. Rev. Lett., 101, 122501

\bibitem[\protect\citeauthoryear{{Diaz-Torres}, {Gasques} \&
  {Wiescher}}{{Diaz-Torres} et~al.}{2007}]{2007PhLB..652..255D}
{Diaz-Torres} A.,  {Gasques} L.~R.,    {Wiescher} M.,  2007, Phys. Lett. B,
  652, 255

\bibitem[\protect\citeauthoryear{{Dillmann}, {Heil}, {K{\"a}ppeler}, {Plag},
  {Rauscher} \& {Thielemann}}{{Dillmann} et~al.}{2006}]{2006AIPC..819..123D}
{Dillmann} I.,  {Heil} M.,  {K{\"a}ppeler} F.,  {Plag} R.,  {Rauscher} T.,
  {Thielemann} F.,  2006, in {A.~Woehr \& A.~Aprahamian} ed., Capture Gamma-Ray
  Spectroscopy and Related Topics Vol.~819 of American Institute of Physics
  Conference Series, {KADoNiS- The Karlsruhe Astrophysical Database of
  Nucleosynthesis in Stars}.
pp 123--127

\bibitem[\protect\citeauthoryear{{Eggenberger}, {Meynet}, {Maeder}, {Hirschi},
  {Charbonnel}, {Talon} \& {Ekstr{\"o}m}}{{Eggenberger}
  et~al.}{2008}]{2008ApSS.316...43E}
{Eggenberger} P.,  {Meynet} G.,  {Maeder} A.,  {Hirschi} R.,  {Charbonnel} C.,
  {Talon} S.,    {Ekstr{\"o}m} S.,  2008, Ap\&SS, 316,
  43

\bibitem[\protect\citeauthoryear{{El Eid}, {Meyer} \& {The}}{{El Eid}
  et~al.}{2004}]{2004ApJ...611..452E}
{El Eid} M.~F.,  {Meyer} B.~S.,    {The} L.,  2004, ApJ, 611,
  452

\bibitem[\protect\citeauthoryear{{El Eid}, {The} \& {Meyer}}{{El Eid}
  et~al.}{2009}]{2009SSRv..147....1E}
{El Eid} M.~F.,  {The} L.-S.,    {Meyer} B.~S.,  2009, Space Sci. Rev.,
  147, 1

\bibitem[\protect\citeauthoryear{{Ferguson}, {Alexander}, {Allard}, {Barman},
  {Bodnarik}, {Hauschildt}, {Heffner-Wong} \& {Tamanai}}{{Ferguson}
  et~al.}{2005}]{2005ApJ...623..585F}
{Ferguson} J.~W.,  {Alexander} D.~R.,  {Allard} F.,  {Barman} T.,  {Bodnarik}
  J.~G.,  {Hauschildt} P.~H.,  {Heffner-Wong} A.,    {Tamanai} A.,  2005,
  ApJ, 623, 585

\bibitem[\protect\citeauthoryear{{Fryer}}{{Fryer}}{2009}]{2009APS..APR.B4001F}
{Fryer} C.,  2009, APS April Meeting Abstr., pp B4001+

\bibitem[\protect\citeauthoryear{{Fuller}, {Fowler} \& {Newman}}{{Fuller}
  et~al.}{1985}]{1985ApJ...293....1F}
{Fuller} G.~M.,  {Fowler} W.~A.,    {Newman} M.~J.,  1985, ApJ, 293, 1

\bibitem[\protect\citeauthoryear{{Fynbo}}{{Fynbo}}{2005}]{2005Natur.433..136F}
{Fynbo} H.~O.~U. e.~a.,  2005, Nat, 433, 136

\bibitem[\protect\citeauthoryear{{Gasques}, {Brown}, {Chieffi}, {Jiang},
  {Limongi}, {Rolfs}, {Wiescher} \& {Yakovlev}}{{Gasques}
  et~al.}{2007}]{2007PhRvC..76c5802G}
{Gasques} L.~R.,  {Brown} E.~F.,  {Chieffi} A.,  {Jiang} C.~L.,  {Limongi} M.,
  {Rolfs} C.,  {Wiescher} M.,    {Yakovlev} D.~G.,  2007, Phys. Rev. C,
  76, 035802

\bibitem[\protect\citeauthoryear{{Grevesse} \& {Noels}}{{Grevesse} \&
  {Noels}}{1993}]{1993oee..conf...15G}
{Grevesse} N.,  {Noels} A.,  1993, in {N.~Prantzos, E.~Vangioni-Flam, \&
  M.~Casse} ed., Origin and Evolution of the Elements {Cosmic abundances of the
  elements.}.
pp 15--25

\bibitem[\protect\citeauthoryear{Gropp, Lusk \& Skjellum}{Gropp
  et~al.}{1999}]{mpi_330577}
Gropp W.,  Lusk E.,    Skjellum A.,  1999, Using MPI (2nd ed.): portable
  parallel programming with the message-passing interface.
MIT Press, Cambridge, MA, USA

\bibitem[\protect\citeauthoryear{{Heger}, {Langer} \& {Woosley}}{{Heger}
  et~al.}{2000}]{2000ApJ...528..368H}
{Heger} A.,  {Langer} N.,    {Woosley} S.~E.,  2000, ApJ,
  528, 368

\bibitem[\protect\citeauthoryear{{Herwig}}{{Herwig} et~al.}{2008}]{2008nuco.confE..23H}
{Herwig} F. et al., 2008, in Nuclei in the Cosmos (NIC X)
  {Nucleosynthesis simulations for a wide range of nuclear production sites
  from NuGrid}

\bibitem[\protect\citeauthoryear{{Hirschi}, {Meynet} \& {Maeder}}{{Hirschi}
  et~al.}{2004}]{2004A&A...425..649H}
{Hirschi} R.,  {Meynet} G.,    {Maeder} A.,  2004, A\&A,
  425, 649

\bibitem[\protect\citeauthoryear{{Hirschi}, {Meynet} \& {Maeder}}{{Hirschi}
  et~al.}{2005}]{2005A&A...433.1013H}
{Hirschi} R.,  {Meynet} G.,    {Maeder} A.,  2005, A\&A,
  433, 1013

\bibitem[\protect\citeauthoryear{{Hix}, {Khokhlov}, {Wheeler} \&
  {Thielemann}}{{Hix} et~al.}{1998}]{1998ApJ...503..332H}
{Hix} W.~R.,  {Khokhlov} A.~M.,  {Wheeler} J.~C.,    {Thielemann} F.,  1998,
  ApJ, 503, 332

\bibitem[\protect\citeauthoryear{{Iapichino} \& {Lesaffre}}{{Iapichino} \&
  {Lesaffre}}{2010}]{2010A&A...512A..27I}
{Iapichino} L.,  {Lesaffre} P.,  2010, A\&A, 512, A27+

\bibitem[\protect\citeauthoryear{{Iliadis}, {D'Auria}, {Starrfield}, {Thompson}
  \& {Wiescher}}{{Iliadis} et~al.}{2001}]{2001ApJS..134..151I}
{Iliadis} C.,  {D'Auria} J.~M.,  {Starrfield} S.,  {Thompson} W.~J.,
  {Wiescher} M.,  2001, ApJS, 134, 151

\bibitem[\protect\citeauthoryear{{Imanishi}}{{Imanishi}}{1968}]{1968PL..27B..267Im}
{Imanishi} B., 1968, Phys. Lett., 27B, 267

\bibitem[\protect\citeauthoryear{{Imbriani}, {Limongi}, {Gialanella}, {Terrasi},
{Straniero} \& {Chieffi}}{{Imbriani} et~al.}{2001}]{2001ApJ...558..903I}
{Imbriani}, G., {Limongi}, M., {Gialanella}, L., {Terrasi}, F., {Straniero}, O.,
 {Chieffi}, A., 2001, ApJ, 558, 903

\bibitem[\protect\citeauthoryear{{Imbriani}}{{Imbriani et~al.}}{2005}]{2005EPJA...25..455I}
{Imbriani et al.} G.,  2005, Eur. Phys. J. A, 25, 455

\bibitem[\protect\citeauthoryear{{Itoh}, {Adachi}, {Nakagawa}, {Kohyama} \&
  {Munakata}}{{Itoh} et~al.}{1989}]{1989ApJ...339..354I}
{Itoh} N.,  {Adachi} T.,  {Nakagawa} M.,  {Kohyama} Y.,    {Munakata} H.,
  1989, ApJ, 339, 354

\bibitem[\protect\citeauthoryear{{Itoh}, {Hayashi}, {Nishikawa} \&
  {Kohyama}}{{Itoh} et~al.}{1996}]{1996ApJS..102..411I}
{Itoh} N.,  {Hayashi} H.,  {Nishikawa} A.,    {Kohyama} Y.,  1996,
  ApJS, 102, 411

\bibitem[\protect\citeauthoryear{{Jaeger}, {Kunz}, {Mayer}, {Hammer}, {Staudt},
  {Kratz} \& {Pfeiffer}}{{Jaeger} et~al.}{2001}]{2001PhRvL..87t2501J}
{Jaeger} M.,  {Kunz} R.,  {Mayer} A.,  {Hammer} J.~W.,  {Staudt} G.,  {Kratz}
  K.~L.,    {Pfeiffer} B.,  2001, Phys. Rev. Lett., 87, 202501

\bibitem[\protect\citeauthoryear{{Jiang}, {Esbensen}, {Back}, {Janssens} \&
  {Rehm}}{{Jiang} et~al.}{2004}]{2004PhRvC..69a4604J}
{Jiang} C.~L.,  {Esbensen} H.,  {Back} B.~B.,  {Janssens} R.~V.,    {Rehm}
  K.~E.,  2004, Phys. Rev. C, 69, 014604

\bibitem[\protect\citeauthoryear{{Jiang}, {Rehm}, {Back} \& {Janssens}}{{Jiang}
  et~al.}{2007}]{2007PhRvC..75a5803J}
{Jiang} C.~L.,  {Rehm} K.~E.,  {Back} B.~B.,    {Janssens} R.~V.~F.,  2007,
  Phys. Rev. C, 75, 015803

\bibitem[\protect\citeauthoryear{{K\"appeler}, {Beer} \& {Wisshak}}{{K\"appeler}
  et~al.}{1989}]{1989RPPh...52..945K}
{\"Kappeler} F.,  {Beer} H.,    {Wisshak} K.,  1989, Rep. Prog. Phys., 52, 945

\bibitem[\protect\citeauthoryear{{Kondo}, {Matsuse}, \& {Abe}
  }{1978}]{1978PTP...59..465}
{Kondo}Y.,  {Matsuse} T.,  {Abe}Y., 1978,
  Prog. Theo. Phys, 59, 465

\bibitem[\protect\citeauthoryear{{Kunz}, {Fey}, {Jaeger}, {Mayer}, {Hammer},
  {Staudt}, {Harissopulos} \& {Paradellis}}{{Kunz}
  et~al.}{2002}]{2002ApJ...567..643K}
{Kunz} R.,  {Fey} M.,  {Jaeger} M.,  {Mayer} A.,  {Hammer} J.~W.,  {Staudt} G.,
   {Harissopulos} S.,    {Paradellis} T.,  2002, ApJ, 567,
  643

\bibitem[\protect\citeauthoryear{{Lamb}, {Howard}, {Truran} \& {Iben}
  Jr.}{{Lamb} et~al.}{1977}]{1977ApJ...217..213L}
{Lamb} S.~A.,  {Howard} W.~M.,  {Truran} J.~W.,    {Iben} Jr. I.,  1977,
  ApJ, 217, 213

\bibitem[\protect\citeauthoryear{{Limongi}, {Straniero} \& {Chieffi}}{{Limongi}
  et~al.}{2000}]{2000ApJS..129..625L}
{Limongi} M.,  {Straniero} O.,    {Chieffi} A.,  2000, ApJS, 129, 625

\bibitem[\protect\citeauthoryear{{Maeder}}{{Maeder}}{2009}]{2009pfer.book.....%
M}
{Maeder} A.,  2009, {Physics, Formation and Evolution of Rotating Stars}.
Springer-Verlag Berlin Heidelberg

\bibitem[\protect\citeauthoryear{{Mauron} \& {Josselin}}{{Mauron} \&
  {Josselin}}{2011}]{2011A&A...526A.156M}
{Mauron} N.,  {Josselin} E.,  2011, A\&A, 526, A156+

\bibitem[\protect\citeauthoryear{{Meynet} \& {Maeder}}{{Meynet} \&
  {Maeder}}{2003}]{2003A&A...404..975M}
{Meynet} G.,  {Maeder} A.,  2003, A\&A, 404, 975

\bibitem[\protect\citeauthoryear{{Michaud} \& {Vogt}}{{Michaud} \&
  {Vogt}}{1972}]{1972PhRvC...5..350M}
{Michaud} G.~J.,  {Vogt} E.~W.,  1972, Phys. Rev. C, 5, 350

\bibitem[\protect\citeauthoryear{{Montes}, {Beers}, {Cowan}, {Elliot},
  {Farouqi}, {Gallino}, {Heil}, {Kratz}, {Pfeiffer}, {Pignatari} \&
  {Schatz}}{{Montes} et~al.}{2007}]{2007ApJ...671.1685M}
{Montes} F.,  {Beers} T.~C.,  {Cowan} J.,  {Elliot} T.,  {Farouqi} K.,
  {Gallino} R.,  {Heil} M.,  {Kratz} K.,  {Pfeiffer} B.,  {Pignatari} M.,
  {Schatz} H.,  2007, ApJ, 671, 1685

\bibitem[\protect\citeauthoryear{{Mukhamedzhanov}}{{Mukhamedzhanov} et~al.}{2003}]{2003PhRvC..67f5804M}
{Mukhamedzhanov} A.~M. et al., 2003, Phys. Rev. C, 67, 065804

\bibitem[\protect\citeauthoryear{{Nugis} \& {Lamers}}{{Nugis} \&
  {Lamers}}{2000}]{2000A&A...360..227N}
{Nugis} T.,  {Lamers} H.~J.~G.~L.~M.,  2000, A\&A, 360,
  227

\bibitem[\protect\citeauthoryear{{Oda}, {Hino}, {Muto}, {Takahara} \&
  {Sato}}{{Oda} et~al.}{1994}]{1994ADNDT..56..231O}
{Oda} T.,  {Hino} M.,  {Muto} K.,  {Takahara} M.,    {Sato} K.,  1994, ADNDT, 56, 231

\bibitem[\protect\citeauthoryear{{Perez-Torres}, {Belyaeva} \&
  {Aguilera}}{{Perez-Torres} et~al.}{2006}]{2006PAN....69.1372P}
{Perez-Torres} R.,  {Belyaeva} T.~L.,    {Aguilera} E.~F.,  2006, Phys. Atomic Nuclei, 69, 1372

\bibitem[\protect\citeauthoryear{{Peters}}{{Peters}}{1968}]{1968ApJ...154..225%
P}
{Peters} J.~G.,  1968, ApJ, 154, 225

\bibitem[\protect\citeauthoryear{{Pignatari}, {Gallino}, {Heil}, {Wiescher},
  {K{\"a}ppeler}, {Herwig} \& {Bisterzo}}{{Pignatari}
  et~al.}{2010}]{2010ApJ...710.1557P}
{Pignatari} M.,  {Gallino} R.,  {Heil} M.,  {Wiescher} M.,  {K{\"a}ppeler} F.,
  {Herwig} F.,    {Bisterzo} S.,  2010, ApJ, 710, 1557

\bibitem[\protect\citeauthoryear{{Prantzos}, {Arnould} \&
  {Arcoragi}}{{Prantzos} et~al.}{1987}]{1987ApJ...315..209P}
{Prantzos} N.,  {Arnould} M.,    {Arcoragi} J.,  1987, ApJ,
  315, 209

\bibitem[\protect\citeauthoryear{{Raiteri}, {Busso}, {Picchio} \&
  {Gallino}}{{Raiteri} et~al.}{1991}]{1991ApJ...371..665R}
{Raiteri} C.~M.,  {Busso} M.,  {Picchio} G.,    {Gallino} R.,  1991,
  ApJ, 371, 665

\bibitem[\protect\citeauthoryear{{Raiteri}, {Busso}, {Picchio}, {Gallino} \&
  {Pulone}}{{Raiteri} et~al.}{1991}]{1991ApJ...367..228R}
{Raiteri} C.~M.,  {Busso} M.,  {Picchio} G.,  {Gallino} R.,    {Pulone} L.,
  1991, ApJ, 367, 228

\bibitem[\protect\citeauthoryear{{Rauscher}, {Heger}, {Hoffman} \&
  {Woosley}}{{Rauscher} et~al.}{2002}]{2002ApJ...576..323R}
{Rauscher} T.,  {Heger} A.,  {Hoffman} R.~D.,    {Woosley} S.~E.,  2002,
  ApJ, 576, 323

\bibitem[\protect\citeauthoryear{{Rauscher} \& {Thielemann}}{{Rauscher} \&
  {Thielemann}}{2000}]{2000ADNDT..75....1R}
{Rauscher} T.,  {Thielemann} F.,  2000, ADNDT,75, 1

\bibitem[\protect\citeauthoryear{{Rauscher} \& {Thielemann}}{{Rauscher} \&
  {Thielemann}}{2001}]{2001ADNDT..79...47R}
{Rauscher} T.,  {Thielemann} F.,  2001, ADNDT, 79, 47

\bibitem[\protect\citeauthoryear{{Rogers}, {Swenson} \& {Iglesias}}{{Rogers}
  et~al.}{1996}]{1996ApJ...456..902R}
{Rogers} F.~J.,  {Swenson} F.~J.,    {Iglesias} C.~A.,  1996, ApJ, 456, 902

\bibitem[\protect\citeauthoryear{{Spillane}}{{Spillane} et~al.}{2007}]{2007PhRvL..98l2501S}
{Spillane} T. et al., 2007, Phys. Rev. Lett., 98, 122501

\bibitem[\protect\citeauthoryear{{Strieder}}{{Strieder}}{2008}]{2008JPhG...35a%
4009S}
{Strieder} F.,  2008, J. Phys. G: Nuclear Phys., 35, 014009

\bibitem[\protect\citeauthoryear{{Strieder}}{{Strieder}}{2010}]{2010JPhCS.202a%
2025S}
{Strieder} F.,  2010, JPhCS, 202, 012025

\bibitem[\protect\citeauthoryear{{Terrasi}}{{Terrasi} et~al.}{2007}]{2007NIMPB.259...14T}
{Terrasi} F. et al, 2007, Nuclear Instruments and Methods in
  Physics Research B, 259, 14

\bibitem[\protect\citeauthoryear{{The}, {El Eid} \& {Meyer}}{{The}
  et~al.}{2007}]{2007ApJ...655.1058T}
{The} L.,  {El Eid} M.~F.,    {Meyer} B.~S.,  2007, ApJ, 655,
  1058

\bibitem[\protect\citeauthoryear{{The}, {El Eid} \& {Meyer}}{{The}
  et~al.}{2000}]{2000ApJ...533..998T}
{The} L.-S.,  {El Eid} M.~F.,    {Meyer} B.~S.,  2000, ApJ,
  533, 998

\bibitem[\protect\citeauthoryear{{Tinsley}}{{Tinsley}}{1979}]{1979ApJ...229.10%
46T}
{Tinsley} B.~M.,  1979, ApJ, 229, 1046

\bibitem[\protect\citeauthoryear{{Travaglio}, {Gallino}, {Arnone}, {Cowan},
  {Jordan} \& {Sneden}}{{Travaglio} et~al.}{2004}]{2004ApJ...601..864T}
{Travaglio} C.,  {Gallino} R.,  {Arnone} E.,  {Cowan} J.,  {Jordan} F.,
  {Sneden} C.,  2004, ApJ, 601, 864

\bibitem[\protect\citeauthoryear{{Truran} \& {Cameron}}{{Truran} \&
  {Cameron}}{1971}]{1971Ap&SS..14..179T}
{Truran} J.~W.,  {Cameron} A.~G.~W.,  1971, Ap\&SS, 14,
  179

\bibitem[\protect\citeauthoryear{{Truran} \& {Iben} Jr.}{{Truran} \&
  {Iben}}{1977}]{1977ApJ...216..797T}
{Truran} J.~W.,  {Iben} Jr. I.,  1977, ApJ, 216, 797

\bibitem[\protect\citeauthoryear{{Tur}, {Heger} \& {Austin}}{{Tur}
  et~al.}{2009}]{2009ApJ...702.1068T}
{Tur} C.,  {Heger} A.,    {Austin} S.~M.,  2009, ApJ, 702,
  1068

\bibitem[\protect\citeauthoryear{{Vink}, {de Koter} \& {Lamers}}{{Vink}
  et~al.}{2001}]{2001A&A...369..574V}
{Vink} J.~S.,  {de Koter} A.,    {Lamers} H.~J.~G.~L.~M.,  2001, A\&A, 369, 574

\bibitem[\protect\citeauthoryear{{Woosley}, {Heger} \& {Weaver}}{{Woosley}
  et~al.}{2002}]{2002RvMP...74.1015W}
{Woosley} S.~E.,  {Heger} A.,    {Weaver} T.~A.,  2002, Rev. Mod. Phys., 74, 1015

\bibitem[\protect\citeauthoryear{{Woosley} \& {Weaver}}{{Woosley} \&
  {Weaver}}{1986}]{1986ARA&A..24..205W}
{Woosley} S.~E.,  {Weaver} T.~A.,  1986, ARA\&A, 24, 205

\bibitem[\protect\citeauthoryear{{Woosley} \& {Weaver}}{{Woosley} \&
  {Weaver}}{1995}]{1995ApJS..101..181W}
{Woosley} S.~E.,  {Weaver} T.~A.,  1995, ApJS, 101, 181

\bibitem[\protect\citeauthoryear{{Xu}, {Qi}, {Liotta}, {Wyss}, {Wang}, {Xu} \&
  {Jiang}}{{Xu} et~al.}{2010}]{2010PhRvC..81e4319X}
{Xu} C.,  {Qi} C.,  {Liotta} R.~J.,  {Wyss} R.,  {Wang} S.~M.,  {Xu} F.~R.,
  {Jiang} D.~X.,  2010, Phys. Rev. C, 81, 054319

\bibitem[\protect\citeauthoryear{{Yakovlev}, {Beard}, {Gasques} \&
  {Wiescher}}{{Yakovlev} et~al.}{2010}]{2010PhRvC..82d4609Y}
{Yakovlev} D.~G.,  {Beard} M.,  {Gasques} L.~R.,    {Wiescher} M.,  2010,
  Phys. Rev. C, 82, 044609

\bibitem[\protect\citeauthoryear{{Young} \& {Fryer}}{{Young} \&
  {Fryer}}{2007}]{2007ApJ...664.1033Y}
{Young} P.~A.,  {Fryer} C.~L.,  2007, ApJ, 664, 1033

\bibitem[\protect\citeauthoryear{{Zickefoose}}{{Zickefoose} et~al.}{2010}]{2010nuco.confE..19Z}
{Zickefoose} J. et al., 2010, Proc. 11th Symp. on Nuclei in the Cosmos. Heidelberg, Germany.

\end{thebibliography}
\end{document}